\documentclass[a4paper,12pt]{report}

\usepackage[english]{babel}
\usepackage[T1]{fontenc}
\usepackage[utf8]{inputenc}

\usepackage[a4paper,left=2.2cm,right=2.2cm]{geometry}

\usepackage[dvips]{graphicx}

\usepackage[nodisplayskipstretch]{setspace}

\usepackage{indentfirst}
\usepackage{fancyhdr}
\usepackage{cite}

\usepackage{relsize}

\usepackage{array}
\usepackage{color}
\usepackage{colortbl}

\usepackage{amssymb}
\usepackage{amsmath} 
\usepackage{latexsym}
\usepackage{amsthm} 
\usepackage{eufrak} 
\usepackage[mathscr]{euscript}
\usepackage{amsfonts}
\usepackage{simplewick}
\usepackage{slashed}

\usepackage[section]{placeins}

\usepackage{graphicx}
\usepackage{cleveref}

\usepackage{bbold}

\usepackage{eso-pic,graphicx}
\makeatletter
\newcommand\BackgroundPicture[2]{
\setlength{\unitlength}{1pt}
\put(0,\strip@pt\paperheight){
\parbox[t][\paperheight]{\paperwidth}{
\vfill
\centering\includegraphics[angle=#2]{#1}
\vfill
}
}
}
\makeatother
\usepackage{subfigure}

\renewcommand{\chaptermark}[1]{\markboth{\thechapter.\ #1}{}} 
 
\def \jp {{{\ensuremath{{J/\psi}}}}}
\def \br {{{{\rm BR}}}}
\def\gp  		{\mbox{$G$-parity}}
\def\jp  		{\ensuremath{{J/\psi}}}
\def\pipi  		{\ensuremath{\pi^+\pi^-}}
\def\ee  		{\ensuremath{e^+ e^-}}

\def\a	 		{\ensuremath{\mathcal{A}}}
\def\b 			{\ensuremath{\mathcal{B}}}
\def\bg 		{\ensuremath{{\rm BR}^\gamma}}
\def\bgg 		{\ensuremath{{\rm BR}^{gg\gamma}}}
\def\bggg 		{\ensuremath{{\rm BR}^{ggg}}}
\def\B 			{\ensuremath{\mathcal{B}}}
\def\be			{\begin{eqnarray}}
\def\en			{\end{eqnarray}}
\def\nen		{\nonumber\end{eqnarray}}
\def\no			{\nonumber}
\def\hh			{\hspace{2mm}}

\def\re			{\ensuremath{{\rm Re}}}
\def\im			{\ensuremath{{\rm Im}}}
\def\epg 		{\ensuremath{{\eta'\gamma}}}
\def\ds			{\displaystyle}

\def\lq			{\ensuremath{\left[}}
\def\rq			{\ensuremath{\right]}}
\def\lt			{\ensuremath{\left(}}
\def\rt			{\ensuremath{\right)}}
\def\f		{\ensuremath{f_1(1285)}}
\def\eg 		{\ensuremath{{\eta\gamma}}}
\def\fg 		{\ensuremath{{f_1\gamma}}}
\def\pdg		{Tanabashi:2018oca}
\def\be		{\begin{eqnarray}}
\def\en		{\end{eqnarray}}
\def\nen	{\nonumber\end{eqnarray}}
\def\no		{\nonumber}
\def\lmo		{\left|}
\def\rmo		{\right|}
\def\lt		{\left(}
\def\rt		{\right)}
\def\lq		{\left[}
\def\rq		{\right]}

\def\ds		{\ensuremath{\displaystyle}}
\def\hh		{\hspace{5 mm}}

\def\jp		{\ensuremath{J\!/\psi}}
\def\psii		{\ensuremath{\psi(2S)}}
\def\mj		{\ensuremath{M_{J\!/\psi}}}

\def\LL		{\ensuremath{\Lambda\overline{\Lambda}}}
\def\Ss		{\ensuremath{\Sigma^0\overline{\Sigma}{}^0}}

\def\ov		{\ensuremath{\overline}}
\def\ee		{\ensuremath{e^+e^-}}
\def\BB		{\ensuremath{B\overline{B}}}

\def\bb		{\ensuremath{\mathcal{B}\overline{\mathcal{B}}}}
\def\bb		{\ensuremath{B \overline{B}}}
\def\BB		{\bb}
\def\b		{\ensuremath{\mathcal{B}}}
\def\aggg	{\ensuremath{\mathcal{A}^{ggg}}}
\def\agg	{\ensuremath{\mathcal{A}^{gg\gamma}}}
\def\ag		{\ensuremath{\mathcal{A}^{\gamma}}}
\def \br {{\rm{BR}}}

\def \colormod {black} 
\def \colormodtwo {black} 
\def \colormodthree {black} 
\def \colormodfour {black} 

\DeclareMathOperator*{\SumInt}{%
\mathchoice%
  {\ooalign{$\displaystyle\sum$\cr\hidewidth$\displaystyle\int$\hidewidth\cr}}
  {\ooalign{\raisebox{.14\height}{\scalebox{.7}{$\textstyle\sum$}}\cr\hidewidth$\textstyle\int$\hidewidth\cr}}
  {\ooalign{\raisebox{.2\height}{\scalebox{.6}{$\scriptstyle\sum$}}\cr$\scriptstyle\int$\cr}}
  {\ooalign{\raisebox{.2\height}{\scalebox{.6}{$\scriptstyle\sum$}}\cr$\scriptstyle\int$\cr}}
}

\begin{document}

\begin{titlepage}
\begin{center}

\includegraphics[width=3cm]{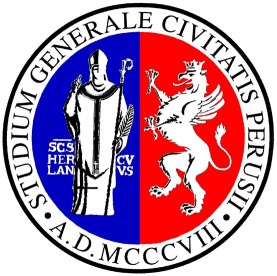}\\[0,7cm]

\textsc{\Large Universit\`a degli Studi di Perugia}\\
\textsc{\large Dipartimento di Fisica e Geologia}\\[0.4cm]
  
\textsc{\Large XXXII Dottorato di Ricerca in Fisica}\\[0,4cm]

\textsc{\small Academic Year 2018/2019}\\[1,5cm]

\textsc{\Large Doctoral Thesis}\\[2.2cm]

{ \huge \bfseries Hadronic decays}\\[0,4cm]
{ \huge \bfseries of the $J/ \psi$ meson}\\[3cm]

\begin{minipage}{0.4\textwidth}
\begin{flushleft} \large
\quad\\
\emph{Candidate:}\\
Alessio Mangoni
\end{flushleft}
\end{minipage}
\begin{minipage}{0.4\textwidth}
\begin{flushright} \large
\emph{Supervisors:} \\
Prof.~Simone Pacetti\\
Prof.~Livio Fanò
\end{flushright}
\end{minipage}

\end{center}
\end{titlepage}

\clearpage\null\thispagestyle{empty}\newpage

%

%
%

%

\clearpage\null\thispagestyle{empty}\newpage

\tableofcontents

%


\chapter*{Introduction} 
\addcontentsline{toc}{chapter}{Introduction}
\chaptermark{Introduction}
This dissertation is devoted to the study of the $J/\psi$ meson and its hadronic decays. Since their discovery $c\overline{c}$ mesons, the so-called charmonia, have been representing unique tools to expand our knowledge on the dynamic of the strong interaction at various energy ranges, for example one of the most challenging questions of our times is to understand the strong interaction in the confinement domain~\cite{Ablikim:2019vaj}.\\
The $J/\psi$ meson high production rate in electron-positron collisions and very large radiative decay rate make it a perfect laboratory for studying exotic hadrons composed of light quarks and gluons, which are the keys to understanding the nature of the strong interaction~\cite{besiiisite}.\\
In this work we propose mainly phenomenological and theoretical models that allow, together with the use of the available experimental data, the calculation of form factors (FFs) and decay amplitudes.\\
{\textcolor{\colormodfour}{In the first chapter we give an overview of the $J/\psi$ meson, including details of its discover and its properties. We calculate and report also some useful results about the $J/\psi$ meson. In this chapter we talk about the BESIII collaboration, as one of the most important experiment for the study of the $J/\psi$ meson.\\
The second and the third chapter are dedicated to the $J/\psi$ decays, respectively, into mesons and baryons, showing our recent results that have been published in refereed journals~\cite{Ferroli:2016jri,Ferroli:2018yad,Ferroli:2019nex}. In particular in the second chapter we focus our attention on the $J/\psi$ decay into a pair of pions, showing that this process does not proceed only electromagnetically as believed so far, due to the presence of a non-negligible mixed strong-EM contribution to the total branching ratio.\\
In the third chapter we consider mainly the decay of the $J/\psi$ meson into a pair of baryon-antibaryon, where we separate, for the first time, the single strong, electromagnetic and strong-EM contributions to the total BR and the relative Feynman amplitudes, obtaining also the relative phase between the strong and the electromagnetic ones.\\
At the end of this thesis there are two appendices, in the first we include notations and some experimental data, while in the second we report some calculations about decay widths and branching ratios.
}}

%

%
%
\chapter[The $J/\psi$ meson]{The $J/\psi$ meson}
\section[The discovery of the $J/\psi$]{The discovery of the $J/\psi$}
The quark model starts in 1964 with the proposal of the quarks existence~\cite{GellMann:1964nj,Zweig:1964jf,Zweig:1965ujx} for a description of the fundamental structure of hadrons in terms of the SU(3) symmetry group. In the same year, various models of strong interaction symmetry were proposed~\cite{Tarjanne:1963zza,Hara:1963gw}. Bjorken and Glashow, in the framework of the ``eightfold way'' idea by Gell-Mann~\cite{GellMann:1961ky,GellMann:1964jm,Neeman:1961jhl,GellMann:1962xb}, proposed the existence of a new quark called ``charm'' ($c$)~\cite{Bjorken:1964gz}. This was the fourth after the ``up'' ($u$), ``down'' ($d$) and ``strange'' ($s$) quarks. The existence of a fourth quark was proposed also in a work of some years later~\cite{Glashow:1970gm}, where it was necessary to explain some anomalies in the kaon decays. A first suggestion for the order of magnitude of the charm quark mass was proposed in 1974 by Gaillard and Lee~\cite{Gaillard:1974hs}, who found a value of around 1-2 GeV.\\
The new quark allowed the possibility of the presence of bound states composed by the quark-antiquark pair. In 1974, in the so-called ``November Revolution'' the first $c \overline c$ bound state, known today as $J/\psi$ meson, was discovered simultaneously by the team of Samuel Ting~\cite{Aubert:1974js} at Brookhaven (they called it $J$), from the reaction
$$
p + {\rm Be} \to e^+ + e^- + {\rm X}
$$
and by the team of Burton Richter~\cite{Augustin:1974xw} at SLAC (they called it $\psi$), with the processes
$$
\ee \to \ee, \, {\rm hadrons}, \, \mu^+ \mu^- \,,
$$
\begin{figure}[ht!]
\centering
\includegraphics[width=0.9\columnwidth]{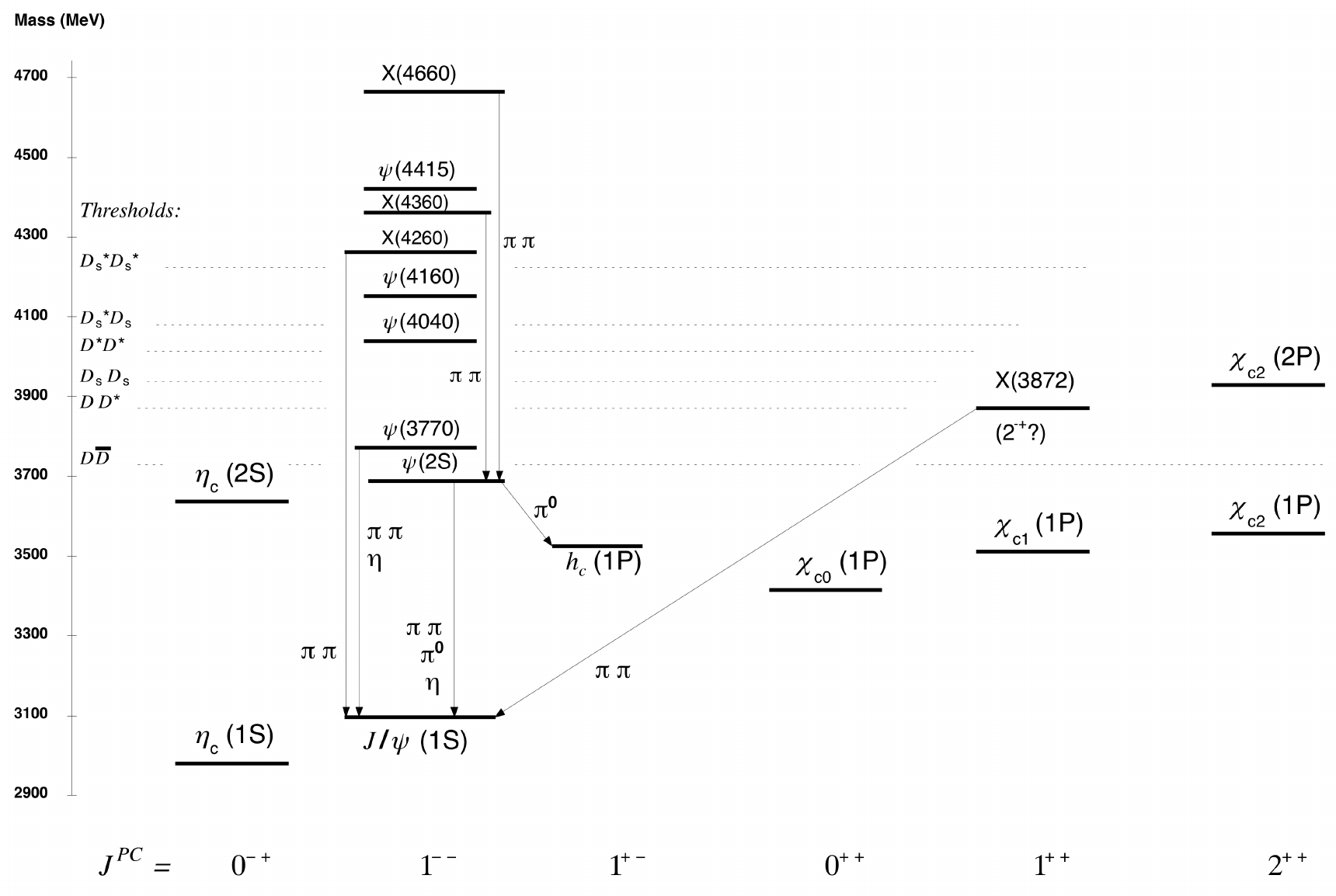}
\caption{The level of charmonia~\cite{Zhu:2012sy}. \label{fig.ccbar_spectroscopy}}
\end{figure}
in both cases as a resonance. There were subsequent confirmations about the anomalous increase in the cross section also at other experiments~\cite{Bacci:1974za,Braunschweig:1974mw}. Samuel Ting and Burton Richter received subsequently the Nobel prize in 1976 for their pioneering work in the discovery~\cite{Nobel1976}. The discovery had a great impact in the community, for many reasons such as its simultaneous discovery in two different laboratories on two entirely different type of machines~\cite{Khare:1999zz}.\\
This resonance, the $\jp$ meson, has a particularly small decay width, of the order of $10^2$ keV, which suggests that decays to lighter hadrons are suppressed. This fact is related to the phenomenological OZI\footnote{by S. Okubo, G. Zweig and J. Iizuka.} rule~\cite{Okubo:1963fa,Zweig:1964jf,Iizuka:1966wu,Okubo:1977rk}, which postulates that processes having Feynman diagrams with disconnected quark lines are suppressed relative to connected ones.\\
\begin{table} [ht!] %
\centering
\caption{\label{tab:charmSpPDG}Principal charmonium states properties~\cite{\pdg}.}
\smallskip
\begin{tabular}{r|r|r|r|r}
\hline\hline\noalign{\smallskip}
 Name & $I^G (J^{PC})$ & $n^{2S+1}L_J$ & Mass (MeV) & Width (MeV) \\
\noalign{\smallskip}\hline\hline\noalign{\smallskip}%
$\eta_c(1S)$ & $0^+(0^{-+})$ & $1^{1}S_0$ & $2983.9 \pm 0.5$ & $32.0 \pm 0.8$\\
\hline
$J/\psi(1S)$ & $0^-(1^{--})$ & $1^{3}S_1$ & $3096.900 \pm 0.006$ & $(92.9 \pm 2.8) \times 10^{-3}$\\
\hline
$\chi_{c0}(1P)$ & $0^+(0^{++})$ & $1^{3}P_0$ & $3414.71 \pm 0.30$ & $10.8 \pm 0.6$\\
\hline
$\chi_{c1}(1P)$ & $0^+(1^{++})$ & $1^{3}P_1$ & $3510.67 \pm 0.05$ & $0.84 \pm 0.04$\\
\hline
$h_c(1P)$ & $?^?(1^{+-})$ & $1^{1}P_1$ & $3525.38 \pm 0.11$ & $0.7 \pm 0.4$\\
\hline
$\chi_{c2}(1P)$ & $0^+(2^{++})$ & $1^{3}P_2$ & $3556.17 \pm 0.07$ & $1.97 \pm 0.09$\\
\hline
$\eta_c(2S)$ & $0^+(0^{-+})$ & $2^{1}S_0$ & $3637.6 \pm 1.2$ & $11.3^{+3.2}_{-2.9}$\\
\hline
$\psi(2S)$ & $0^-(1^{--})$ & $2^{3}S_1$ & $3686.097 \pm 0.025$ & $(294 \pm 8) \times 10^{-3}$\\
\hline
$\psi(3770)$ & $0^-(1^{--})$ & $1^{3}D_1$ & $3773.13 \pm 0.35$ & $27.2 \pm 1.0$\\
\hline
$\chi_{c1}(3872)$ & $0^+(1^{++})$ &  & $3871.69 \pm 0.17$ & $<1.2$\\
\hline
$\chi_{c2}(3930)$ & $0^+(2^{++})$ & $2^{3}P_2$ & $3927.2 \pm 2.6$ & $24 \pm 6$\\
\hline
$\psi(4040)$ & $0^-(1^{--})$ & $3^{3}S_1$ & $4039 \pm 1$ & $80 \pm 10$\\
\hline
$\psi(4160)$ & $0^-(1^{--})$ & $2^{3}D_1$ & $4191 \pm 5$ & $70 \pm 10$\\
\hline
$\psi(4360)$ & $0^-(1^{--})$ & & $4368 \pm 13$ & $96 \pm 7$\\
\hline
$\psi(4415)$ & $0^-(1^{--})$ & $4^{3}S_1$ & $4421 \pm 4$ & $62 \pm 20$\\
\hline
$\psi(4660)$ & $0^-(1^{--})$ & & $4643 \pm 9$ & $72 \pm 11$\\
\noalign{\smallskip}\hline\hline
\end{tabular}
\end{table}%
The $c \overline c$ bound states are called ``charmonium''~\cite{Appelquist:1974yr} in analogy with positronium, the electron-positron ($\ee$) bound state, whose bound-state level structure was similar and the $J/\psi$ was the first charmonium to be discovered.\\
In 1976 a second sharp peak in the cross section for the $\ee \to {\rm hadrons}$ process was found~\cite{Abrams:1974yy}, this was the discovery of the $\psi'(3685)$ charmonium. The spectroscopy of the charmonium family is shown in Fig.~\ref{fig.ccbar_spectroscopy}, while in Table \ref{tab:charmSpPDG} are reported some values of the principal charmonium states~\cite{Metreveli:2007sj}.\\
At that time, the charm quark was found in charm-anticharm bound states. Subsequently, also the lightest charmed mesons, named $D$ mesons, were discovered~\cite{Goldhaber:1994wr}. The $D^0$ meson was found as a resonance in $K^\pm \pi^\mp$ decays~\cite{Goldhaber:1976xn}, while soon after were also discovered the $D^+$, $D^-$ and the excited state $D^*$~\cite{Lansberg:2006dh}.

\section[$J/\psi$ properties]{$J/\psi$ properties}
The $J/\psi$ meson has the following quantum numbers
$$
I^G(J^{PC}) = 0^- ( 1^{--} ) \,,
$$
the parity and charge conjugation eigenvalues are related to the orbital angular momentum ($L$) and spin ($S$) as follows
$$
P=(-1)^{L+1} \,, \ \ \ \ \ C=(-1)^{L+S} \,.
$$
The $\jp$ meson has the same quantum numbers of the photon, as was early investigated~\cite{Braunschweig:1976zw}, in particular with the study of the $\mu^+ \mu^-$ leptonic final state, where it was seen an interference with the non-resonant amplitude. Moreover the observation that decays into an odd number of pions were preferred led to the determination of a negative $G$-parity~\cite{JeanMarie:1975gg}.\\
The $\jp$ mass and decay width are determined experimentally and have the following values
$$
M_{J/\psi} = (3.096916 \pm 0.000011) \ {\rm GeV} \, , \ \ \ \ \ \ \Gamma_{J/\psi} = (92.9 \pm 2.8) \ {\rm keV} \,,
$$
taken from the Particle Data Group (PDG)~\cite{\pdg}. For this reasons the $J/\psi$ is indicated also with the $J/\psi(3100)$ notation, where the particle mass (in MeV) is indicated in parenthesis, or also $J/\psi(1S)$. The $J/\psi$ meson is hindered to decay into mesons which contain a $c$ quark due to its lightness, moreover, as anticipated, the small value of its decay width leads to a suppression of the decays into lighter hadrons.\\
Its mass has been measured firstly with high precision around 1980~\cite{Zholents:1980qu,VanDerMeer:1981gj,Baglin:1987rn}, with methods that overcome the limitations due to the calibration of the absolute energy scale~\cite{Derbenev:1978hv,Jackson:1975vf}. On the other hand the $\jp$ decay width has not been measured directly due to the high energy spread of $\ee$ and $p \overline p$ accelerators, but should be inferred from the integrated leptonic reaction rate and the leptonic BR, assuming lepton universality. The first measures of the $\jp$ decay width had a relative error of about 10\%~\cite{Konigsmann:1987yb,Konigsmann:1986tv}.\\
The charmonium can be considered a non-relativistic system, contrary to what happens in a meson formed by light quarks. In a $c \overline c$ bound state the velocity $v$ of the quark charm, in the center of mass (CM) system, is such that~\cite{Metreveli:2007sj} $v^2/c^2 \sim 0.2$ or~\cite{Brambilla:2004wf} $v^2/c^2 \sim 0.3$, where $c$ is the speed of light in vacuum\footnote{$c = 299 \ 792 \ 458$ m/s~\cite{\pdg}.}. Consequently, the charmonium spectrum can be simply described by the Schr\"{o}dinger equation
$$
\bigg[-{\hbar^2 \over 2\mu} \vec \nabla^{\,2} + V(\vec r) \bigg] \Psi(\vec r) = E \Psi(\vec r) \,,
$$
where $\mu \sim 2 m_c$, being $m_c$ the mass of the charm quark, and with a conventional quarkonium potential composed by the standard color Coulomb potential plus a linear term~\cite{Eichten:1978tg} of the type
\be
\label{eq.V.strong}
V(r) = - {a \over r} + kr \,,
\en
called ``Cornell potential'', where $a$ and $k$ can be found using a fitting procedure on the available data. In Ref.~\cite{Eichten:1980mw} authors found the following best fit values
$$
a=0.520 \,, \ \ \ \ \ k=0.183 \ {\rm GeV^2} \,.
$$
Corrections to the potential can also include, for example, terms for the fine structure or hyperfine interactions~\cite{Barnes:2005pb}. A potential model which incorporates the asymptotic freedom and linear quark confinement in a unified manner with the feature of a minimal number of parameters can be found in Ref.~\cite{Richardson:1978bt}, while various potential models, from many authors, are reported in Refs.~\cite{Bhanot:1978mj,Eichten:1976jk,Celmaster:1977vh,Celmaster:1978jt,Eichten:1974af,Lichtenberg:1978vq,Margolis:1978hc}. The standard potential shown in Eq.~\eqref{eq.V.strong} is similar to that of positronium ($e^+e^-$ bound state), except for the presence of the confinement term, linear in $r$.
\section[Experiments]{Experiments}
The investigation of the properties and decays of the $ J/\psi $ meson is strongly related to its production in a particle accelerator. In particular, the simplest and cleanest way to produce $J/\psi$ mesons is using an $\ee$ collider where the CM energy is fixed at the one corresponding to the $J/\psi$ mass. The first $\ee$ colliders are SPEAR at Stanford~\cite{Richter:1970gf,Allen:1974tv,Biddick:1977sv,BarbaroGaltieri:1977ti}, ADONE at Frascati~\cite{Amman:1965iea,Amman:1969ex,Ash:1974zz}, DORIS at Hamburg~\cite{Steffen:1969zza,LeDuff:1974tw,Besch:1981ka} and DCI at Orsay~\cite{Group:1971zz,Marin:1974ula}. Some of the detectors that provided measurements of the $\jp$ decays are~\cite{Augustin:1974xq,Brandelik:1979hy,Bartel:1976tg,Burmester:1976mn,Oreglia:1981fx,Abrams:1979xk,Augustin:1980ad}:
\begin{itemize}
	\item Mark I (SPEAR);
	\item DASP (DORIS);
	\item DESY (DORIS);
	\item PLUTO (DORIS);
	\item Crystal Ball (SPEAR);
	\item Mark II (SPEAR);
	\item DM2 (DCI);
	\item Mark III (SPEAR);
	\item BES (BEPC).
\end{itemize}
\subsection[The BESIII experiment]{The BESIII experiment}
One of the most important experiment for the study of the $\jp$ meson operates at the Beijing Electron–Positron Collider II (BEPCII), an $\ee$ collider located in Beijing, People's Republic of China at the Institute of High Energy Physics (IHEP). It uses the third generation of the Beijing Electron Spectrometer (BESIII) for the studies of light quarks, charm quarks and $\tau$ physics, being considered a so-called ``charm-tau'' factory~\cite{Zweber:2009qf}. %
\begin{figure} [ht!]
	\begin{center}
		\includegraphics[width=.95\columnwidth]{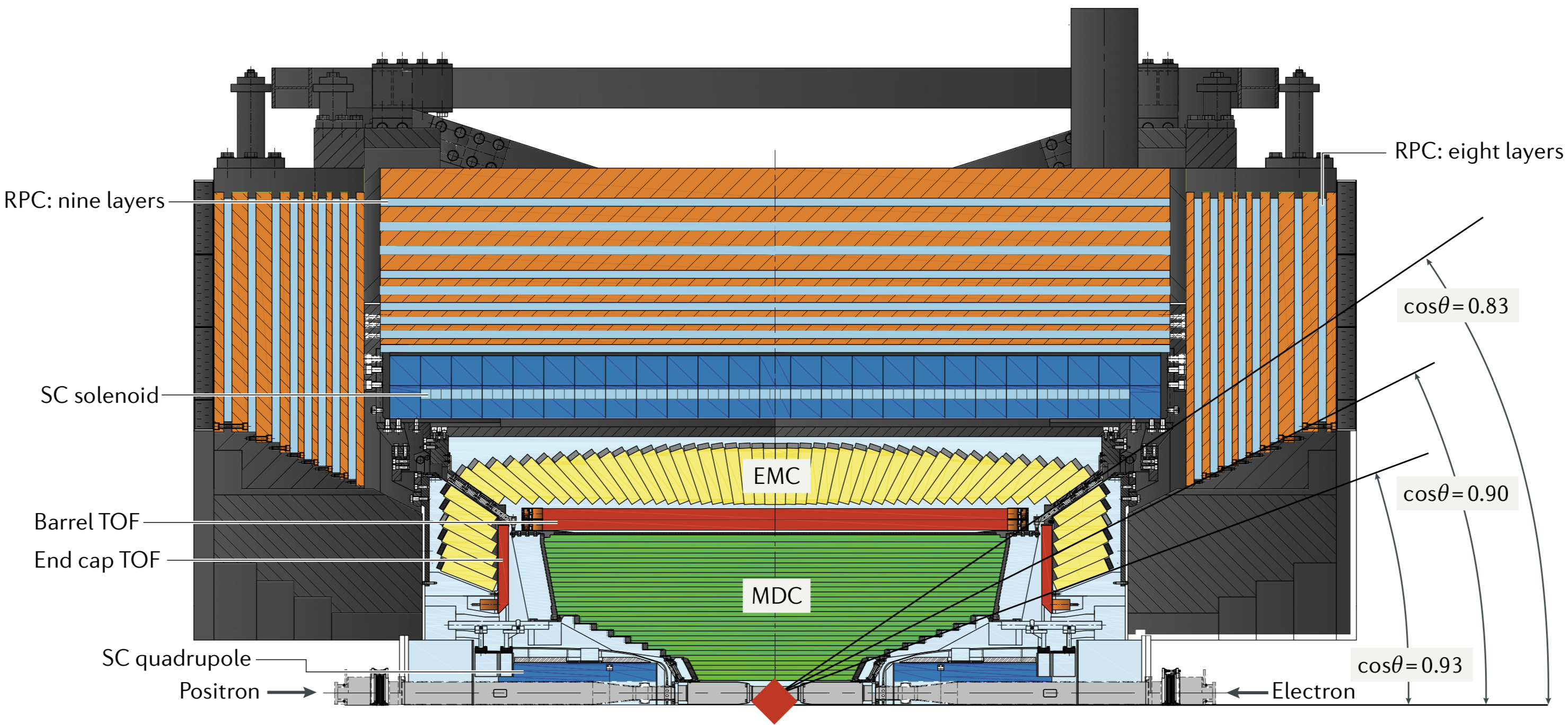}
		\caption{The BESIII detector~\cite{Yuan:2019zfo}.}\label{fig:besiiidetect}
	\end{center}
\end{figure}
In Fig.~\ref{fig:besiiidetect} is reported a scheme of the BESIII detector.\\
The physics program of the BESIII experiment includes: tests of electroweak interactions, studies of light hadron spectroscopy and decay properties, studies of the production and decay properties of the main charmonia, studies of charm and $\tau$ physics, search for glueballs, quark-hybrids, multi-quark states and other exotic states, precision measurements of QCD parameters and CKM parameters and search for new physics~\cite{Ablikim:2009aa}. The BESIII experiment has the capability of adjusting the $\ee$ CM energy to the peaks of resonances and to just above or below the energy thresholds for particle–antiparticle pair formation~\cite{Yuan:2019zfo}. In 2019 the collaboration has about 500 members from 72 institutions in 15 countries.\\
An $\ee$ collider has the advantage that the virtual photon produced by $\ee$ annihilation has the same quantum number of the $\jp$, allowing its direct production and, therefore, precise measurements of its mass and widths. BEPCII is a double ring machine with a design luminosity of about $1 \times 10^{33} \ \rm cm^{-2} \, s^{-1}$ and a CM energy in the range $(2.0 \,\mbox{-}\, 4.6)$ GeV~\cite{Ablikim:2009aa}. Some of its parameters are reported in Table~\ref{tab:BPCIIpar}.\\%
{\textcolor{\colormodthree}{%
In 2019 the BESIII detector finished accumulating a sample of 10 billion $\jp$ that is the world's largest data sample produced directly from $\ee$ annihilations. With 1.3 billion $\jp$ events collected in 2009 and 2012, BESIII has reported many studies and the latest improvements have considerably boosted the sensitivity~\cite{besiiisite}.
}}
\begin{table} [ht!]
\centering
\caption{\label{tab:BPCIIpar}Principal parameters of BEPCII~\cite{Ablikim:2009aa}.}
\smallskip
\begin{tabular}{r|r}
\hline\hline\noalign{\smallskip}
Parameters & BEPCII \\
\noalign{\smallskip}\hline\hline\noalign{\smallskip}%
Center of mass Energy & $(2.0 \div 4.6)$ GeV \\
\hline
Peak luminosity at $2 \times 1.89$ GeV & $\sim 10^{33} \ \rm cm^{-2} \, s^{-1}$ \\
\hline
Circumference & $237.5$ m \\
\hline
Number of rings & $2$ \\
\hline
RF frequency & $499.8$ MHz \\
\hline
Number of bunches & $2 \times 93$ \\
\hline
Beam current & $2 \times 0.91$ A \\
\hline
Bunch spacing & $2.4/8 \ \rm m \, ns^{-1}$ \\
\hline
Bunch length ($\sigma_z$) & $1.5$ cm \\
\hline
Bunch width ($\sigma_x$) & $\sim 380 \ \rm \mu m$ \\
\hline
Bunch height ($\sigma_y$) & $\sim 5.7 \ \rm \mu m$ \\
\hline
Relative energy spread & $5 \times 10^{-4}$ \\
\hline
Crossing angle & $\pm 11$ mrad \\
\noalign{\smallskip}\hline\hline
\end{tabular}
\end{table}
%
\section[$J/\psi$ production in $e^+e^-$ annihilation]{$J/\psi$ production in $e^+e^-$ annihilation}
Concerning the baryon-antibaryon production from an $e^+e^-$ collider, we can find a relation between the continuum and the resonant amplitudes, where the considered intermediate state is the $\jp$ meson. Consider firstly the generic process
\begin{figure} [ht!]
	\centering
	\includegraphics[width=.6\columnwidth]{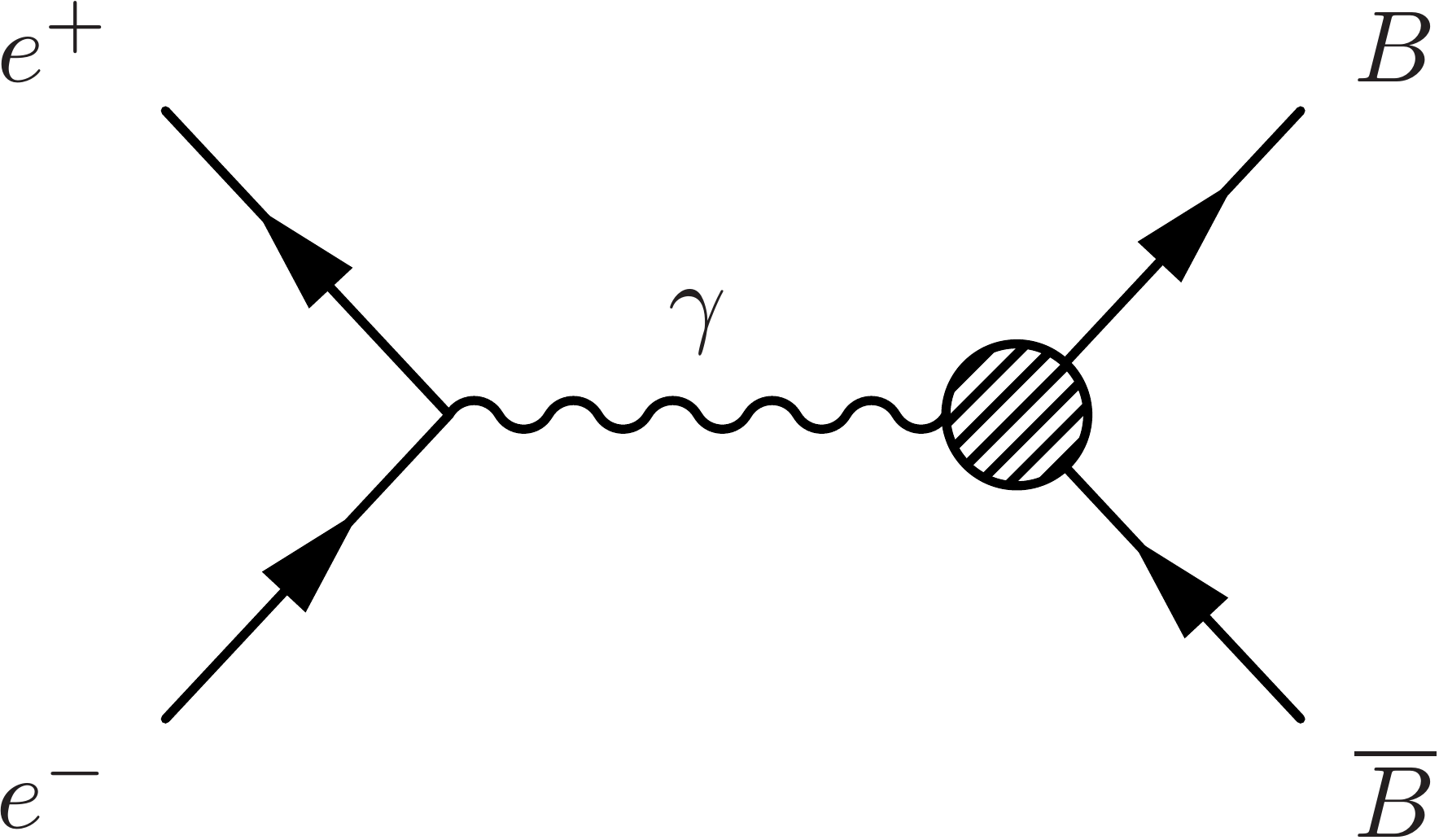}
	\caption{\label{fig.bb.dec2} Feynman diagram for the process $e^+e^- \to B \overline B$.}
\end{figure}
$$
e^- (k_1) e^+ (k_2) \to B (p_1) \overline B (p_2) \,,
$$
where in parenthesis are shown the particle four-momenta and in Fig.~\ref{fig.bb.dec2} is reported the corresponding Feynman diagram. The amplitude can be written in terms of two FFs, $\mathcal F_1^B$ and $\mathcal F_2^B$, that are functions of $q^2$, being $q=p_1+p_2=k_1+k_2$, as
$$
-i \mathcal A (\ee \to \bb) = \overline v (k_2) ie \gamma^\mu u(k_1) \left({-i \eta_{\mu \nu} \over q^2} \right) \overline u(p_1) ie\Gamma^\nu v(p_2) \,,
$$
where $u$ ($v$) is the $e^-$ ($e^+$) spinor, $\gamma^\mu$ are the Dirac matrices, $\Gamma^\mu$ is the baryonic vertex term and $\eta$ is the Minkowski metric tensor with the $(+,-,-,-)$ signature. We have
\be
\mathcal A (\ee \to \bb) &=& -e^2 \overline v (k_2) \gamma^\mu u(k_1) \left({\eta_{\mu \nu} \over q^2} \right) \overline u(p_1) \Gamma^\nu v(p_2) \no \\
&=& -{e^2 \over q^2} \overline v (k_2) \gamma^\mu u(k_1) \overline u(p_1) \Big( \gamma_\mu \mathcal F_1^B + i {\sigma_{\mu \nu} q^\nu \over 2M_B} \mathcal F_2^B \Big) v(p_2) \,,
\nen
where $\ov{v}(k_2)\gamma_\mu u(k_1)$ is the leptonic four-current and $\ov{u}(p_1)\Gamma^\mu v(p_2)$ is the baryonic four-current
\begin{equation}
\label{eq.gammamuB.1.4}
\Gamma^\mu = \gamma^\mu \mathcal F_1^B + i {\sigma^{\mu \nu} q_\nu \over 2M_B} \mathcal F_2^B \,,
\end{equation}
being $M_B$ the baryon mass. In the case of the non-resonant process
$$
\ee \to \gamma^* \to \bb
$$
the two functions $\mathcal F_1^B$ and $\mathcal F_2^B$ are, respectively, the Dirac and Pauli FFs, called $f_1^B$ and $f_2^B$. The previous amplitude becomes
$$
\mathcal A (\ee \to \bb) = -e^2 \overline v (k_2) \gamma^\mu u(k_1) \left({\eta_{\mu \nu} \over q^2} \right) \overline u(p_1) \Big( \gamma^\mu f_1^B + i {\sigma^{\mu \nu} q_\nu \over 2M_B} f_2^B \Big) v(p_2) \,.
$$
In the case of the resonant process with the $\jp$ meson, i.e.,
$$
\ee \to \gamma^* \to \jp \to \bb \,,
$$
the $\mathcal F$ functions include the $\jp$ propagator
$$
\mathcal F_j^B = -i{G_j^{\gamma*} G_j^{\bb} \over q^2 - \mj^2 + i \mj \Gamma_{\jp}} \,, \ \ \ \ \ j=1,2 \,,
$$
where $G_j^\gamma$ and $G_j^{\bb}$ represent, respectively, the coupling constant of the $\jp$ with the photon and of the $\jp$ with the baryon-antibaryon pair. Finally in the case of the process
$$
\ee \to \gamma^* \to \jp \to \gamma^* \to \bb \,,
$$
the FFs are
\be
\mathcal F_j^{B \gamma} &=& \left(-i{G_j^{\gamma*} G_j^\gamma f_j \over q^2 - \mj^2 + i \mj \Gamma_{\jp}} \right)\left({-i \over q^2}\right) \nonumber \\
&=& -{|G_j^{\gamma}|^2 f_j \over q^2(q^2 - \mj^2 + i \mj \Gamma_{\jp})} \,, \ \ \ \ \ j=1,2 \,.
\nen
Summing up all the three contributions (the continuum, the strong-resonant and the EM-resonant) we obtain the total FFs
\be
\mathcal F^{\, \rm tot}_j &=& f_j^B + \mathcal F_j^B + \mathcal F_j^{B \gamma} = f_j^B -i \Bigg({G_j^{\gamma*} G_j^{\bb} -i|G_j^{\gamma}|^2 f_j/q^2 \over q^2 - \mj^2 + i \mj \Gamma_{\jp}} \Bigg) \no \\
&=& f_j^B \Bigg[1 - \Bigg({|G_j^{\gamma}|^2/q^2 +iG_j^{\gamma*} G_j^{\bb}/f_j^B \over q^2 - \mj^2 + i \mj \Gamma_{\jp}} \Bigg) \Bigg] \,, \ \ \ \ \ j=1,2 \,.
\nen
Around the mass of the $\jp$ meson we can write
\be
\mathcal F^{\, \rm tot}_j &=& f_j^B \Bigg[1 - \Bigg({|G_j^{\gamma}|^2/M_{\jp}^2 +iG_j^{\gamma*} G_j^{\bb}/f_j^B \over q^2 - \mj^2 + i \mj \Gamma_{\jp}} \Bigg) \Bigg] \no \\
&=& f_j^B \Bigg[1 - \Bigg({|C_j^\gamma| +|C_j| e^{i \varphi} \over q^2 - \mj^2 + i \mj \Gamma_{\jp}} \Bigg) \Bigg] \,,
\nen
with
$$
|C_j^\gamma| \equiv {|G_j^{\gamma}|^2 \over q^2} \,, \ \ \ \ \ C_j \equiv |C_j| e^{i \varphi} \equiv {iG_j^{\gamma*} G_j^{\bb} \over f_j^B} \,, \ \ \ \ \ \varphi = \arg{\left( {iG_j^{\gamma*} G_j^{\bb} \over f_j^B} \right)} \,,
$$
where $\varphi$ is the relative phase between $C_j^\gamma$ and $C_j$, being $C_j^\gamma$ a real quantity. The Feynman amplitude for the complete process (continuum, EM-resonant and strong-resonant) can be written as
$$
\mathcal A = -{e^2 \over q^2} \overline v (k_2) \gamma^\mu u(k_1) \overline u(p_1) \Big( \gamma_\mu \mathcal F_1^{\, \rm tot} + i {\sigma_{\mu \nu} q^\nu \over 2M_B} \mathcal F_2^{\, \rm tot} \Big) v(p_2) \,.
$$
In the high energy limit, $q^2 >> \Lambda_{\rm QCD}^2$, the baryonic four-current $\ov{u}(p_1)\Gamma^\mu v(p_2)$ tends to $\ov{u}(p_1)\gamma^\mu G_M v(p_2)$, where $G_M$ is the Sachs magnetic FF. The Feynman amplitude becomes
$$
\mathcal A = -{e^2 \over q^2} \overline v (k_2) \gamma^\mu u(k_1) \overline u(p_1) \gamma_\mu v(p_2) G_M \Bigg[1 - \Bigg({|C_j^\gamma| +|C_j| e^{i \varphi} \over q^2 - \mj^2 + i \mj \Gamma_{\jp}} \Bigg) \Bigg] \,,
$$
and we can conclude that the continuum amplitude and the EM amplitude have opposite signs.

%
\section[$J/\psi$ decay mechanisms]{$J/\psi$ decay mechanisms}
\begin{figure} [ht!]
	\centering
\subfigure[The purely strong, $(ggg)^*$, intermediate state.]{%
  \includegraphics[width=.6\textwidth]{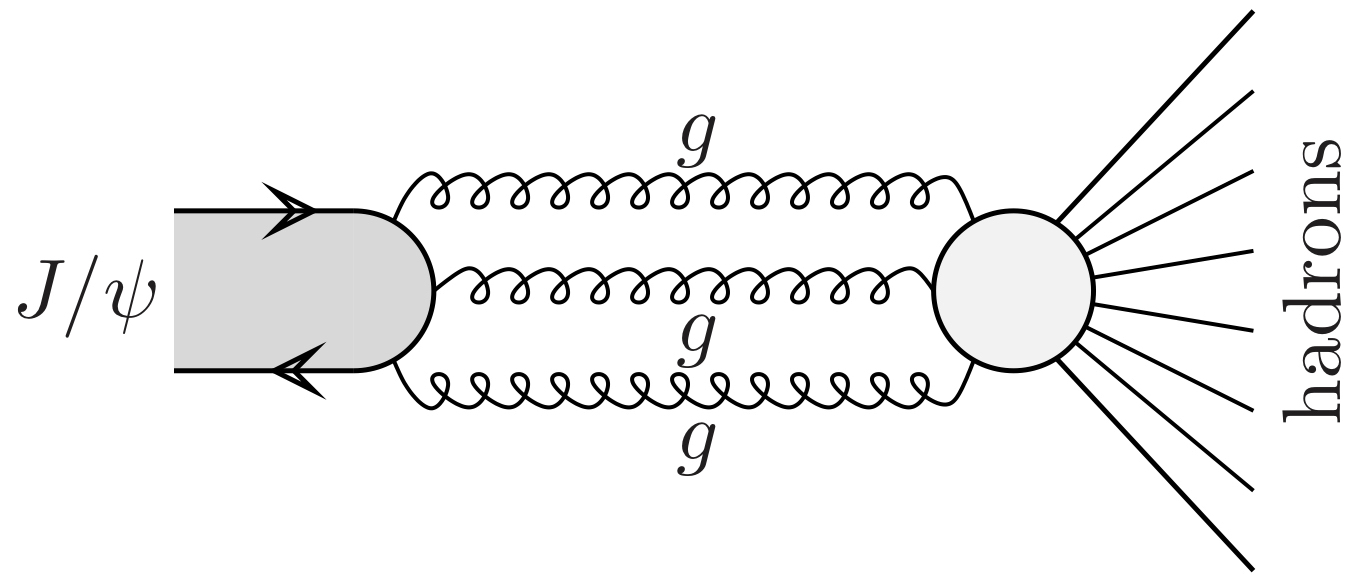}}\\
\subfigure[The mixed strong-EM, $(gg\gamma)^*$, intermediate state.]{%
  \includegraphics[width=.6\textwidth]{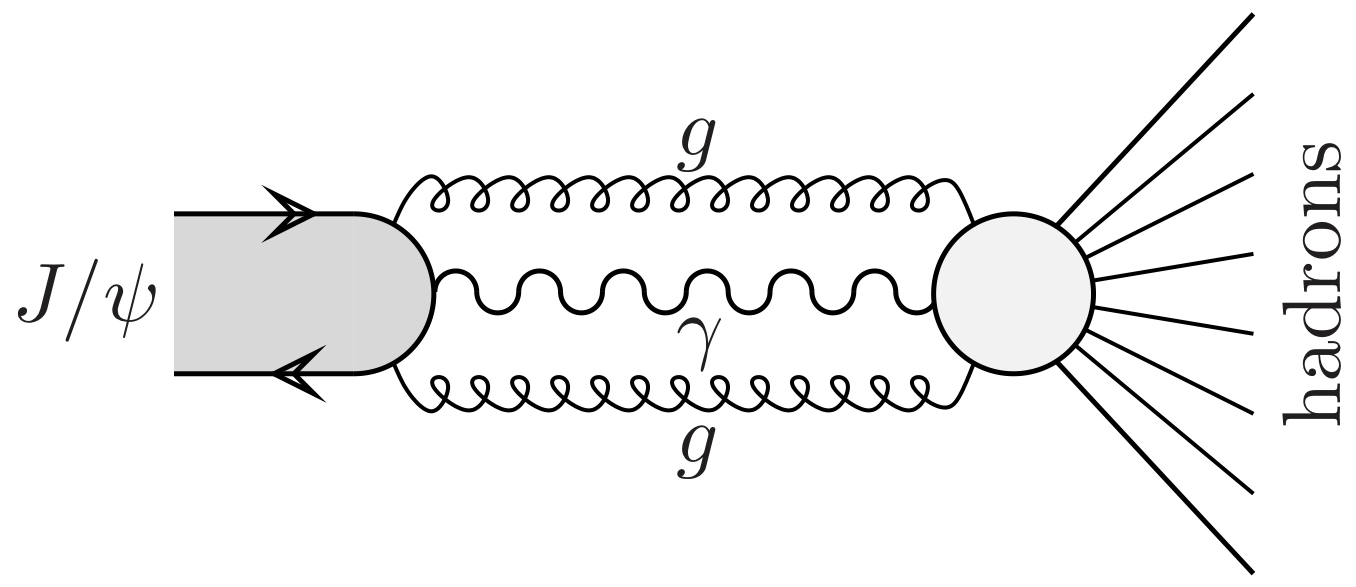}}\\
\subfigure[The purely EM, $(\gamma)^*$, intermediate state.]{%
  \includegraphics[width=.6\textwidth]{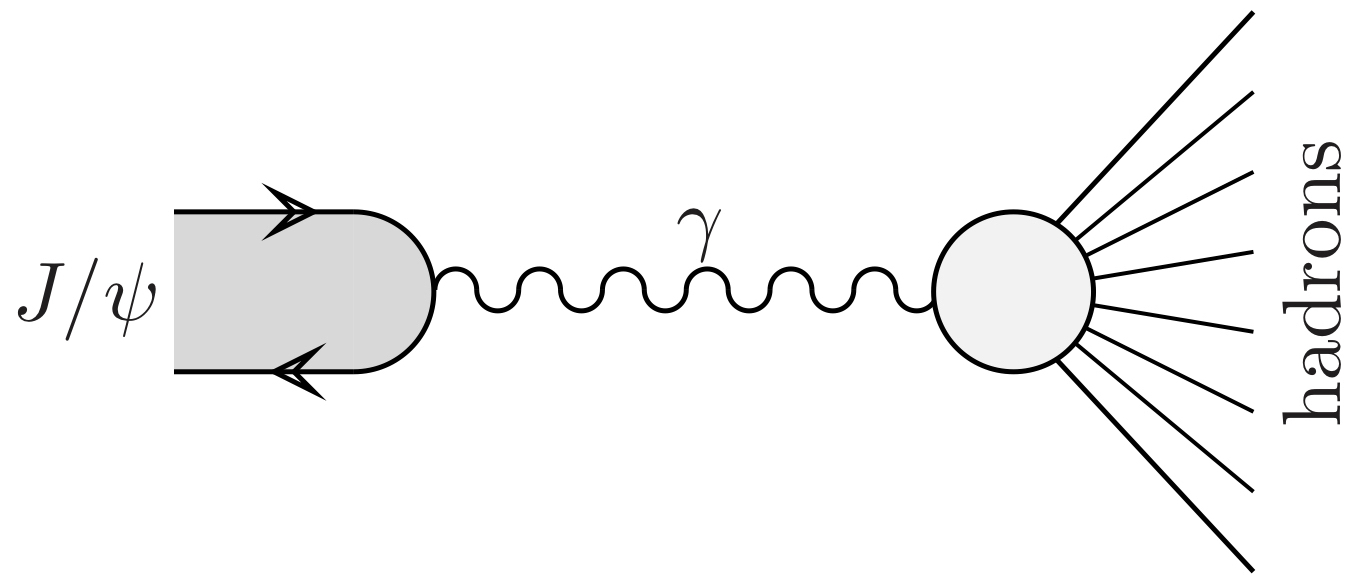}}
	\caption{\label{fig.3.contr} The three principal contributions to the decay $J/\psi \to \mbox{hadrons}$.}
\end{figure}

The theory that describes the strong interaction, the quantum chromodynamics (QCD), is a very powerful theory at high energy. In the region of low and medium energy, calculations based on first principles are very difficult due to the non-perturbative contribution and often models are needed. In particular, charmonium states are on the boundary between perturbative and non-perturbative regimes so their decays, especially the hadronic ones, could be used to study QCD. Generally, the hadronic decays of the $J/\psi$ meson can be parametrized using three principal contributions, whose Feynman diagrams are shown in Fig.~\ref{fig.3.contr}. At leading order, these are characterized by: a three-gluon ($ggg$), a two-gluon-plus-one-photon ($gg \gamma$) and a one-photon ($\gamma$) intermediate states, respectively~\cite{Korner:1986vi,Kopke:1988cs}. The former is related to the purely strong amplitude, while the latter is related to the purely electromagnetic (EM) one. It can be easily observed that in the case of the $gg \gamma$ contribution, see panel (b) of Fig.~\ref{fig.3.contr}, a photon replaces one of the three gluons of the $ggg$ one, shown in panel (a) of the same figure.\\
Consider the generic decay
$$
J/\psi \to \mbox{hadrons} \,,
$$
we can write its Feynman amplitude as the sum~\cite{Claudson:1981fj,Bini:1999nr}
\begin{equation}
\label{eq.3ampl.J}
\mathcal A(J/\psi \to \mbox{hadrons}) = \aggg + \agg + \ag \,.
\end{equation}%
The intermediate state with three photons, that has the same structure of the three-gluon one, is neglected being of order $\alpha^2$ with respect to that with a single photon, being $\alpha$ the fine-structure constant\footnote{$\alpha = 7.297 \ 352 \ 5664(17) \times 10^{-3}$~\cite{\pdg}.}. These three contributions can also interfere, being, in general, complex quantities. %
{\textcolor{\colormodthree}{
Consider, for example, the case of $G$-parity conserving or violating decays. $G$-parity is a multiplicative quantum number defined as the product of charge conjugation and the isospin rotation by $\pi$ radians around the $y$-axis, therefore the corresponding operator can be written as
$$
\hat G = e^{-i \pi \hat I_y} \hat C \,.
$$
}}%
In the case of $G$-parity conserving decays, the first term on the right side of Eq.~\eqref{eq.3ampl.J}, i.e. $\mathcal A^{ggg}$, could be dominant ($\propto \alpha_S^3$, being $\alpha_S$ the QCD coupling constant), while it would be suppressed in presence of $G$-parity violation.\\
We report some experimental values of the $J/\psi$ main decays, from PDG~\cite{\pdg},
$$
{\rm BR}(J/\psi \to {\rm hadrons}) = (87.7 \pm 0.5)\% \, ,
$$
$$
{\rm BR}(J/\psi \to e^+e^-) = (5.971 \pm  0.032)\% \, ,
$$
$$
{\rm BR}(J/\psi \to \mu^+\mu^-) = (5.961 \pm 0.033)\% \, .
$$
The three-gluon and two-gluon-plus-one-photon decay widths are~\cite{Appelquist:1974zd,Appelquist:1974yr,Chanowitz:1975ee,Okun:1976dv,Appelquist:1975ya,Appelquist:1978aq,Mackenzie:1981qa,Kwong:1987mj,Kwong:1987ak}
$$
\Gamma (J/\psi \to ggg) = {160 \over 81} (\pi^2 - 9) {\alpha_S^3 \over M_{J/\psi}^2} \,|\psi_{J/\psi}(0)|^2 \left( 1 + 4.9 \, {\alpha_S \over \pi} \right) \,,
$$
$$
\Gamma (J/\psi \to gg\gamma) = {128 \over 9} (\pi^2 - 9) {\alpha_S^2 \alpha Q_c^2 \over M_{J/\psi}^2} \,|\psi_{J/\psi}(0)|^2 \left( 1 - 0.9 \, {\alpha_S \over \pi} \right) \,.
$$
The EM decay width into leptons is~\cite{Barbieri:1981gj,VanRoyen:1967nq}
\be
\Gamma (J/\psi \to l^+ l^-) = 16 \pi \, {\alpha^2 Q_c^2 \over M_{J/\psi}^2} \,|\psi_{J/\psi}(0)|^2 \left( 1 - {16 \over 3} {\alpha_S \over \pi} \right) \,,
\label{eq.J_l+l-_c}
\en
while the three-photons decay width is
\be
\Gamma(J/\psi \to \gamma \gamma \gamma) = {64 \over 3} (\pi^2 - 9) {\alpha^3 Q_c^6 \over M_{J/\psi}^2} \,|\psi_{J/\psi}(0)|^2 \left( 1 - 12.6 \, {\alpha_S \over \pi} \right) \,.
\label{eq.J_gagaga_c}
\en
In the previous expressions the corrections to $\alpha_S$ at the first order are also included.\\
It is common to define the ratio, called $R$, between the mixed strong-EM amplitude related to the $gg \gamma$ contribution shown in Fig.~\ref{fig.3.contr}, panel (b), and the purely strong one, panel (a). This ratio can be calculated in the framework of perturbative QCD (pQCD), in fact it scales as the ratio of the EM to the strong coupling constant~\cite{Korner:1986vi} and the result is
\be
\label{eq.ratio1}
\lim_{q^2 \to +\infty} R(q^2)
= -{4 \over 5} {\alpha \over \alpha_S(q^2)} \,.	
\en
%
%
%
\section[Value of $|\psi_{J/\psi}(0)|^2$]{Value of $|\psi_{J/\psi}(0)|^2$}
The value of the modulus squared of the radial wave function of the $\jp$ at the origin, $|\psi_{J/\psi}(0)|^2$, can be calculated from Eq.~\eqref{eq.J_l+l-_c}, since it is the expression where quantities are measured with more accuracy. Using numerical values from Ref.~\cite{\pdg}, one obtains
$$
|\psi_{J/\psi}(0)|^2 = (0.0447 \pm 0.0014) \ {\rm GeV^3} \, .
$$
%
\section[$J/\psi$ decays into leptons]{$J/\psi$ decays into leptons}
The value of the BR for the decays of the $J/\psi$ meson into a pair of leptons, $l^+l^-$, from PDG~\cite{\pdg}, is
$$
{\rm BR}(\jp \to l^+l^-) = (11.932 \pm 0.032) \% \,.
$$
We can perform this computation by considering the generic process $c \overline c \to l^+ l^-$
\begin{figure} [ht!]
	\centering
	\includegraphics[width=.6\columnwidth]{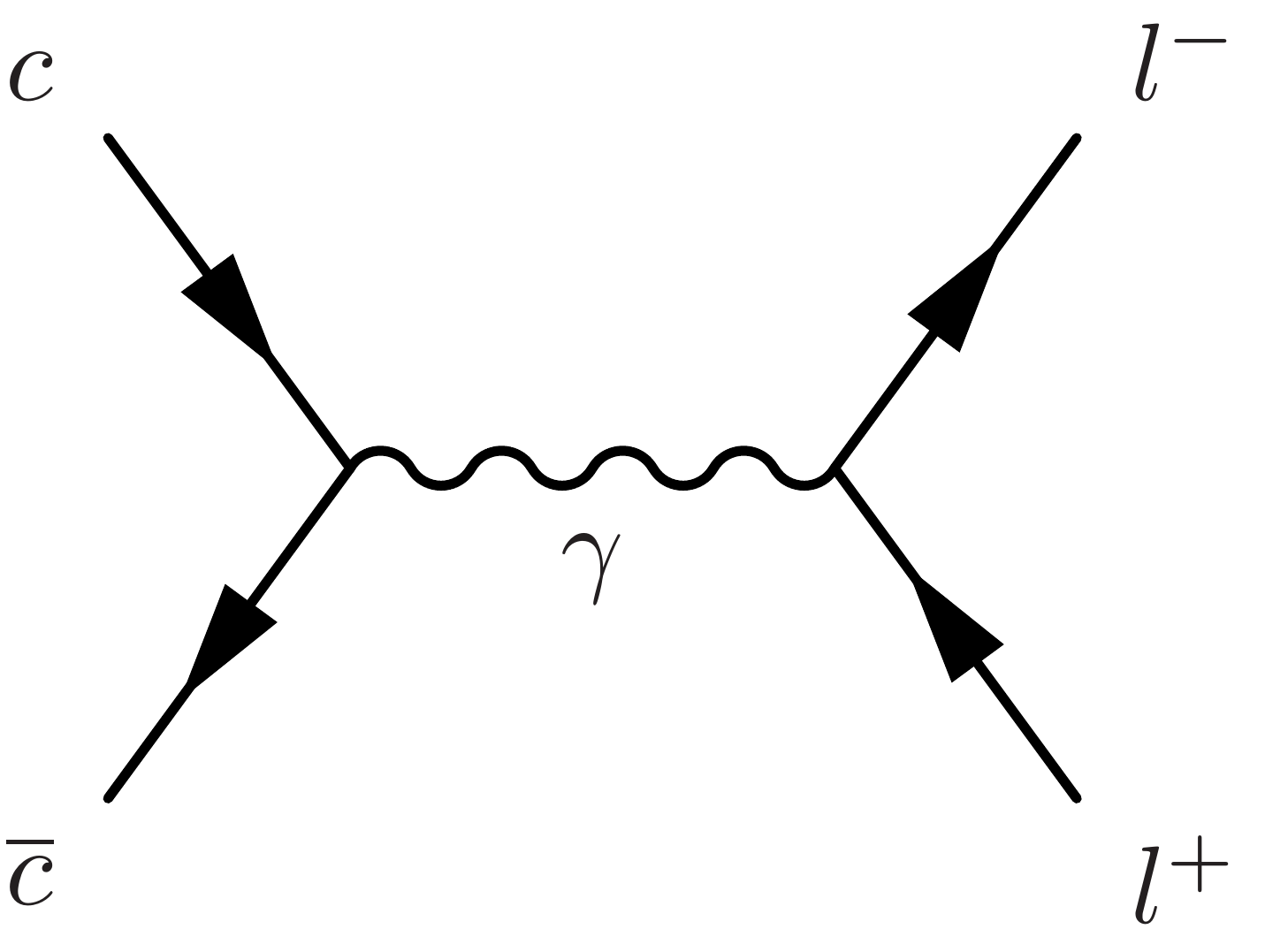}
	\caption{\label{fig.ll.dec} Feynman diagram for the EM scattering process $c \overline c \to l^+ l^-$.}
\end{figure}
and write the amplitude of the Feynman diagram showed in Fig.~\ref{fig.ll.dec},
$$
\mathcal A = i e^2 Q_c \overline u (k_1) \gamma^\mu v (k_2) {\eta_{\mu \nu} \over q^2} \overline v (p_2) \gamma^\nu u (p_1) \,,
$$
where $k_1$, $k_2$ and $p_1$, $p_2$ are, respectively, the four-momenta of the leptons and of the charm quarks, in the CM frame where $p_1=p_2 \equiv p =(m_c,0,0,0)$ with $m_c$ mass of the charm quark, while $q$ is the photon four-momentum, $e$ is the elementary charge\footnote{$e=1.602 \ 176 \ 6208(98) \times 10^{-19}$ C~\cite{\pdg}.} and $Q_c$ is the charm quark charge in units of $e$ ($Q_c=2/3$), with $q^2 =(k_1+k_2)^2 = (2p)^2 = 4m_c^2$. The mean squared modulus of the amplitude is
$$
\overline{| \mathcal A |^2} = {\pi^2 \alpha^2 Q_c^2 \over 4m_c^4} \, {\rm Tr}\Big( \gamma_\mu (\slashed{p}+m_c) \gamma_\nu (\slashed{p}-m_c) \Big) Tr\Big( \gamma^\mu  (\slashed{k}_2-m) \gamma^\nu (\slashed{k}_1+m) \Big) \,,
$$
where $m$ is the mass of the lepton\footnote{$m_e=0.510 \ 998 \ 9461(31)$ MeV, $m_\mu=938.272 \ 0813(58)$ MeV~\cite{\pdg}.} and we have used the expression $e^2 = 4\pi \alpha$. The traces are
\be
{\rm Tr} \Big( \gamma_\mu (\slashed{p}+m_c) \gamma_\nu (\slashed{p}-m_c) \Big) = 8 \Big( p_\mu p_\nu - m_c^2 \eta_{\mu \nu} \Big) \,,
\nen
\be
{\rm Tr}\Big( \gamma^\mu  (\slashed{k}_2-m) \gamma^\nu (\slashed{k}_1+m) \Big) = 4 \Big( k_1^\mu k_2^\nu + k_2^\mu k_1^\nu - k_1 \cdot k_2 \eta^{\mu \nu} - m^2 \eta^{\mu \nu} \Big) \,.
\nen
Neglecting $m$ with respect to $m_c$, we obtain
\be
\overline{| \mathcal A |^2} = {8\pi^2 \alpha^2 Q_c^2 \over m_c^4} \, \Big(2(k_1 \cdot p)(k_2 \cdot p) +m_c^2(k_1 \cdot k_2) \Big) = 32\pi^2 \alpha^2 Q_c^2 \,. \no
\en
where we have used the following expressions
$$
k_1 \cdot k_2 = 2m_c^2 \, , \ \ \ \ \ \ \ k_1 \cdot p = m_c^2 \, , \ \ \ \ \ \ \ k_2 \cdot p = m_c^2 \, .
$$
From Eq.~\eqref{eq.DW.J.cc2mfin} we have, finally,
\be
\Gamma(J/\psi \to l^+l^-) = {64 \pi \over 9} {\alpha^2 \over M_{J/\psi}^2} \, |\psi_{J/\psi}(0)|^2 \,,
\nen
with $m_c=M_{\jp}/2$, as reported in Eq.~\eqref{eq.J_l+l-_c}.
%
\section[$J/\psi$ decays into three photons]{$J/\psi$ decays into three photons}
The $\jp \to \gamma \gamma \gamma$ decay was studied by various experiments, for example the Crystal Ball~\cite{Bloom:1983pc}, and its BR can be calculated from a theoretical point if view. As done for the leptons final state, we can calculate the decay width of $\jp \to \gamma \gamma \gamma$ by considering the process $c \overline c \to \gamma \gamma \gamma$. In the CM frame we have the four-momenta of the charm quarks $p_1=p_2 \equiv p=(m_c,0,0,0)$ with $m_c$ mass of the charm quark. We denote with $k_1, k_2, k_3$ the four-momenta of the final state photons. %
\begin{figure}[ht!]
\centering
\subfigure[$i\mathcal A_1$]{%
  \includegraphics[width=4.5cm]{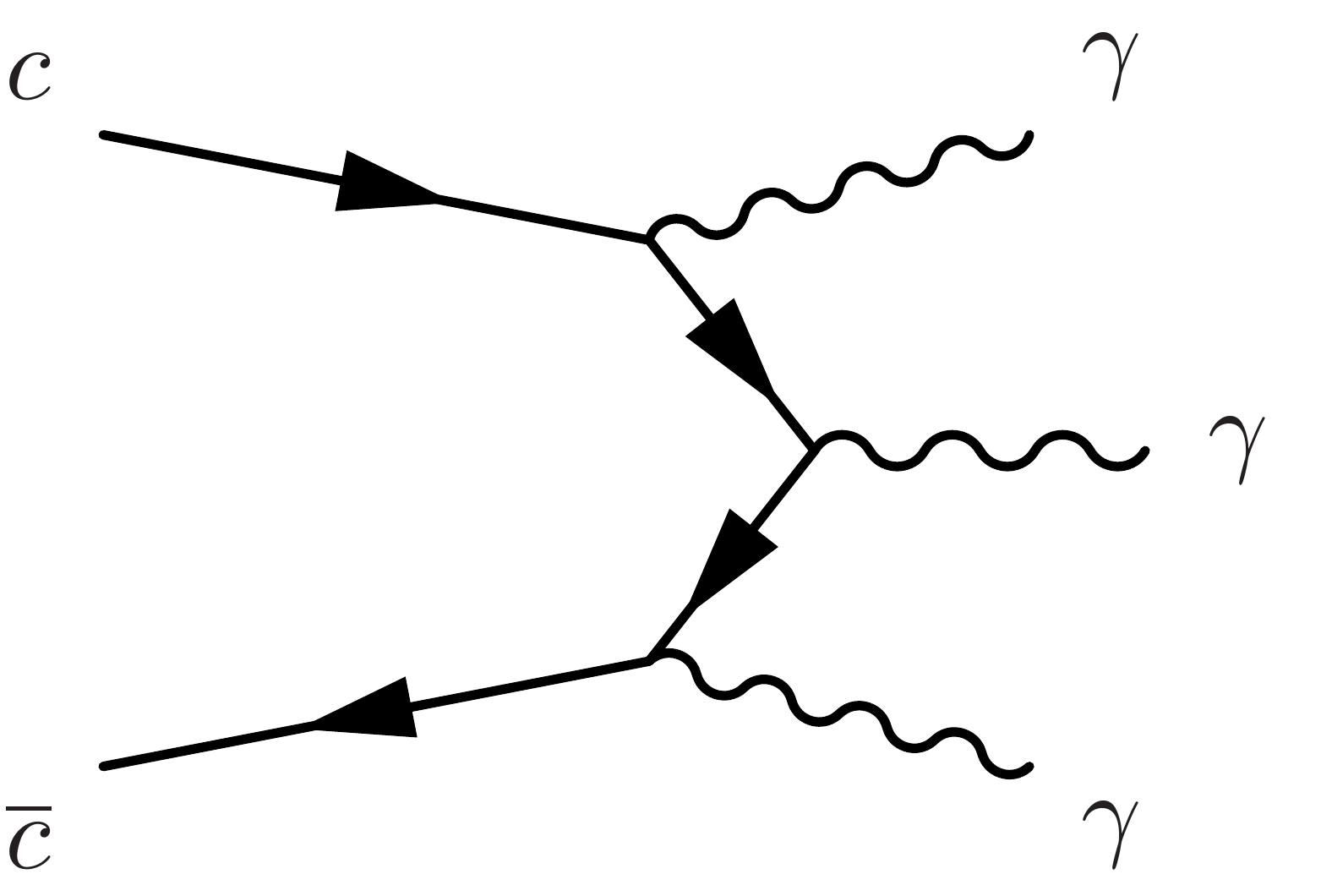}
  \label{fig.feyn.ccbar.to.3gamma.6diag1.sub1}}
  \put(-43,6){{\tiny{$k_3$}}}
  \put(-34,34){{\tiny{$k_2$}}}
  \put(-40,63){{\tiny{$k_1$}}}
\quad
\subfigure[$i\mathcal A_2$]{%
  \includegraphics[width=4.5cm]{figures/ccbar_to_3gamma2.pdf}
  \label{fig.feyn.ccbar.to.3gamma.6diag1.sub2}}
  \put(-43,6){{\tiny{$k_1$}}}
  \put(-34,34){{\tiny{$k_2$}}}
  \put(-40,63){{\tiny{$k_3$}}}
\quad
\subfigure[$i\mathcal A_3$]{%
  \includegraphics[width=4.5cm]{figures/ccbar_to_3gamma2.pdf}
  \label{fig.feyn.ccbar.to.3gamma.6diag1.sub3}}
  \put(-43,6){{\tiny{$k_1$}}}
  \put(-34,34){{\tiny{$k_3$}}}
  \put(-40,63){{\tiny{$k_2$}}}
\\
\subfigure[$i\mathcal A_4$]{%
  \includegraphics[width=4.5cm]{figures/ccbar_to_3gamma2.pdf}
  \label{fig.feyn.ccbar.to.3gamma.6diag1.sub4}}
  \put(-43,6){{\tiny{$k_2$}}}
  \put(-34,34){{\tiny{$k_3$}}}
  \put(-40,63){{\tiny{$k_1$}}}
\quad
\subfigure[$i\mathcal A_5$]{%
  \includegraphics[width=4.5cm]{figures/ccbar_to_3gamma2.pdf}
  \label{fig.feyn.ccbar.to.3gamma.6diag1.sub5}}
  \put(-43,6){{\tiny{$k_3$}}}
  \put(-34,34){{\tiny{$k_1$}}}
  \put(-40,63){{\tiny{$k_2$}}}
\quad
\subfigure[$i\mathcal A_6$]{%
  \includegraphics[width=4.5cm]{figures/ccbar_to_3gamma2.pdf}
  \label{fig.feyn.ccbar.to.3gamma.6diag1.sub6}}
  \put(-43,6){{\tiny{$k_2$}}}
  \put(-34,34){{\tiny{$k_1$}}}
  \put(-40,63){{\tiny{$k_3$}}}
\caption{Feynman diagrams for the process $c \overline c \to \gamma \gamma \gamma$. \label{fig.feyn.ccbar.to.3ph.6diag1_noapp}}
\end{figure}
Since there are three identical photons in the final state we have, at leading order, the $3!=6$ Feynman diagrams shown in Fig.~\ref{fig.feyn.ccbar.to.3ph.6diag1_noapp}.\\
The squared moduli of the total amplitude is
\be
\overline{| \mathcal A |^2} &=& {1 \over 3!}{1 \over 4} \sum_{\rm spin} \sum_{\rm pol} |\mathcal A|^2 = {1024 \pi^3 Q_c^6 \alpha^3 \over 3x^2(2m_c-x-z)^2z^2} \Big(2 m_c^4-6 m_c^3 (x+z) \notag \\
&+& m_c^2 \left(7 x^2+13 x z+7 z^2\right) - m_c \left(4 x^3+9 x^2 z+9 x z^2+4 z^3\right) \notag \\
&+& \left(x^2+x z+z^2\right)^2\Big) \,,
\nen
where $x$, $y$ and $z$ are the energy of the photons, $Q_c=2/3$ is the charm quark charge in units of the elementary charge $e$. The decay width is, using Eq.~\eqref{eq.drho3.sp.fasi.3.inE1E2} and $m_c=M_{\jp}/2$,
$$
\Gamma(J/\psi \to \gamma \gamma \gamma) = {4096 \over 2187} (\pi^2 - 9) {\alpha^3 \over M_{J/\psi}^2} |\psi_{J/\psi}(0)|^2 \,,
$$
as reported in Eq.~\eqref{eq.J_gagaga_c}.
%
\section[$J/\psi$ decays into baryons]{$J/\psi$ decays into baryons}
We consider the decay of a $c \overline c$ vector meson $\psi$ (for example the $\jp$ meson), produced via $\ee$ annihilation, into a baryon-antibaryon pair, \BB, i.e.,
\be
e^-(k_1)+ e^+(k_2)\to \psi(q)\to B(p_1)+\ov{B}(p_2) \,,
\label{eq:ee-jpsi-BB}
\en
where in parentheses are shown the four-momenta.
\begin{figure}
	\begin{center}
		\includegraphics[width=.7\columnwidth]{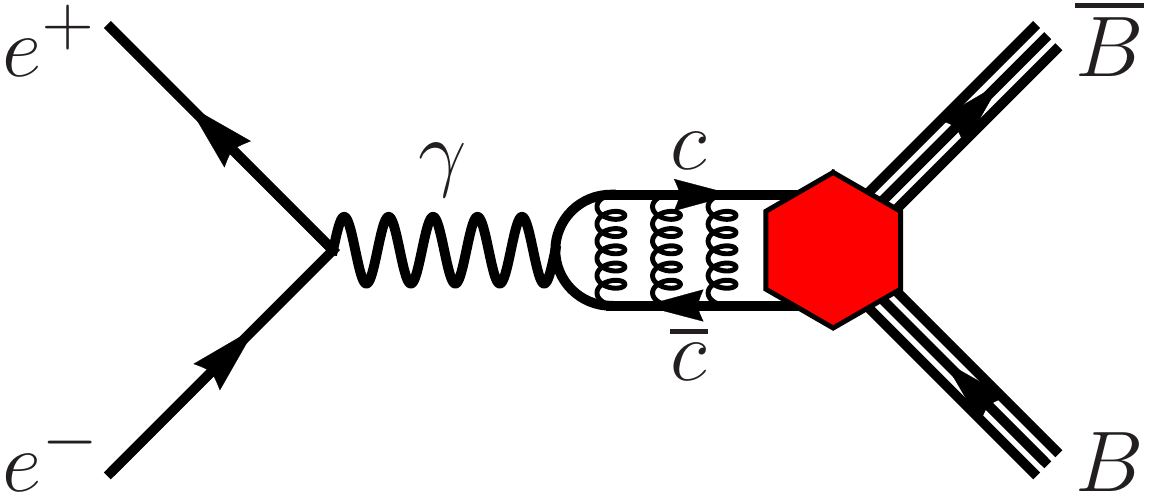}
		\caption{\label{fig:ee-jpsi-bb}Feynman diagram of the process $\ee\to\psi\to\BB$, the red hexagon represents the $\psi\BB$ coupling.}
	\end{center}
\end{figure}
The amplitude of the related Feynman diagram, shown in Fig.~\ref{fig:ee-jpsi-bb}, is  
\be
\mathcal{A}(\ee\to\psi\to\BB)=-ie^2 J_B^\mu 
\ D_{\psi}\!\lt q^2\rt
\, \ov{v}(k_2)\gamma_\mu u(k_1) \,,
\nen
where
$$
J_B^\mu=\ov{u}(p_1)\Gamma^\mu v(p_2) \,,
$$
is the baryonic four-current, $D_{\psi}\lt q^2\rt$ is the $\psi$ propagator, which includes the $\gamma$-$\psi$ EM coupling, and $\ov{v}(k_2)\gamma_\mu u(k_1)$ is the leptonic four-current.
The matrix $\Gamma^\mu$, following Eq.~\eqref{eq.gammamuB.1.4}, is~\cite{Salzman:1955zz}
\be
\Gamma^\mu =\gamma^\mu f_{1}^B +\frac{i\sigma^{\mu\nu}q_\nu}{2M_B} \,f_2^B \,,
\label{eq:Gamma}
\en
where $M_B$ is the baryon mass and, $f_1^B$ and $f_2^B$ are constant FFs called, respectively, strong Dirac and Pauli couplings. The two FFs weight the vector and tensor part of the $\psi \BB$ vertex, where the latter contains also the anomalous magnetic moment. The strong electric and magnetic Sachs couplings~\cite{Ernst:1960zza}, that have the same structure of the EM Sachs FFs, are
\be
g_E^B=f_1^B+\frac{M_{\psi}^2}{4M_B^2}f_2^B \,, \hh \hh \hh
g_M^B=f_1^B+f_2^B \,,
\nen
where $M_\psi$ is the mass of the charmonium state\footnote{The four quantities $f_1^B$, $f_2^B$, $g_E^B$ and $g_M^B$ are in general complex numbers.}. The differential cross section of the process $\ee \to \psi \to B \overline B$, in the $\ee$ CM frame, can be expressed in terms of the two Sachs couplings as follows
\be
\frac{d\sigma}{d\cos\theta}=
\frac{\pi\alpha^2 \beta}{2M_\psi^2} \lt \lmo g_M^B\rmo^2 + \frac{4M_B^2}{M_\psi^2}\lmo g_E^B\rmo^2 \rt \! \Big(1 + \alpha_B \cos^2 \theta \Big) \,,
\nen
where $\beta$ is the velocity\footnote{We recall that $\beta=v/c$ therefore, in natural units where $c=1$, $\beta=v$.} of the baryon in final state, at the $\psi$ mass, that can be written as
\be
\beta=\sqrt{1 - {4 M_B^2 \over M_{\psi}^2}} \,,
\nen
being $\theta$ the scattering angle. Moreover, the polarization parameter $\alpha_B$ depends only on the modulus of the ratio ${g_E^B}/{g_M^B}$ and is given by
\be
\label{Eq.polpar1}
\alpha_B
=\frac{M_\psi^2\lmo g_M^B\rmo^2-4M_B^2|g_E^B|^2}{M_\psi^2\lmo g_M^B\rmo^2+4M_B^2\lmo g_E^B\rmo^2} = \frac{M_\psi^2-4M_B^2 |g_E^B/g_M^B|^2}{M_\psi^2+4M_B^2 |g_E^B/g_M^B|^2} \,,
\en
with $\alpha_B\in[-1,1]$.\\
The behavior of the polarization parameter $\alpha_B$ (as a function of the ratio $|g_E|/|g_M|$) and that of the ratio $|g_E|/|g_M|$ (as a function of $\alpha_B$), in the particular case of $\psi=\jp$, are shown, respectively, in Fig.~\ref{fig:pol.par.J} and Fig.~\ref{fig:pol.par.J_inv}.\\
\begin{figure}[ht!]
\begin{center}
	\includegraphics[width=.7\columnwidth]{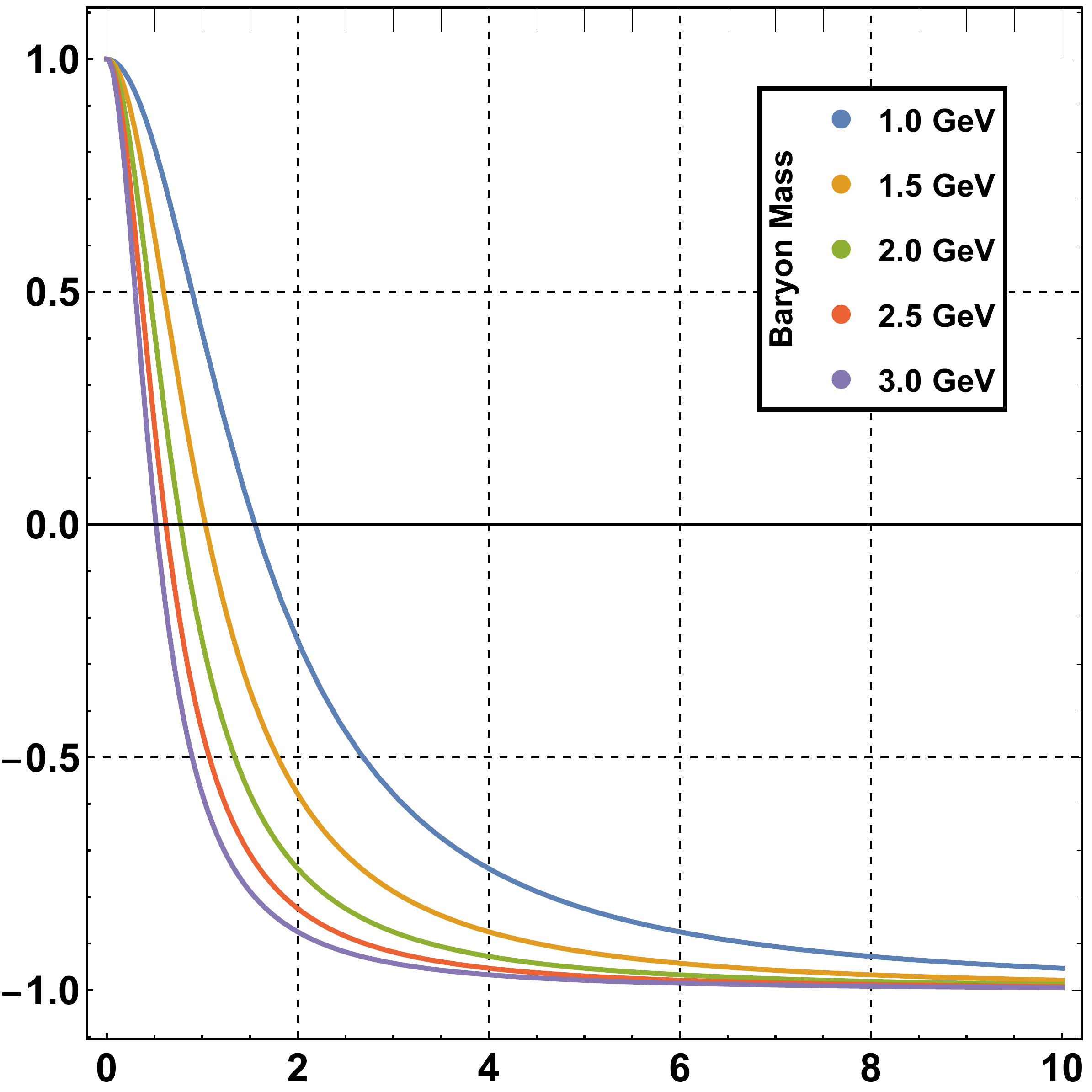}
	\put(-60,-15){\Large $|g_E/g_M|$}
	\put(-346,315){\Large $\alpha_B$}
\caption{\label{fig:pol.par.J} Polarization parameter $\alpha$ for $\psi=J/\psi$ and for various baryon masses from 1.0 GeV to 3.0 GeV, as a function of the ratio $|g_E|/|g_M|$.}
\end{center}
\end{figure}
\begin{figure}[ht!]
\begin{center}
	\includegraphics[width=.7\columnwidth]{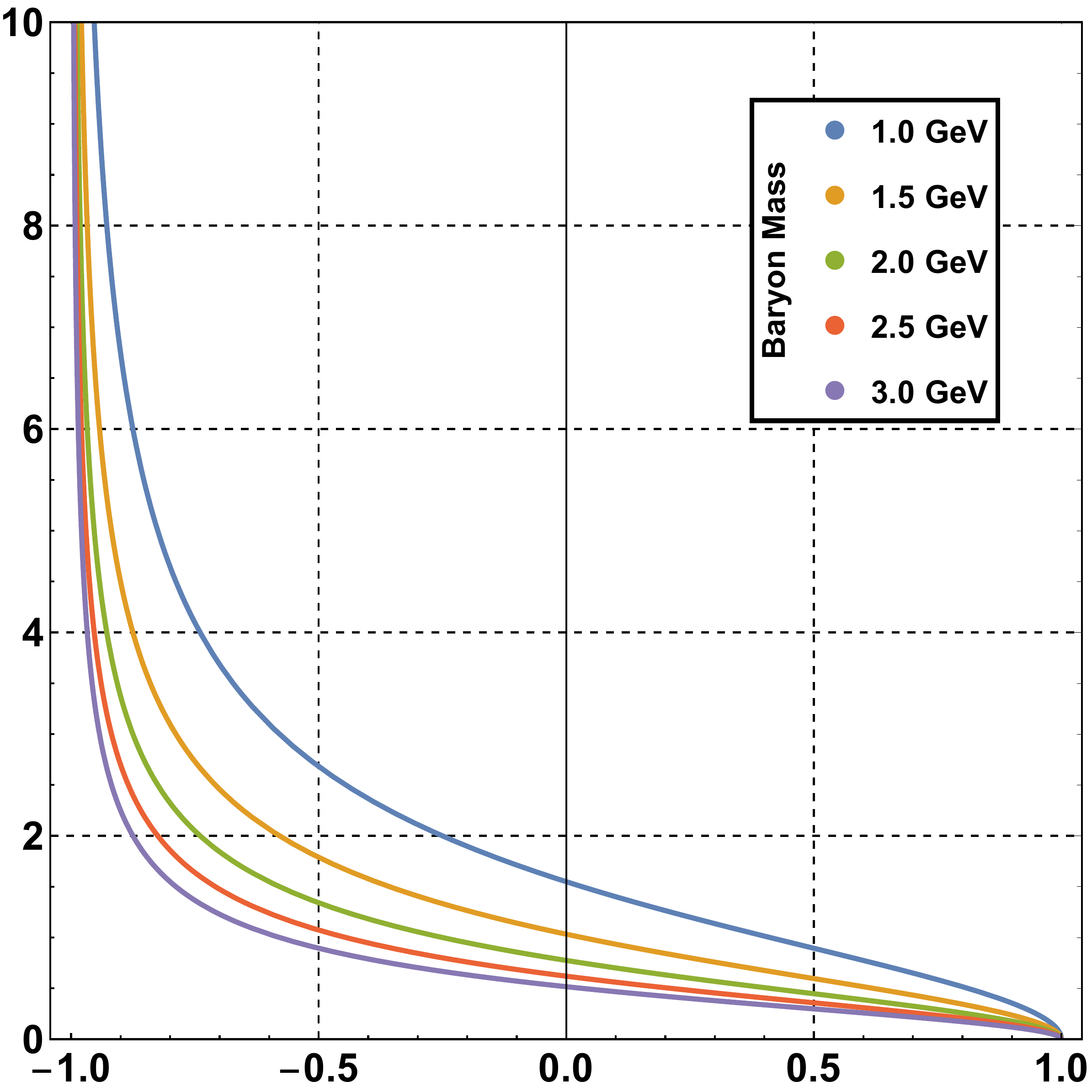}
	\put(-20,-14){\Large $\alpha_B$}
	\put(-362,310){\Large ${|g_E| \over |g_M|}$}
\caption{\label{fig:pol.par.J_inv} Ratio of the moduli of $g_E$ and $g_M$ for $\psi=J/\psi$ and for various baryon masses from 1.0 GeV to 3.0 GeV, as a function of the polarization parameter $\alpha$.}
\end{center}
\end{figure}
There three remarkable cases that can be discussed as examples:
\begin{itemize}
	\item Maximum positive polarization ($\alpha_B=1$)\\
the strong electric Sachs coupling vanishes, i.e.,
$$
\alpha_B=1 \ \ \to \ \ g_E^B=0 \,, \ \ f_1^{B} = -\frac{M_\psi^2}{4M_B^2}f_2^{B} \,,
$$
$$
g_M^B=f_1^B \left( 1 -\frac{4M_B^2}{M_\psi^2} \right) = f_2^B \left( 1 -\frac{M_\psi^2}{4M_B^2} \right) \,,
$$
the relative phase between $f_1^B$ and $f_2^B$ is $i\pi$, and the ratio of the moduli is $M_\psi^2/(4M_B^2)$.
	\item Maximum negative polarization ($\alpha_B=-1$)\\
$$
\alpha_B=-1 \ \ \to \ \ g_M^B=0 \,, \ \
f_1^{B}=-f_2^{B} \,,
$$
$$
g_E^B=f_1^B \left( 1 -\frac{M_\psi^2}{4M_B^2} \right) = f_2^B \left(\frac{M_\psi^2}{4M_B^2} -1 \right) \,,
$$
in this case the strong magnetic Sachs coupling vanishes,
 the relative phase between $f_1^B$ and $f_2^B$ is $-i\pi$ and the ratio of the moduli is one.
	\item No polarization ($\alpha_B=0$)\\
the modulus of the ratio between the Sachs couplings become
\be
\alpha_B=0 \ \ \to \ \ \frac{|g_E^{B}|}{|g_M^{B}|}=\frac{M_\psi}{2M_B} \,.
\nen
\end{itemize}
The Feynman amplitude for the decay $\psi \to \BB$ can be written in terms of the strong magnetic and Dirac FFs as
\be
\mathcal{A}(\psi\to\BB)=-i\epsilon_{\psi}^\mu \, 
\ov{u}(p_1)\Gamma_\mu v(p_2)
\nen
where the matrix $\Gamma_\mu$ is defined in Eq.~\eqref{eq:Gamma}, 
 $\epsilon_{\psi}^\mu$ is the polarization vector of the $\psi$ meson, and the four-momenta follow the labelling of Eq.~\eqref{eq:ee-jpsi-BB}. The branching ratio (BR) is given by the standard form for the two-body decay
\be
\label{eq.BRBBbar11}
\mathcal{B}(\psi\to\BB)=\frac{1}{8\pi \Gamma_\psi}\ov{\lmo\mathcal{A}(\psi\to\BB)\rmo^2}\,\frac{\lmo\vec{p}_1\rmo}{M_{\psi}^2}\,,
\en
where $\Gamma_\psi$ is the total width of the $\psi$ meson. Using the mean value of the modulus squared of the amplitude, written in terms of the Sachs couplings,
\be
\overline{\lmo\mathcal{A}(\psi\to\BB)\rmo^2}&\!=\!&
\frac{4}{3}M_{\psi}^2\lt
|g_M^B|^2+\frac{2M_B^2}{M_{\psi}^2}|g_E^B|^2\rt\,.
\nen
we obtain the BR
\be
\mathcal{B}(\psi\to\BB)=\frac{M_{\psi} \beta}{12 \pi \Gamma_\psi} \lt |g_M^B|^2 + \frac{2M_B^2}{M_{\psi}^2}|g_E^B|^2 \rt .
\label{eq:BR.1}
\en
Since it does not depend on $\alpha_B$, it cannot be used to determine the polarization parameter.\\
The previous expression for the BR can be written as the sum of the moduli squared of two amplitudes 
\be
{{\rm BR}}(\psi\to\BB)=\lmo A^B_M\rmo^2+\lmo A^B_E\rmo^2\,,
\label{eq:rate.A}
\en
where, comparing with Eq.~\eqref{eq:BR.1},
\be
A^B_M=\sqrt{\frac{M_{\psi} \beta}{12 \pi \Gamma_{\psi}}}\, g_M^B
\,,\ \
A^B_E=\sqrt{\frac{M_{\psi} \beta}{6 \pi \Gamma_{\psi}}}\frac{M_B}{M_{\psi}}\, g_E^B
\,. \no
\label{eq:amp-ff}
\en
It follows that the polarization parameter of Eq.~\eqref{Eq.polpar1} can be also written as
\be
\alpha_B = {1-2|A_E^B|^2/|A_M^B|^2 \over 1+2|A_E^B|^2/|A_M^B|^2} \,.
\nen
\subsection{Effective Lagrangian for $\jp \to \BB$} \label{efflagr}
The spin-1/2 baryons are the proton ($p$), the neutron ($n$), the sigma baryons ($\Sigma^\pm,\Sigma^0$), the lambda particle ($\Lambda$) and the $\Xi$ baryons ($\Xi^-, \Xi^0$). Some of their properties are shown in Table~\ref{tab:spin1.2.mass.pdg}. %
\begin{table} [ht!]
\vspace{-2mm}
\centering
\caption{Mass and properties of spin-1/2 baryons from PDG~\cite{Tanabashi:2018oca}.}
\smallskip
\label{tab:spin1.2.mass.pdg} 
\begin{tabular}{l|l|l} 
\hline\hline\noalign{\smallskip}
Baryon & Mass $M$ & Quark content \\
\noalign{\smallskip}\hline\hline\noalign{\smallskip}%
$p$ & $938.2720813(58) \ {\rm MeV}$ & $u \ u \ d$ \\
\hline
$n$ & $939.5654133(58) \ {\rm MeV}$ & $u \ d \ d$ \\
\hline
$\Sigma^0$ & $1192.642(24) \ {\rm MeV}$ & $u \ d \ s$ \\
\hline
$\Sigma^+$ & $1189.37(7) \ {\rm MeV}$ & $u \ u \ s$ \\
\hline
$\Sigma^-$ & $1197.449(30) \ {\rm MeV}$ & $d \ d \ s$ \\
\hline
$\Lambda$ & $1115.683(6) \ {\rm MeV}$ & $u \ d \ s$ \\
\hline
$\Xi^0$ & $1314.86(20) \ {\rm MeV}$ & $u \ s \ s$ \\
\hline
$\Xi^-$ & $1321.71(7) \ {\rm MeV}$ & $d \ s \ s$ \\
\noalign{\smallskip}\hline\hline
\end{tabular}
\end{table}%
They are organized into an octet of SU(3) and we can consider the following baryon matrix
$$
\b=\begin{pmatrix}
\Lambda/\sqrt{6}+\Sigma^0/\sqrt{2} & \Sigma^+ & p\\
\Sigma^- & \Lambda/\sqrt{6}-\Sigma^0/\sqrt{2} & n\\
\Xi^- & \Xi^0 & -2 \Lambda/\sqrt{6}\end{pmatrix} \,.
$$
Since the $J/\psi$ meson is a $c \overline c$ bound state it behaves as a singlet with respect to the SU(3) symmetry group, therefore the leading order Lagrangian density for the decay $J/\psi \to\bb$ should have the invariant form~\cite{Zhu:2015bha}
$$
\mathcal L^0 \propto {\rm Tr}\left(\bb \right) \,.
$$
Terms describing SU(3) symmetry breaking effects can be included to obtain a more complete Lagrangian density~\cite{LopezCastro:1994xw}. We consider, in particular, two types of symmetry breaking sources: the quark mass difference and the EM interaction. The first one can be parametrized by introducing the ``spurion'' matrix~\cite{Zhu:2015bha,Haber:1985cv,Morisita:1990cg}
$$
S_m={g_m \over 3}\begin{pmatrix}
1 & 0 & 0\\
0 & 1 & 0\\
0 & 0 & -2\\
\end{pmatrix} \,,
$$
where $g_m$ is the effective coupling constant. This matrix describes the mass breaking effect due to the $s$ and $u,d$ quarks mass difference related to the term
$$
{2 m_u+m_s \over 3} \overline q q + {m_u-m_s \over \sqrt{3}} \overline q \lambda_8 q \,,
$$
where the SU(2) isospin symmetry is assumed, so that: $m_u=m_d$. The $S_m$ matrix is proportional to the $8\mbox{-th}$ Gell-Mann matrix. The EM breaking effect is related to the fact that the photon-quark coupling constant is proportional to the electric charge, related to the term~\cite{LopezCastro:1994xw}
$$
\mathcal H_{\rm em}={e \over 2}A^\mu \overline q \gamma_\mu \left( \lambda_3 + {\lambda_8 \over \sqrt 3} \right) q \,,
$$
where $e$ is the elementary charge, $A^\mu$ is the electromagnetic four-potential and $\lambda_3$ is the $3\mbox{-rd}$ Gell-Mann matrix. This effect can be parametrized using the spurion matrix
$$
S_e={g_e \over 3}\begin{pmatrix}
2 & 0 & 0\\
0 & -1 & 0\\
0 & 0 & -1\\
\end{pmatrix} \,,
$$
where $g_e$ is the EM effective coupling constant.
\\
The most general SU(3)-invariant effective Lagrangian density, which accounts for these effects, is
\begin{eqnarray}
\label{eq.dens.lagr}
\mathcal L &=& g \, {\rm Tr}(\bb) + d \, {\rm Tr}\big(\{\b, \overline \b \} S_e \big)+f \, {\rm Tr}\big([\b, \overline \b ] S_e \big) \nonumber \\
&&+ d' \, {\rm Tr}\big(\{\b, \overline \b \} S_m \big)+f' \, {\rm Tr}\big([\b, \overline \b ] S_m \big) \,, 
\end{eqnarray}
where $g,d,f,d',f'$ are coupling constants. 
\\
By considering single $\bb$ final states, the complete Lagrangian density can be written as the sum of seven contributions
\be
\mathcal L = \mathcal L_{\Sigma^0 \Lambda} + \mathcal L_{p} + \mathcal L_{n} + \mathcal L_{\Sigma^+} + \mathcal L_{\Sigma^-} + \mathcal L_{\Xi^0} + \mathcal L_{\Xi^-} \,,
\nen
with the following sub-Lagrangian density, for the $\Sigma^0$ and $\Lambda$ hyperons,
\be
\label{eq.Lagr.dens.sig.lam}
\mathcal L_{\Sigma^0 \Lambda} &=& \left(g+{1 \over 3}dg_e+{2 \over 3}d'g_m \right) \Sigma^0 \overline \Sigma{}^0 + \left(g-{1 \over 3}dg_e-{2 \over 3}d'g_m \right) \Lambda \overline \Lambda \no \\
&+& \left({\sqrt{3} \over 3} d g_e \right) \Sigma^0 \overline \Lambda + \left({\sqrt{3} \over 3} d g_e \right) \Lambda \overline \Sigma{}^0 \,,
\en
for the nucleons
\begin{gather}
\mathcal L_{p} = \left( g+{1 \over 3}d g_e+fg_e-{1 \over 3}d'g_m+f'g_m \right) p \overline p \,, \no \\
\mathcal L_{n} = \left( g-{2 \over 3}d g_e-{1 \over 3}d'g_m+f'g_m \right) n \overline n \,, \no 
\end{gather}
for the charged $\Sigma$ hyperons
\begin{gather}
\mathcal L_{\Sigma^+} = \left( g+{1 \over 3}d g_e+fg_e+{2 \over 3}d'g_m \right) \Sigma^+ \overline \Sigma{}^- \,, \no \\
\mathcal L_{\Sigma^-} = \left( g+{1 \over 3}d g_e-fg_e+{2 \over 3}d'g_m \right) \Sigma^- \overline \Sigma{}^+ \,, \no
\end{gather}
and, finally, for the $\Xi$ hyperons
\begin{gather}
\mathcal L_{\Xi^0} = \left( g-{2 \over 3}d g_e-{1 \over 3}d'g_m - f'g_m \right) \Xi^0 \overline \Xi{}^0 \,, \no \\
\mathcal L_{\Xi^-} = \left( g+{1 \over 3}d g_e-fg_e-{1 \over 3}d'g_m - f'g_m \right) \Xi^- \overline \Xi{}^+ \,. \no
\end{gather}
%

%
%
%
%
\chapter[$J/\psi$ decays into mesons]{$J/\psi$ decays into mesons}
%
In this chapter we present our results concerning the decays of the $\jp$ into mesons. In Table~\ref{tab:BRJpsi.mesons.pdg} we report some of the larger BR of $\jp \to {\rm mesons}$ from PDG~\cite{Tanabashi:2018oca}.
\begin{table} [ht!]
\centering
\caption{Branching ratios data from PDG~\cite{Tanabashi:2018oca} for some of the larger BR of the $\jp$ decays into mesons.}
\smallskip
\label{tab:BRJpsi.mesons.pdg} 
\begin{tabular}{r|r|r} 
\hline\hline\noalign{\smallskip}
Decay process & Branching ratio & Error \\
\noalign{\smallskip}\hline\hline\noalign{\smallskip}%
$J/\psi \to 2(\pipi) \pi^0$ & $(3.37 \pm 0.26) \times 10^{-2}$ & $7.72 \%$ \\
\hline
$J/\psi \to 3(\pipi) \pi^0$ & $(2.9 \pm 0.6) \times 10^{-2}$ & $20.69 \%$ \\
\hline
$J/\psi \to \pipi \pi^0 \pi^0 \pi^0$ & $(2.71 \pm 0.29) \times 10^{-2}$ & $10.70 \%$ \\
\hline
$J/\psi \to \pipi \pi^0$ & $(2.10 \pm 0.08) \times 10^{-2}$ & $3.81 \%$ \\
\hline
$J/\psi \to 2(\pipi \pi^0)$ & $(1.61 \pm 0.21) \times 10^{-2}$ & $13.04 \%$ \\
\hline
$J/\psi \to \pipi \pi^0 K^+ K^-$ & $(1.20 \pm 0.30) \times 10^{-2}$ & $25.00 \%$ \\
\noalign{\smallskip}\hline\hline
\end{tabular}
\end{table}
%
%
%
%
\section[The $J/\psi \to \pi^+\pi^-$ decay]{The $J/\psi \to \pi^+\pi^-$ decay} \label{secpipi}
%
\subsection[Introduction]{Introduction}
The decay of the $J/\psi$ meson into a pair of pions is an example of a $G$-parity violating decay. The $J/\psi$ meson, having negative $C$-parity,  $C_{\jp} = -1$, and isospin $I_{\jp}=0$, has negative \gp\, $G_{\jp}$, being $G_{\jp} = C_{\jp} (-1)^{I_{\jp}}=-1$. The \gp\ of a system of two pions, being a multiplicative quantum number, is positive $\big(G_{\pi \pi}=(-1)^2\big)$. Therefore, the decay $\jp\to \pi^+ \pi^-$ does not conserve \gp. Strong interaction preserves \gp\ as a consequence of its charge conjugation and isospin conservation. Electromagnetic and weak interactions can violate \gp, being not invariant under $G$ transformations. %
{\textcolor{\colormodthree}{
In this case of $G$-parity violation only two out of three contributions of Eq.~\eqref{eq.3ampl.J} appear in the decay amplitude and we can write
\be
\mathcal A ({\jp \to \pi^+ \pi^-}) = \agg + \ag \,.
\label{eq.2contrgg.p}
\en
}%
This fact can be generalized: when a decay violates isospin the purely strong amplitude is suppressed by the small dimensionless factor
$$
{m_u-m_d \over \sqrt{q^2}} \,,
$$
where $m_u$ and $m_d$ are the masses of $u$ and $d$ quarks and $q^2$ is the typical square momentum in the process.\\
The BR of \jp\ decay into a pair of pions can be decomposed as
\be
{\rm BR}(\jp \to \pipi) &=& \bggg(\jp \to \pipi) + \bgg(\jp \to \pipi) \no \\
&+& \bg(\jp \to \pipi)  + \mbox{interference terms} \,,
\label{eq:br-contributions}
\en
where
$$
\mathcal{B}_{X}(\jp \to \pipi)\propto |\mathcal{A}_{X}(\jp \to \pipi)|^2 \,,
$$
with $X=ggg$, $gg\gamma$, $\gamma$.
%
%
\subsection[Electromagnetic branching ratio]{Electromagnetic branching ratio}
The EM contribution $\bg(\jp \to \pipi)$, corresponding to the intermediate state of the third Feynman diagram of Fig.~\ref{fig.3.contr}, can be computed in terms of the {``dressed'' $\ee\to \pipi$ and ``bare'' $\ee\to\mu^+\mu^-$ cross sections, evaluated at the mass of the $\jp$, as~\cite{Milana:1993wk,Ferroli:2016zqf}
\be
\bg(\jp \to \pipi)= \B(\jp\to\mu^+ \mu^-) \, {\sigma(\ee \to \pipi) \over \sigma^0(\ee \to \mu^+ \mu^-)} \Bigg|_{q^2 = M_{\jp}^2}\,,
\label{eq:b-gamma}
\en
\begin{figure} [ht!]
\centering
\includegraphics[width=.7\columnwidth]{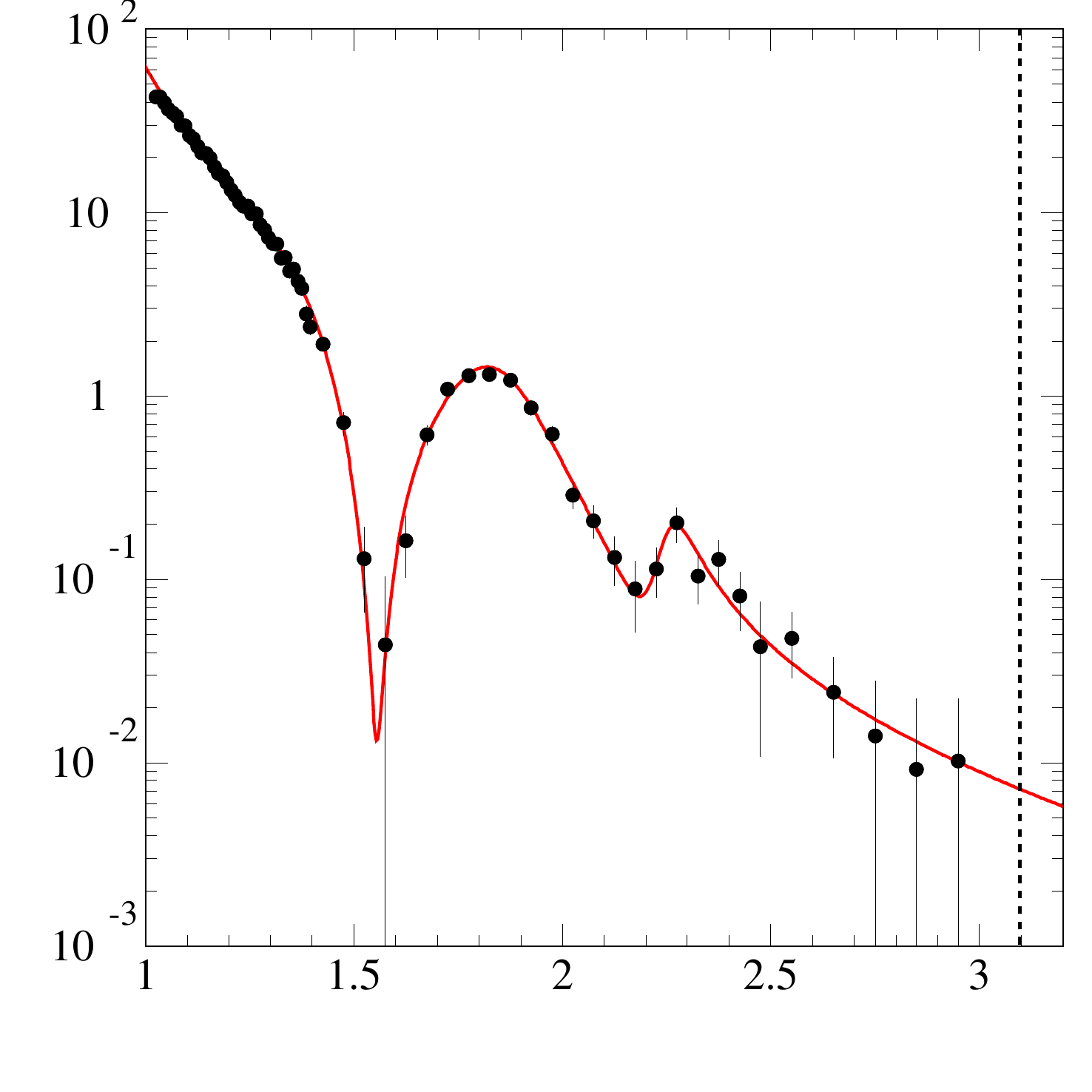}
\put(-68,10){$\sqrt{q^2} \ (\rm GeV)$}
\put(-315,190){\rotatebox{90}{$\sigma(e^+e^- \to \pi^+\pi^-) \ (\rm nb)$}}
\caption{\textsc{BaBar} data on the $e^+e^- \to \pi^+\pi^-$ cross section and the fit (red line) from~\cite{Lees:2012cj}. The vertical dashed line shows the \jp\ mass.}
\label{fig:babar}
\end{figure}%
where $\sigma^0$ stands for the bare cross section, i.e., the cross section corrected for the vacuum-polarization contribution.
It is a common belief the hypothesis of $\bg(h)$-dominance in the \jp\ decays that violate {\gp}, i.e., the fact that these decays proceed almost purely electromagnetically, with a negligible mixed strong-EM BR contribution, $\bgg(\jp \to \pipi)$. Under this hypothesis the total BR should be given by the only EM one. This fact can be verified, for example, for all the hadronic final states, $h$, with even numbers of pions as $h=2(\pipi)$, $h=2(\pipi\pi^0)$ and $h=3(\pipi)$, for which $\bg(h)\simeq \b_{\rm PDG}(h)$, as discussed in Ref.~\cite{Ferroli:2016zqf} using data from Refs.~\cite{Aubert:2005eg,Aubert:2006jq}.\\
For the decay $J/\psi \to \pipi$, using the value of the cross section $\sigma(\ee\to\pipi)$ at the \jp\ mass, extrapolated from the \textsc{BaBar} data~\cite{Lees:2012cj} with a fit based on the Gounaris-Sakurai formula~\cite{Gounaris:1968mw} (see Fig.~\ref{fig:babar}) the BR due to the one-photon exchange mechanism is
\be
\bg(\pipi)=(4.7\pm 1.7)\times 10^{-5}\,,
\label{eq:bg-pipi}
\en
to be compared with~\cite{\pdg}}
\be
\b_{\rm PDG}(\pipi) = (14.7 \pm 1.4) \times 10^{-5}\,.
\label{eq:bpdg-pipi}
\en
In this case the purely electromagnetic BR, Eq.~\eqref{eq:bg-pipi}, differs from the PDG value, Eq.~\eqref{eq:bpdg-pipi}, by almost 4.3 standard deviations. This result unavoidably means that there must be a further contribution beyond the purely EM one. Since the purely strong three-gluon amplitude, \aggg, is suppressed by \gp\ conservation, the remaining amplitude that, contrary to what commonly expected, could play an important role is the one related to the second diagram of Fig.~\ref{fig.3.contr}, i.e., \agg. Moreover, having two sizable amplitudes, $\ag$ and $\agg$, there could also be a constructive interference term that would help in reconciling the prediction and the measured value for the total BR. The amplitude is the one in Eq.~\eqref{eq.2contrgg.p} where the two terms, purely EM and mixed strong-EM, are to be considered both relevant. The total BR, from Eq.~\eqref{eq:br-contributions}, becomes
\be
\b(\jp \to \pipi)&=&\bg(\jp \to \pipi)+\bgg(\jp \to \pipi)\no \\
&+& \mathcal{I}(\jp \to \pipi)\,,
\label{eq:bpi-2amp}
\en
where $\mathcal{I}(\jp \to \pipi)$ accounts for the interference term.\\
%
%
\subsection[Theoretical background]{Theoretical background}\label{thbkg}
The calculation of the amplitude $\agg(\jp \to \pipi)$ in the framework of QCD is quite difficult because the hadronization of the two-gluon plus one-photon intermediate state into \pipi\ occurs at the few-GeV energy regime where QCD is still not perturbative. We find a lower limit for $\bgg(\jp \to \pipi)$, as reported in Ref.~\cite{Ferroli:2016jri}, and show that, within the errors, it is of the same order of $\bg(\jp \to \pipi)$.\\
First of all we can decompose the BR contribution \bgg\ as
\be
\bgg = \bgg_{\re} + \bgg_{\im} &=& {1 \over 2M_{J/\psi} \Gamma_{J/\psi}} \times \bigg(\int \!\!d\rho_2 \, \overline{\big(\re(\agg)\big)^2} \no \\
&+& \int\!\! d\rho_2 \, \overline{\big(\im(\agg)\big)^2} \,\bigg) \,,
\label{eq:b-gg}
\en
in order to highlight the contributions due to the real and imaginary parts of \agg, where $d\rho_2$ is the element of the two-body phase space, $M_{\jp}$ and $\Gamma_{\jp}$ are the mass and the width of the \jp\ meson.\\
It is possible to use the Cutkosky rule~\cite{Cutkosky:1960sp} to calculate the imaginary part of the amplitude $\agg$ by considering all possible on-shell intermediate states that can contribute to the decay chain $\jp\to (gg \gamma)^* \to \pipi$. Taking into account the mechanism where the two gluons hadronize into a set  $\mathcal{P}$ of {$C=+1$} mesons $h_j$, the decay proceeds as
\be
\jp \to \sum_{h_j\in \mathcal{P}} (h_j \gamma)^* \to \pi^+ \pi^- \,.
\nen
The elements $h_j$ of the set $\mathcal{P}$ are only light unflavored mesons, that then couple strongly (OZI-allowed process) with the \pipi\ final state.\\
Using the Cutkosky rule~\cite{Cutkosky:1960sp}, sketched in Fig.~\ref{fig.feyn.cutkosky.rule.1}, the imaginary part of \agg\ is given in terms of a series on the intermediate states $h_j\gamma$, i.e.,
\be
\im(\agg) = \frac{1}{2} \sum_j \SumInt\! d\rho \, \a^*(\jp \to h_j \gamma) \a(\pipi \to h_j \gamma) \,,
\label{eq:cut0}
\en
\begin{figure}[ht!]
\centering
\includegraphics[width=14.5cm]{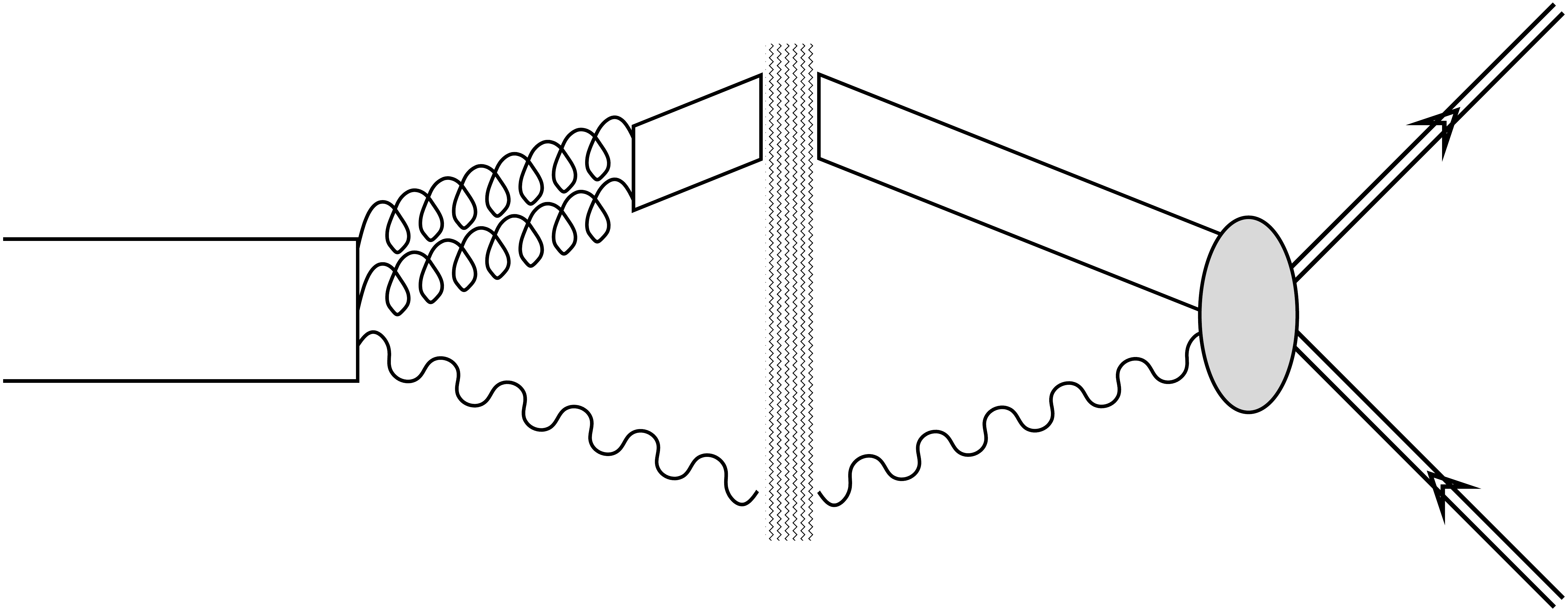}
\put(-380,76){$J/\psi$}
\put(-232,119){\rotatebox{20}{$h_j$}}
\put(-188,122){\rotatebox{-22}{$h_j$}}
\put(-275,33){$\gamma$}
\put(-140,33){$\gamma$}
\put(-10,24){$\pi^-$}
\put(-10,130){$\pi^+$}
\caption{Application of the Cutkosky rule for the decay $J/\psi \to (gg \gamma)^* \to \pi^+ \pi^-$. \label{fig.feyn.cutkosky.rule.1}}
\end{figure}
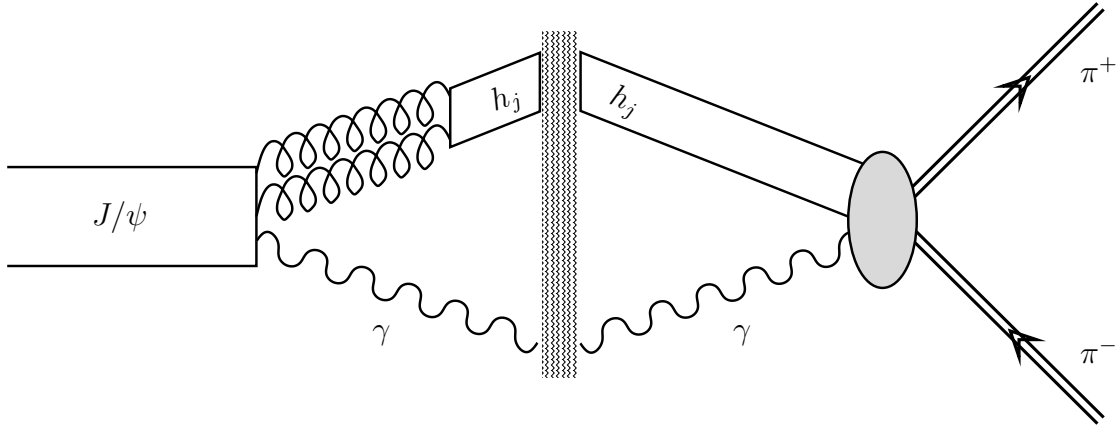%
where the internal-sum runs over the photon polarizations and the integration is on the phase space
\be
\label{eq:PS}
d\rho = \frac{p^0}{4 \pi M_{J/\psi}} \, \frac{d\Omega}{4 \pi} \, ,
\en
being $p^\mu$ the four-momentum of the photon. The selection of all the possible intermediate channels is experimentally driven. %
\begin{table}[ht!]
\centering
\caption{\label{tab:1BRpipi}Branching ratios of a selection of intermediate decays~\cite{\pdg}.}
\smallskip
\begin{tabular}{r|r|r|r}
\hline\hline\noalign{\smallskip}
Meson $M$ & $J^{PC}$ & $10^3 \times \b(\jp\to h_j\gamma)$ & $10^3 \times \b(h_j\to\pipi\gamma)$ \\
\noalign{\smallskip}\hline\hline\noalign{\smallskip}%
$\eta$ & $0^{-+}$  & $1.104\pm 0.034$ & $42.2\pm 0.8$\\
\hline
$\eta'(958)$ & $0^{++}$ &  $5.13\pm 0.17$ & $289\pm 50$\\
\hline
$f_2(1270)$ & $2^{++}$  & $1.64\pm 0.12$ & no data\\
\hline
$f_1(1285)$ & $1^{++}$  & $0.61\pm 0.08$ & ($\rho^0$) $53\pm 12$\\
\hline
$f_0(1500)$ & $0^{++}$  & $0.109\pm 0.024$ & no data\\
\hline
$f'_2(1525)$ & $2^{++}$  & $0.57^{+0.08}_{-0.05}$ & no data\\
\hline
$f_0(1710)$ & $0^{++}$  & $0.38\pm 0.05$ & no data\\
\hline
$f_4(2050)$ & $4^{++}$ & $2.7\pm 0.7$ & no data\\
\hline
$f_0(2100)$ & $0^{++}$  & $0.62\pm 0.10$ & no data\\
\hline
$\eta(2225)$ & $0^{-+}$  & $0.314^{+0.050}_{-0.019}$ & no data\\
\noalign{\smallskip}\hline\hline
\end{tabular}
\end{table}
As a first estimate of the contribution that each channel can give, one could consider the product of the BRs, i.e., $\b(\jp\to h_j\gamma)\times\b(\pipi\to h_j\gamma)$. Table~\ref{tab:1BRpipi} reports all the BRs listed in Ref.~\cite{\pdg}. While there are ten candidates on the \jp\ side, only three sets of data are available on the \pipi\ side. The most prominent contribution is the one due to the $\eta'$ meson, followed by that due to the $\eta$. A further contribution that could be considered is the one due to the axial vector meson \f, for which the combined strength is compatible with that of the $\eta$ meson. In light of that, the imaginary part of the amplitude $\agg$, from Eq~\eqref{eq:cut0}, has three main contributions, i.e.,
{\be
\im(\agg)& \simeq &
\frac{1}{2} \SumInt d\rho \,  \a^*(\jp \to \eta\gamma) \a(\pipi \to \eta\gamma) \nonumber \\
&+& \frac{1}{2} \SumInt d\rho \,  \a^*(\jp \to \epg) \a(\pipi \to \epg) \no \\
&+&
\frac{1}{2} \SumInt d\rho \,  \a^*(\jp \to f_1\gamma) \a(\pipi \to f_1\gamma) \nonumber \\
&\simeq & \im(\a_{\eta'\gamma})+\im(\a_{\eta\gamma})+\im(\a_{f_1\gamma})\,,
\label{eq:ImAgg}
\en}%
where $f_1$ stands for the \f\ meson and the approximate identity is due to the truncation of the series.\\
The first amplitude in the right-hand-side of Eq~\eqref{eq:cut0}, considering the decay
\be
\jp(P) \to h_j(k) + \gamma(p)\,,
\nen
where in parentheses are reported the particle four-momenta, can be written as~\cite{Pacetti:2009pg,Rudenko:2017bel}
\be
\begin{array}{rcl}
\mathcal A(\jp \to \eg) &=& g_{\eg}^{\jp}  p_{\tau} P_{\lambda} \epsilon_\delta (\jp) \epsilon_\sigma (\gamma) \varepsilon^{\tau \lambda \delta \sigma} \, ,\\
&&\\
\mathcal A(\jp \to \fg) &=& g_{\fg}^{\jp} p_{\tau} \epsilon_\lambda(f_1) \epsilon_\delta (\jp) \epsilon_\sigma (\gamma) \varepsilon^{\tau \lambda \delta \sigma} \, ,\\
\end{array}
\label{eq:vertexPV}
\en
where $g_{\eg}^{J/\psi}$ and $g_{\fg}^{J/\psi}$ are the coupling constants, $\epsilon_\delta (\jp)$, $\epsilon_\sigma (\gamma)$ and $\epsilon_\lambda (f_1)$ are the \jp, photon and axial vector polarization vectors, and $\varepsilon^{\tau \lambda \delta \sigma}$ is the Levi-Civita symbol.\\
The second amplitude in the right-hand-side of Eq.~\eqref{eq:cut0} concerns the \pipi\ annihilation process
\be
\pi^+(k_1)+\pi^-(k_2) \to h_j(k) +\gamma(p)\,.
\label{eq:pipi-geta}
\en
\begin{figure}[h!]
\begin{center}
	  \includegraphics[width=.7\columnwidth]{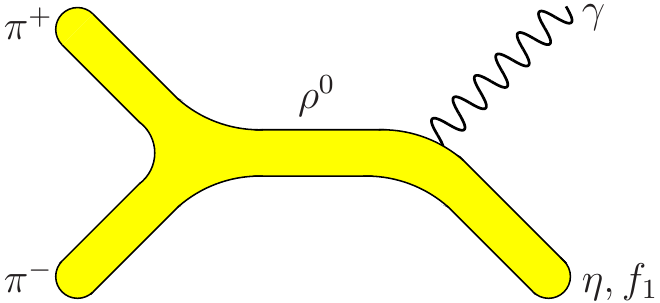}
\end{center}
\caption{Feynman diagram for $\pi^+ \pi^- \to \eg$ and $\pi^+ \pi^- \to \fg$ mediated by the $\rho^0$ meson.}
\label{fig:pi.to.eta.gamma}
\end{figure}\\
The amplitude for this process can be computed in terms of effective meson fields, as described by the Feynman diagram of Fig.~\ref{fig:pi.to.eta.gamma}. Here the coupling between the \pipi\ initial state and the $h_j\gamma$ final state is assumed to be mediated by the $\rho^0$ vector meson. Such an assumption is supported by the strong affinity of the two-pion system with quantum numbers $J^{PC}=1^{--}$ and the $\rho^0$, experimentally confirmed by the BR $\b(\rho^0 \to\pipi)= 1$~\cite{\pdg}. It follows that the amplitudes read~\cite{Pacetti:2009pg,Rudenko:2017bel}
\be
\begin{array}{rcl}
\a(\pi\pi \to \eg) &\!\!=\!\!& \ds g_{\eg}^{\pi\pi} {d_{\alpha} p_{\beta}  \epsilon_\mu (\gamma) q_\nu \varepsilon^{ \alpha\beta \mu \nu} \over M_{\rho}^2 - q^2-iM_{\rho} \Gamma_{\rho}} \, ,\\
&&\\
\a(\pi\pi \to \fg) &\!\!=\!\!& \ds
 g_{\fg}^{\pi\pi} 
{d_{\alpha} p_{\beta}  \epsilon_\mu (\gamma)\epsilon_\nu(f_1) \varepsilon^{\alpha\beta \mu \nu} \over M_{\rho}^2 - q^2-iM_{\rho} \Gamma_{\rho}} \, ,\\
\end{array}
\label{eq:vertex-pp-pg}
\en
where $g^{\pipi}_{\eta(f_1)\gamma}$ is the $\pipi$-$\eta(f_1) \gamma$ coupling constant, $d=k_1-k_2$, while $q= k_1+k_2$, $M_\rho$ and $\Gamma_\rho$ are the four-momentum, the mass and the width of the $\rho^0$ meson.\\
The imaginary term at denominator, $iM_{\rho} \Gamma_{\rho}$, can be omitted, because its contribution to the resulting BR is of the order of $0.01\%$ and then it is negligible with respect to the experimental uncertainty. Moreover, the negligibility of this term allows to recover the reality of $\im(\agg)$, by also validating the truncation of the Cutkosky series.\\
In the CM system we can write the four-momenta
\begin{gather}
\label{gth.fm}
P = q = (M_{J/\psi},0,0,0) \,, \ \ \ p = (p^0,\vec p)=p^0(1, \sin (\theta), 0, \cos (\theta)) \,, \no \\
k = (k^0,-\vec p) \,, \ \ \ k_{1,2} = (M_{J/\psi}/2,0,0, \pm\omega) \,,
\end{gather}
with the scalar products
\begin{gather}
\label{gth.fm.sp}
p^2=0 \, , \ \ \ \ P^2=M_{J/\psi}^2 \, , \ \ \ \ (k_1-k_2) \cdot P = 0 \, , \ \ \ \ (k_1-k_2) \cdot p = -2 p^0 \omega \cos \theta \, , \no \\
P \cdot \epsilon(J/\psi) = 0 \, , \ \ \ \ p \cdot P = p^0 M_{J/\psi} \,,
\end{gather}
where $\theta$ is the scattering angle of the photon and $\omega$ is the modulus of the pion three-momenta. We have also the following relations for the energies
\begin{gather}
\label{eq.energ.eta'.gamma.1}
E_{\gamma}=p^0=|\vec p| = {M_{J/\psi} \over 2} \left( 1-{M_{\eta}^2 \over M^2_{J/\psi}} \right) \, , \no \\
E_{h_j}=k^0 = \sqrt{M_{h_j}^2+(p^0)^2} = {M_{J/\psi} \over 2} \left( 1+{M_{h_j}^2 \over M^2_{J/\psi}} \right) \, .\no
\end{gather}
We calculate explicitly only the case of $h=\eta$, since the others can be obtained by replacing the masses and the coupling constants with the corresponding new ones.\\
We define the quantity
\begin{equation}
\label{eq.A.jpsi.to.pi+pi-1}
Z_{\eta} \equiv \sum_{\rm pol} \mathcal A(\pi^+\pi^- \to \eta \gamma) \mathcal A^*(J/\psi \to \eta \gamma) \,, \no
\end{equation}
and, using the amplitudes of Eq.~\eqref{eq:vertexPV} and Eq.~\eqref{eq:vertex-pp-pg}, we compute
\begin{eqnarray}
Z_{\eta} &=& {g_{\eta \gamma}^{\pi\pi} g_{\eta \gamma}^{J/\psi} \over M_{\rho}^2 -M_{J/\psi}^2} d_{\alpha} p_{\beta} q_\nu P_{\lambda} p_{\tau} \epsilon_\delta (J/\psi) \varepsilon^{\beta \alpha \nu \mu} \varepsilon^{\tau \lambda \delta \sigma} \sum_{\rm pol} \epsilon_\mu (\gamma) \epsilon_\sigma (\gamma) \notag \\
&=& -{g_{\eta \gamma}^{\pi\pi} g_{\eta \gamma}^{J/\psi} \over M_{\rho}^2 -M_{J/\psi}^2} (k_1-k_2)_{\alpha} p_{\beta} q_\nu P_{\lambda} p_{\tau} \epsilon_\delta (J/\psi) \varepsilon^{\beta \alpha \nu \mu} \varepsilon^{\tau \lambda \delta \sigma} \eta_{\mu \sigma} \notag \\
&=& {g_{\eta \gamma}^{\pi\pi} g_{\eta \gamma}^{J/\psi} \over M_{\rho}^2 -M_{J/\psi}^2} (k_1-k_2)_{\alpha} p_{\beta} P_\nu P_{\lambda} p_{\tau} \epsilon_\delta (J/\psi)
\begin{vmatrix} 
\eta^{\beta \tau} & \eta^{\alpha \tau} & \eta^{\nu \tau} \\
\eta^{\beta \lambda} & \eta^{\alpha \lambda} & \eta^{\nu  \lambda} \\
\eta^{\beta \delta} & \eta^{\alpha  \delta} & \eta^{\nu  \delta} 
\end{vmatrix} \, ,
\notag
\end{eqnarray}
where we have used the well-known relation
$$
\sum_{\rm pol} \epsilon_\mu (\gamma) \epsilon_\sigma (\gamma) = -\eta_{\mu \sigma} \, .
$$
Moreover
\begin{eqnarray}
Z_{\eta} &=& {g_{\eta \gamma}^{\pi\pi} g_{\eta \gamma}^{J/\psi} \over M_{\rho}^2 -M_{J/\psi}^2} \Bigg\{ p^2 \Big[(k_1-k_2) \cdot P\Big] \Big[P \cdot \epsilon (J/\psi)\Big] - p^2 P^2 \Big[(k_1-k_2) \cdot \epsilon (J/\psi)\Big] \notag \\
&+& P^2 \Big[(k_1-k_2) \cdot p\Big] \Big[p \cdot \epsilon (J/\psi)\Big] - (p \cdot P) \Big[(k_1-k_2) \cdot p\Big] \Big[P \cdot \epsilon (J/\psi)\Big] \notag \\
&+& (p \cdot P)^2 \Big[(k_1-k_2) \cdot \epsilon (J/\psi)\Big] - (p \cdot P) \Big[(k_1-k_2) \cdot P\Big] \Big[p \cdot \epsilon (J/\psi)\Big] \Bigg\} \, . \notag
\end{eqnarray}
By using the definitions of Eq.~\eqref{gth.fm} and the results of Eq.~\eqref{gth.fm} we obtain
\begin{eqnarray}
Z_{\eta} &=& {g_{\eta \gamma}^{\pi\pi} g_{\eta \gamma}^{J/\psi} \over M_{\rho}^2 -M_{J/\psi}^2} \bigg[ M_{J/\psi}^2 \Big(-2 p^0 \omega \cos \theta \Big) \Big(p^0 \epsilon_0 (J/\psi) + p^1 \epsilon_1 (J/\psi) + p^3 \epsilon_3 (J/\psi)\Big) \notag \\
&+& (p^0)^2 M_{J/\psi}^2 \Big(2\omega \epsilon_3 (J/\psi)\Big) \bigg] \, , \notag
\end{eqnarray}
\begin{eqnarray}
Z_{\eta} &=& {2g_{\eta \gamma}^{\pi\pi} g_{\eta \gamma}^{J/\psi} M_{J/\psi}^2 (p^0)^2 \omega \over M_{J/\psi}^2 - M_{\rho}^2} \bigg[ \epsilon_0 (J/\psi) \cos \theta + \epsilon_1 (J/\psi) \sin \theta \cos \theta \notag \\ 
&+& \epsilon_3 (J/\psi) (1-\cos^2 \theta) \bigg] \,. \no
\end{eqnarray}
By integrating over the solid angle
\begin{eqnarray}
\int d\Omega \, Z_\eta &=& {2g_{\eta \gamma}^{\pi\pi} g_{\eta \gamma}^{J/\psi} M_{J/\psi}^2 (p^0)^2 \omega \over M_{J/\psi}^2 - M_{\rho}^2} \left( \epsilon_3 (J/\psi) \int d\Omega \, \sin^2 \theta \right) \notag \\
&=& {16\pi g_{\eta \gamma}^{\pi\pi} g_{\eta \gamma}^{J/\psi} M_{J/\psi}^2 (p^0)^2 \epsilon_3 (J/\psi) \over 3\big(M_{J/\psi}^2 - M_{\rho}^2 \big)} \sqrt{{M_{J/\psi}^2 \over 4}-M_{\pi}^2} \,, \notag
\end{eqnarray}
where $\epsilon_3(\jp)=\epsilon_3^{(\sigma)}(\jp)$ is the numerical third component ($\mu=3$) of the generic $\sigma$-th polarization four-vector of the $\jp$ meson.\\
Finally, the imaginary parts can be written as
\be
\im(\a^{\eg}) &\!\!\!=\!\!\!&
\sqrt{{M_{J/\psi}^2 \over 4}-M_{\pi}^2}\,\,\,
 \frac{g_{\eg}^{\pi\pi} g_{\eg}^{\jp} M_{\jp}^4 \epsilon_3(\jp)}{48\pi\lt M_{\jp}^2-M_{\rho}^2\rt} \lt 1-\frac{M_{\eta}^2}{M^2_{\jp}} \rt^{\!3}\,,
\no\\
\no\\
\im(\a^{\fg}) &\!\!\!=\!\!\!& \sqrt{{M_{J/\psi}^2 \over 4}-M_{\pi}^2} \,
 \frac{g_{\fg}^{\pi\pi} g_{\fg}^{\jp} M_{\jp}^4 \epsilon_3(\jp)}{48\pi M_{f_1}^2\lt M_{\jp}^2-M_{\rho}^2\rt} \lt 1\!-\!\frac{M_{f_1}^2}{M^2_{\jp}} \rt^{\!3} \!\!
\lt 1+\frac{M_{f_1}^2}{M^2_{\jp}} \rt\,. \no
\en
These expressions and that due to the \epg\ intermediate state, which has the same structure of $\im(\a^{\eg})$, have to be summed up to obtain the complete imaginary part of \agg, see Eq.~\eqref{eq:ImAgg}. The corresponding contribution to the BR, $\bgg_{\im}$, as given in Eq.~\eqref{eq:b-gg}, is
\be
\label{eq:bim-gg}
{\rm BR}^{\rm Im}_{gg \gamma} &\!\!=\!\!& \frac{\sqrt{M^2_{J/\psi}-4M_{\pi}^2}}{16 \pi M^2_{J/\psi} \Gamma_{J/\psi}} \, \overline{|\im(\agg)|^2} = \frac{\left(M^2_{\jp}-4M_{\pi}^2\right)^{3/2}}{4(48\pi)^3 M^6_{\jp} \Gamma_{J/\psi}} 
\frac{\left|
\ds\sum_{h=\eta,\eta',f_1} g_{h\gamma}^{\pi\pi} g_{h \gamma}^{\jp} K_h  \right|^{2}}{\left(M^2_{J/\psi} - M_{\rho}^2\right)^{2}}\,, \no \\
\en
where the average over the polarization states of the $\jp$ meson has been performed and the kinematical factor $K_h$ reads
\be
K_h=\left\{
\begin{array}{ll}
\ds	\lt M_{\jp}^2-M_h^2\rt^3 & h=\eta,\eta'\\
\ds \frac{\lt M_{\jp}^2-M_h^2\rt^3}{M_h^2}\lt1+\frac{M_{h}^2}{M_{\jp}^2}\rt & h=f_1\\
\end{array}
\right..\,\,\,\,\,\,\,\,\,\,
\label{eq:Kh}
\en%
The quantity of Eq.~\eqref{eq:bim-gg} represents a lower limit for \bgg, because the contribution due to the real part of the amplitude, $\bgg_{\re}$, as shown in Eq.~\eqref{eq:b-gg}, is positive. The values of the six coupling constants $g^{\pi\pi}_{h \gamma}$ and $g^{\jp}_{h\gamma}$ ($h=\eta$, $\eta', f_1$) have to be extracted from the data.

\subsection[The coupling constants $g_{\eg}^{\jp}$, $g_{\eta'\gamma}^{\jp}$ and $g_{\fg}^{\jp}$]{The coupling constants $g_{\eg}^{\jp}$, $g_{\eta'\gamma}^{\jp}$ and $g_{\fg}^{\jp}$}
The experimental value of the modulus of the coupling constant $g_{h\gamma}^{J/\psi}$, with $h=\eta$, $\eta'$, $f_1$, can be extracted from the rate of the corresponding radiative decay
\be
\jp(P) \to h(k) + \gamma(p)\,,
\nen
where, in parentheses, are reported the four-momenta, consistently with previous definitions.\\
We show explicitly the calculus for the $h=\eta$ case, since the other cases are quite similar. From Eq.~\eqref{eq:vertexPV} we can write
\begin{eqnarray}
\overline{|\mathcal A(J/\psi \to \eta \gamma)|^2} &=& {1 \over 3} \sum_{\rm pol} |g_{\eta \gamma}^{J/\psi}|^2 \left|P_{\lambda} p_{\tau} \epsilon_\delta (J/\psi) \epsilon_\sigma (\gamma) \varepsilon^{\tau \lambda \delta \sigma} \right|^2 \notag \\
&=& {1 \over 3} |g_{\eta \gamma}^{J/\psi}|^2 P_{\lambda} P_{\lambda'} p_{\tau} p_{\tau'} \sum_{\rm pol} \epsilon_\delta (J/\psi) \epsilon_{\delta'} (J/\psi) \no \\
& \times & \sum_{\rm pol} \epsilon_\sigma (\gamma) \epsilon_{\sigma'} (\gamma) \varepsilon^{\tau \lambda \delta \sigma} \varepsilon^{\tau' \lambda' \delta' \sigma'} \, , \notag
\end{eqnarray}
and
\begin{eqnarray}
\overline{|\mathcal A(J/\psi \to \eta \gamma)|^2} &=& {1 \over 3} |g_{\eta \gamma}^{J/\psi}|^2 P_{\lambda} P_{\lambda'} p_{\tau} p_{\tau'} \varepsilon^{\tau \lambda \delta \sigma} \varepsilon^{\tau' \lambda'}_{\ \ \ \, \delta \sigma} = {2 \over 3} |g_{\eta \gamma}^{J/\psi}|^2 P_{\lambda} P_{\lambda'} p_{\tau} p_{\tau'} \no \\
& \times & (\eta^{\tau \lambda'} \eta^{\lambda \tau'} - \eta^{\tau \tau'} \eta^{\lambda \lambda'}) = {2 \over 3} |g_{\eta \gamma}^{J/\psi}|^2 (P_{\lambda} P^{\tau} p_{\tau} p^{\lambda} - P_{\lambda} P^{\lambda} p_{\tau} p^{\tau}) \no \\
&=& {2 \over 3} |g_{\eta \gamma}^{J/\psi}|^2 \Big((p \cdot P)^2 - p^2 P^2 \Big) = {2 \over 3} |g_{\eta \gamma}^{J/\psi}|^2 \Big((p \cdot P)^2 - p^2 P^2 \Big) \, . \notag
\end{eqnarray}
By using the following scalar products
\begin{gather}
p \cdot P = p^0 M_{J/\psi} = {M_{J/\psi}^2 \over 2} \left( 1-{M_{\eta}^2 \over M^2_{J/\psi}} \right) = {1 \over 2} \left( M_{J/\psi}^2-M_{\eta}^2 \right) \,, \no \\
P^2 = M_{J/\psi}^2  \, , \ \ \ p^2=0 \,, \no
\end{gather}
we obtain
$$
\overline{|\mathcal A(J/\psi \to \eta \gamma)|^2} = {1 \over 6} |g_{\eta \gamma}^{J/\psi}|^2 \left( M_{J/\psi}^2-M_{\eta}^2 \right)^2 \, .
$$
Using Eq.~\eqref{eq.DW.cmfin} we arrive, after some further calculations, to the obtained radiative decay width
\be
\Gamma (\jp \to h \gamma) &=& \frac{K_h}{96\pi M_{\jp}^3} 
 \big|g_{h\gamma}^{J/\psi}\big|^2 \, ,
\nen
where $K_h$ is the kinematical factor defined in Eq.~\eqref{eq:Kh}.\\
It follows that the modulus of the coupling constant can be extracted as
\be
\big|g_{h \gamma}^{J/\psi}\big| = \sqrt{\frac{96 \pi M_{\jp}^3 \Gamma (J/\psi \to h\gamma)}{K_h}}\,.
\nen%
Finally, by using the experimental values of radiative decay widths $\Gamma(\jp\to\eta\gamma)$, $\Gamma(\jp\to\epg)$ and  
 $\Gamma(\jp\to\fg)$~\cite{\pdg}, reported in Table~\ref{tab:Gamma.pdg.A}, the coupling constants are
\be
|g_{h \gamma}^{J/\psi}|\! =\!
\left\{\!\begin{array}{ll}
(1.070 \!\pm\! 0.023) \!\times \!10^{-3} \ {\rm GeV^{-1}} &\hh h=\eta \\
&\\
(2.563 \!\pm\! 0.055) \!\times \!10^{-3} \ {\rm GeV^{-1}} & \hh h=\eta'\\
&\\
{(1.191 \!\pm\! 0.080) \!\times \!10^{-3} }& \hh h=f_1\\
\end{array}\right. \!.\,\,\,\,\,\,\,\,\,
\label{eq:cc-eg}
\en
\begin{table}
\centering
\caption{Decay widths from Ref.~\cite{\pdg}.}
\smallskip
\label{tab:Gamma.pdg.A}
\begin{tabular}{l|l}
\hline\hline\noalign{\smallskip}
Decay processes & Decay widths $\Gamma$ $({\rm GeV})$  \\
\noalign{\smallskip}\hline\hline\noalign{\smallskip}%
$J/\psi \to \eta \gamma$ & $(1.026 \pm 0.044) \times 10^{-7}$ \\ \hline
$J/\psi \to \eta' \gamma$ & $(4.78 \pm 0.14) \times 10^{-7}$ \\ \hline
{$J/\psi \to f_1 \gamma$} & {$(5.67 \pm 0.76) \times 10^{-8}$} \\
\noalign{\smallskip}\hline\hline
\end{tabular}
\end{table}%
We observe that, as a consequence of the structure of the amplitudes shown in Eq.~\eqref{eq:vertexPV}, the coupling constant of the axial vector is adimensional, while those of the pseudoscalar mesons have the dimension of inverse energy. Looking at their structure we see that they differ only by the interchange of the \jp\ four-momentum $P_\lambda$ with the adimensional polarization vector of $f_1$.
%
%
\subsection[The coupling constants $g_{\eg}^{\pi\pi}$, $g_{\eta' \gamma}^{\pi\pi}$ and $g_{\fg}^{\pi\pi}$]{The coupling constants $g_{\eg}^{\pi\pi}$, $g_{\eta' \gamma}^{\pi\pi}$ and $g_{\fg}^{\pi\pi}$}
There are no data on the cross section of the process $\pipi\to h\gamma$, with $h=\eta$, $\eta'$ and $f_1$, so that the coupling constant $g_{h\gamma}^{\pi\pi}$, appearing in Eq.~\eqref{eq:vertex-pp-pg}, can not be directly measured. Nevertheless, the same coupling constants must regulate the amplitudes of the decay 
\be
h(k) \to \pi^+(k_1) + \pi^-(k_2) + \gamma(\tilde p)\,,\hh\hh
h=\eta,\eta', f_1\,.
\nen
as a consequence of the crossing symmetry. This decay is obtained by moving the photon from the final to the initial state of the original reaction of Eq.~\eqref{eq:pipi-geta}, with the Feynman diagram of Fig.~\ref{fig:pi.to.eta.gamma}, and then by making a time-reversal transformation.\\
For the case $h=\eta$, the other cases are very similar, we can write
\begin{eqnarray}
\label{eq.dgamma.etap.pi+pi-gamma1}
d\Gamma(\eta \to \pi^+ \pi^- \gamma) &=& {1 \over 2M_{\eta}} \overline{|\mathcal A(\eta \to \pi^+ \pi^- \gamma)|^2} \, d\rho_3 \notag \\
&=& {1 \over (2\pi)^3} {1 \over 32M_{\eta}^3} \overline{|\mathcal A(\pi^+ \pi^- \to \eta \gamma)|^2} \, dq^2 dq_1^2 \,,
\end{eqnarray}
where $d\rho_3$ is the three-body phase space and with
\begin{equation}
\label{eq.def.q.q1.cost.acc4}
q \equiv k-\tilde p=k_1+k_2 \,, \ \ \ q_1 \equiv k_1+\tilde p=k-k_2 \,.
\end{equation}
For the amplitude $\overline{|\mathcal A(\pi^+ \pi^- \to \eta \gamma)|^2}$ we consider the process
\be
\eta(k) + \gamma(p) \to \rho^0(\tilde q) \to \pi^+(k_1) + \pi^-(k_2) \,,
\nen
where $p=-\tilde p$ and $\tilde q=k+p=q$. From Eq.~\eqref{eq:vertex-pp-pg} we can write
\begin{eqnarray}
\overline {|\mathcal A(\pi\pi \to \eta \gamma)|^2} &=& \sum_{\rm pol} |g_{\eta \gamma}^{\pi\pi}|^2 \left|d_{\alpha} {1 \over M_{\rho}^2 - q^2 -iM_{\rho} \Gamma_{\rho}} p_{\beta} q_\nu \epsilon_\mu (\gamma) \varepsilon^{\beta \alpha \nu \mu}\right|^2 \notag \\
&=& {|g_{\eta \gamma}^{\pi\pi}|^2 d_{\alpha} d_{\alpha'} p_{\beta} p_{\beta'} q_\nu q_{\nu'} \over \big(q^2 - M_{\rho}^2\big)^2 +\Gamma_{\rho}^2 M_{\rho}^2} \sum_{\rm pol} \epsilon_\mu (\gamma) \epsilon_{\mu'} (\gamma) \varepsilon^{\beta \alpha \nu \mu} \varepsilon^{\beta' \alpha' \nu' \mu'} \notag \\
&=& -{|g_{\eta \gamma}^{\pi\pi}|^2 (k_1-k_2)_{\alpha} (k_1-k_2)_{\alpha'} p_{\beta} p_{\beta'} q_\nu q_{\nu'} \varepsilon^{\beta \alpha \nu \mu} \varepsilon^{\beta' \alpha' \nu' \mu'} \eta_{\mu \mu'} \over \big(q^2 - M_{\rho}^2\big)^2 +\Gamma_{\rho}^2 M_{\rho}^2} \notag \\
&=& {|g_{\eta \gamma}^{\pi\pi}|^2 (k_1-k_2)_{\alpha} (k_1-k_2)_{\alpha'} p_{\beta} p_{\beta'} q_\nu q_{\nu'} \over \big(q^2 - M_{\rho}^2\big)^2 +\Gamma_{\rho}^2 M_{\rho}^2} \begin{vmatrix} 
\eta^{\beta \beta'} & \eta^{\alpha \beta'} & \eta^{\nu \beta'} \\
\eta^{\beta \alpha'} & \eta^{\alpha \alpha'} & \eta^{\nu  \alpha'} \\
\eta^{\beta \nu'} & \eta^{\alpha  \nu'} & \eta^{\nu  \nu'} 
\end{vmatrix} \,, \no
\end{eqnarray}
from which we obtain
\begin{eqnarray}
\label{eq.M.pi2.to.etap.gamma.per.dec4}
\overline {|\mathcal A(\pi\pi \to \eta \gamma)|^2} &=& {|g_{\eta \gamma}^{\pi\pi}|^2 \over \big(q^2 - M_{\rho}^2\big)^2 +\Gamma_{\rho}^2 M_{\rho}^2} \Bigg\{ p^2 q^2 (k_1-k_2)^2 - p^2 \Big[(k_1-k_2) \cdot q\Big]^2 \notag \\
&+& (p \cdot q) \Big[(k_1-k_2) \cdot p\Big] \Big[(k_1-k_2) \cdot q\Big] - q^2 \Big[(k_1-k_2) \cdot p\Big]^2 \notag \\
&+& (p \cdot q) \Big[(k_1-k_2) \cdot p\Big] \Big[(k_1-k_2) \cdot q\Big] - (p \cdot q)^2 (k_1-k_2)^2 \Bigg\} \,. \no
\end{eqnarray}
Using the definitions of Eq.~\eqref{eq.def.q.q1.cost.acc4}, we have the following relations
\begin{gather}
k_1^2=k_2^2=M_\pi^2 \,, \ \ \ k^2=M_{\eta}^2 \,, \ \ \ p^2=\tilde p^2=0 \,, \no \\
p \cdot (k_1-k_2)=M_\pi^2-q_1^2+{1 \over 2}(M_{\eta}^2-q^2) \,, \ \ \ q \cdot (k_1-k_2)=0 \,, \no \\
p \cdot q = {1 \over 2}(q^2-M_{\eta}^2) \,, \ \ \ (k_1-k_2)^2=4 M_\pi^2-q^2 \,, \no
\end{gather}
therefore Eq.~\eqref{eq.M.pi2.to.etap.gamma.per.dec4} becomes
\begin{eqnarray}
\overline {|\mathcal A(\pi\pi \to \eta \gamma)|^2} &=& {|g_{\eta \gamma}^{\pi\pi}|^2 \left( -q^2 \Big((k_1-k_2) \cdot p\Big)^2-(p \cdot q)^2 (k_1-k_2)^2 \right) \over \big(q^2 - M_{\rho}^2\big)^2 +\Gamma_{\rho}^2 M_{\rho}^2} \notag \\
&=& |g_{\eta \gamma}^{\pi\pi}|^2 {(q^2-4M_{\pi}^2)\big(q^2-M_{\eta}^2\big)^2 - q^2 \big( q^2 + 2q_1^2 - 2M_{\pi}^2 - M_{\eta}^2 \big)^2 \over 4 \Big[ \big(q^2 - M_{\rho}^2\big)^2 +\Gamma_{\rho}^2 M_{\rho}^2 \Big]} \, . \notag
\end{eqnarray}
{Finally, with an analogous procedure for $h=\eta',f_1$, from Eq.~\eqref{eq.dgamma.etap.pi+pi-gamma1}, we obtain the decay width
\be
\label{eq:Gpipig}
\Gamma(h \to \pi^+ \pi^- \gamma) 
&=& 
\int 
 \overline{|\mathcal A(h \to \pi^+ \pi^- \gamma)|^2} \,d\rho_3 \no \\
&=& \frac{1}{(2\pi)^3} \frac{|g_{h \gamma}^{\pi \pi}|^2 }{128M_{h}^3} \int_{q^2_{\rm min}}^{q^2_{\rm max}} dq^2 \,
\int_{{q_1^2}_{\rm min}(q^2)}^{{q_1^2}_{\rm max}(q^2)} dq_1^2\, I_{h}(q^2,q_1^2)\,,\no \\
\en
where the integration variables and the corresponding limits are: $q^2 \equiv (k-\tilde p)^2=(k_1+k_2)^2$, $q^2_1 \equiv (k_1+\tilde p)^2=(k-k_2)^2$, 
\be
&
q^2_{\rm min} = 4M_{\pi}^2\,,\hh\hh q^2_{\rm max} = M_{h}^2\,,&\no\\
&
{q_1^2}_{\rm min,max}(q^2) =\ds {M_{h}^4 \over 4q^2} - \left( \sqrt{{q^2 \over 4}-M_{\pi}^2} \pm \frac{M_{h}^2-q^2}{2\sqrt{q^2}} \right)^2\,,&
\nen
with $h=\eta$, $\eta'$, $f_1$. The functions $I_h(q^2,q_1^2)$ have two different forms, for the case of pseudoscalar mesons we have
$$
I_h(q^2,q_1^2)=\frac{(q^2\!-\!4M_{\pi}^2)\big(q^2\!-\!M_{h}^2\big)^2\!}{\big(q^2\!-\! M_{\rho}^2\big)^2 \!+\!\Gamma_{\rho}^2 M_{\rho}^2} -\ds\frac{q^2\big(q^2\!+\! 2q_1^2 \!-\! 2M_{\pi}^2\! -\! M_{h}^2 \big)^2}{\big(q^2\!-\! M_{\rho}^2\big)^2 \!+\!\Gamma_{\rho}^2 M_{\rho}^2}\,,\hh h=\eta,\eta'\,,
$$
while for the axial vector meson it reads
\be
I_{f_1}(q^2,q_1^2)
=\!\frac{1}{3M_{f_1}^2}\!\!\lq\frac{(q^2\!-\!4M_{\pi}^2)\big(q^2\!-\!M_{f_1}^2\big)^2\!}{\big(q^2\!-\! M_{\rho}^2\big)^2 \!+\!\Gamma_{\rho}^2 M_{\rho}^2} \right. \left.\!\!-\frac{\big( q^2-2M_{f_1}^2\big)\big(q^2\!+\! 2q_1^2 \!-\! 2M_{\pi}^2\! -\! M_{f_1}^2 \big)^2}{\big(q^2\!-\! M_{\rho}^2\big)^2 \!+\!\Gamma_{\rho}^2 M_{\rho}^2}\rq \!\,.
\nen}%
The phase-space integrals are
\be
\tilde I_h=\int_{q^2_{\rm min}}^{q^2_{\rm max}} \!\!\!\!\!dq^2 \!\!\!
\int_{{q_1^2}_{\rm min}(q^2)}^{{q_1^2}_{\rm max}(q^2)} \!\!\!\!\!dq_1^2\, I_{h}(q^2,q_1^2)=\left\{\begin{array}{ll}	
	\!\!(5.840 \pm 0.011) \times 10^{-5} \, \rm GeV^6 &\hh h=\eta\\
	&\\
	\!\!(2.719 \pm 0.019) \times 10^{-1}\,\rm GeV^6 &\hh h=\eta'\\
	&\\
	\!\!(6.403 \pm 0.052) \,\rm GeV^4 &\hh h=f_1\\
\end{array}
\right.\,,
\nen
{and also in this case the contribution due to the axial vector meson has a different dimension, $E^4$ instead of $E^6$, as a consequence of the different structure of the amplitude, see Eq.~\eqref{eq:vertex-pp-pg}}. Finally, the corresponding coupling constants can be extracted by means of
\be
|g_{h\gamma}^{\pi\pi}| = (2\pi M_{h})^{3/2} \sqrt{\frac{128 \Gamma(h\to \pi^+ \pi^- \gamma)}{\tilde I_{h}}}=\left\{\begin{array}{ll}	
	\!\!(2.223 \pm 0.047) \ {\rm GeV^{-1}} &\hh h=\eta\\
	&\\
	\!\!(2.431 \pm 0.060) \ {\rm GeV^{-1}} &\hh h=\eta'\\
	&\\
	\!\!3.55 \pm 0.41  &\hh h=f_1\\
\end{array} \no
\right.\!\,.
\en
where we have used the experimental data show in Table~\ref{tab:Gamma.pdg.B}.
\begin{table}
\centering
\caption{Decay widths from Ref.~\cite{\pdg}.}
\smallskip
\label{tab:Gamma.pdg.B}
\begin{tabular}{l|l} 
\hline\hline\noalign{\smallskip}
Decay processes & Decay widths $\Gamma$ $({\rm GeV})$  \\
\noalign{\smallskip}\hline\hline\noalign{\smallskip}%
$\eta \to \pi^+ \pi^- \gamma$ & $(5.53 \pm 0.32) \times 10^{-8}$ \\ \hline
$\eta' \to \pi^+ \pi^- \gamma$ & $(5.76 \pm 0.10) \times 10^{-5}$ \\ \hline
{$f_1 \to \pi^+ \pi^- \gamma$} & {$(1.20 \pm 0.28) \times 10^{-4}$} \\
\noalign{\smallskip}\hline\hline
\end{tabular}
\end{table}%
%
%
%
\subsection[The imaginary part of ${A}^{gg \gamma}$]{The imaginary part of ${A}^{gg \gamma}$}
We calculate the contribution to the BR due to the imaginary part of the $gg \gamma$ amplitude, $\bgg_{\im}$, using Eq.~\eqref{eq:bim-gg} which contains the sum of the three amplitudes related to the intermediate mesons $\eta$, $\eta'$ and $f_1$. We notice that there are also effects due to interference terms having three amplitudes\footnote{The possibility of ortogonality, i.e., amplitudes with a $\pm \pi/2$ relative phase, is predicted by the fact that there are three contributions.}.\\
We assume that the relative phase of the amplitudes of the two pseudoscalar contributions, being due to the $\eta$ meson and to its first excitation $\eta'$, is zero, so they are simply summed up with constructive interference. On the other hand, the relative phase between the axial vector and pseudoscalar mesons amplitudes is unknown.
\\
The single contributions obtained from Eq.~\eqref{eq:bim-gg} are
\be
\bgg_{\im}(\eta) &=& (1.176 \pm 0.080) \times 10^{-6}\,,\no\\
\bgg_{\im}(\eta') &=& (5.34 \pm 0.38) \times 10^{-6}\,,\no\\
\bgg_{\im}(f_1) &=& (0.74 \pm 0.20) \times 10^{-6}\,.
\nen
It is useful to note that they follow the same hierarchy of the relative BRs shown in Table.~\ref{tab:1BRpipi}, with the main contribution given by the pseudoscalar meson $\eta'$.
The total pseudoscalar contribution, due to the $\eta$ and $\eta'$ particles, is
\be
\bgg_{\im}(\eta+\eta') = \lt\sqrt{\bgg_{\im}(\eta)}+\sqrt{\bgg_{\im}(\eta')}\rt^2 = (1.152 \pm 0.066) \times 10^{-5}\,. \no \\
\label{eq:bgg-value}
\en
Concerning the introduction of the $f_1$ contribution we obtain the following two extreme cases  
\be
\begin{array}{rcl}
\bgg_{\im}\lt \eta+\eta'- f_1\rt
&=&
(0.643\pm 0.074)\times 10^{-5}\,,\\
\bgg_{\im}\lt \eta+\eta'+ f_1\rt
&=&
(1.81\pm 0.12)\times 10^{-5}\,,\\
\end{array}
\label{eq:f1+-}
\en
due to destructive and constructive interference respectively.\\
These values represent lower limits for $\bgg$ and they represent the 13\% and the 37\% of the purely EM BR contribution that is
$$
\bg(\pipi) = (4.7 \pm 1.7) \times 10^{-5} \,,
$$
as given in Eq.~\eqref{eq:bg-pipi}. This fact leaves open the possibility that the total \bgg\ contribution would be of the same order of \bg.\\
Finally, using Eq.~\eqref{eq:bpi-2amp} and the value of Eq.~\eqref{eq:bgg-value} for $\bgg_{\im}$ that represents an average of the two possibilities of Eq.\eqref{eq:f1+-}, together with the experimental datum for \bg, as given in Eq.~\eqref{eq:bg-pipi}, we obtain
\be
\b(\pipi)&=&\bg(\pipi)+\bgg(\pipi)+\mathcal{I}(\pipi)\label{eq:Bth}\\
&=&(5.9 \pm 1.7) \!\times\! 10^{-5} \!+\!\bgg_{\re}(\pipi)\!+\!\mathcal{I}(\pipi)\,,
\nen
to be compared with the PDG datum~\cite{\pdg} of Eq.~\eqref{eq:bpdg-pipi}
\be
{\rm BR}_{\rm PDG}(\pi^+\pi^-) = (14.7 \pm 1.4) \times 10^{-5} \,.
\nen
%
\section[The decay $J/\psi \to K^+ K^-$]{The decay $J/\psi \to K^+ K^-$}
The method discussed in the previous section, for the case of the $\jp$ decay into pions, can be used for any other similar processes. One of particular interest is the decay
$$
\jp \to K^+ K^- \,.
$$
In fact in this case we can use simply $h_1=\eta$ and $h_2=\eta'$ as intermediate states for the application of the Cutkosky rule, see Eq.~\eqref{eq:cut0}. The procedure is identical to that described in the previous section, in this case we consider the following decay chain
$$
\jp \to \sum_j \eta_j \gamma \to K^+ K^- \,,
$$
where $\eta_1=\eta$ and $\eta_2=\eta'$. The coupling constant that appears in the amplitude of the first decay, $\jp \to \eta_j \gamma$, is obtained by using the measured decay rate $\Gamma(\jp \to \eta_j \gamma)$ and is the same of the $\pipi$ case. Concerning the amplitude of the second part of the process, the $\eta_j \gamma \to K^+ K^-$, we assume that the $K^+ K^-$ final state system, having $J^{PC} = 1^{--}$, resonates almost completely in the $\phi$ vector meson, that plays the same role of the $\rho^0(770)$ in the case of pions, therefore we consider the scattering
$$
\eta_j \gamma \to \phi \to K^+ K^- \,.
$$
The obtained value for $\bgg_{\im}$ is
$$
\bgg_{\im} (\jp \to K^+K^-) = (2.3 \pm 0.1) \times 10^{-7} \,,
$$
to be compared with the total BR from PDG~\cite{\pdg}
\be
{\rm BR}_{\rm PDG}(K^+K^-) = (28.6 \pm 2.1) \times 10^{-5} \,
\nen
and with the EM ${\rm BR}_\gamma$. The latter can be calculated using data from Ref.~\cite{Lees:2015iba}, and it is
$$
{\rm BR}_{\gamma}(K^+K^-) = (13.8 \pm 0.7) \times 10^{-5} \,.
$$
We can conclude that the contribution to the total BR of $\jp \to K^+ K^-$ due to the imaginary part of the mixed strong-EM amplitude is totally negligible, being three order of magnitude lower than both the purely EM and the total BRs.
%
%
\chapter[$J/\psi$ decays into baryons]{$J/\psi$ decays into baryons}
%
In this chapter we present our results concerning the decays of the $\jp$ into baryons. In Table~\ref{tab:BRJpsi.baryons.pdg} we report some of the larger BR of $\jp \to {\rm baryons}$ from PDG~\cite{Tanabashi:2018oca}.
\begin{table} [ht!]
\vspace{-2mm}
\centering
\caption{Branching ratios data from PDG~\cite{Tanabashi:2018oca} for some of the larger BR of the $\jp$ decays into baryons.}
\smallskip
\label{tab:BRJpsi.baryons.pdg} 
\begin{tabular}{r|r|r} 
\hline\hline\noalign{\smallskip}
Decay process & Branching ratio & Error \\
\noalign{\smallskip}\hline\hline\noalign{\smallskip}%
$J/\psi \to p \overline p$ & $(2.121 \pm 0.029) \times 10^{-3}$ & $1.37 \%$ \\
\hline
$J/\psi \to n \overline n$ & $(2.09 \pm 0.16) \times 10^{-3}$ & $7.66 \%$ \\
\hline
$J/\psi \to \Lambda \overline \Lambda$ & $(1.89 \pm 0.09) \times 10^{-3}$ & $4.76 \%$ \\
\hline
$J/\psi \to \Sigma^+ \overline \Sigma{}^-$ & $(1.50 \pm 0.24) \times 10^{-3}$ & $16.00 \%$ \\
\hline
$J/\psi \to \Sigma^0 \overline \Sigma{}^0$ & $(1.172 \pm 0.032) \times 10^{-3}$ & $2.73 \%$ \\
\hline
$J/\psi \to \Xi^0 \overline \Xi{}^0$ & $(1.17 \pm 0.04) \times 10^{-3}$ & $3.42 \%$ \\
\noalign{\smallskip}\hline\hline
\end{tabular}
\end{table}
%
%
\section[The $J/\psi \to B \overline B$ decay]{The $J/\psi \to B \overline B$ decay}
\subsection[Introduction]{Introduction}
{\textcolor{\colormodfour}{
Particles directly produced at $\ee$ colliders decay with relatively high probability into a baryon–antibaryon, $\bb$, pair~\cite{Kopke:1988cs}. In this kind of collider, electrons and positrons annihilate, producing a resonance, such as the $\jp$ meson, that could decay into entangled $\bb$ pairs. For example in the case of $\ee \to \jp \to \LL$, see Fig.~\ref{fig:feynee}, the $\jp$, produced at rest in a single photon annihilation process, subsequently decays into a $\LL$ pair~\cite{Ablikim:2018zay}. %
\begin{figure} [ht!]
	\begin{center}
		\includegraphics[width=.65\columnwidth]{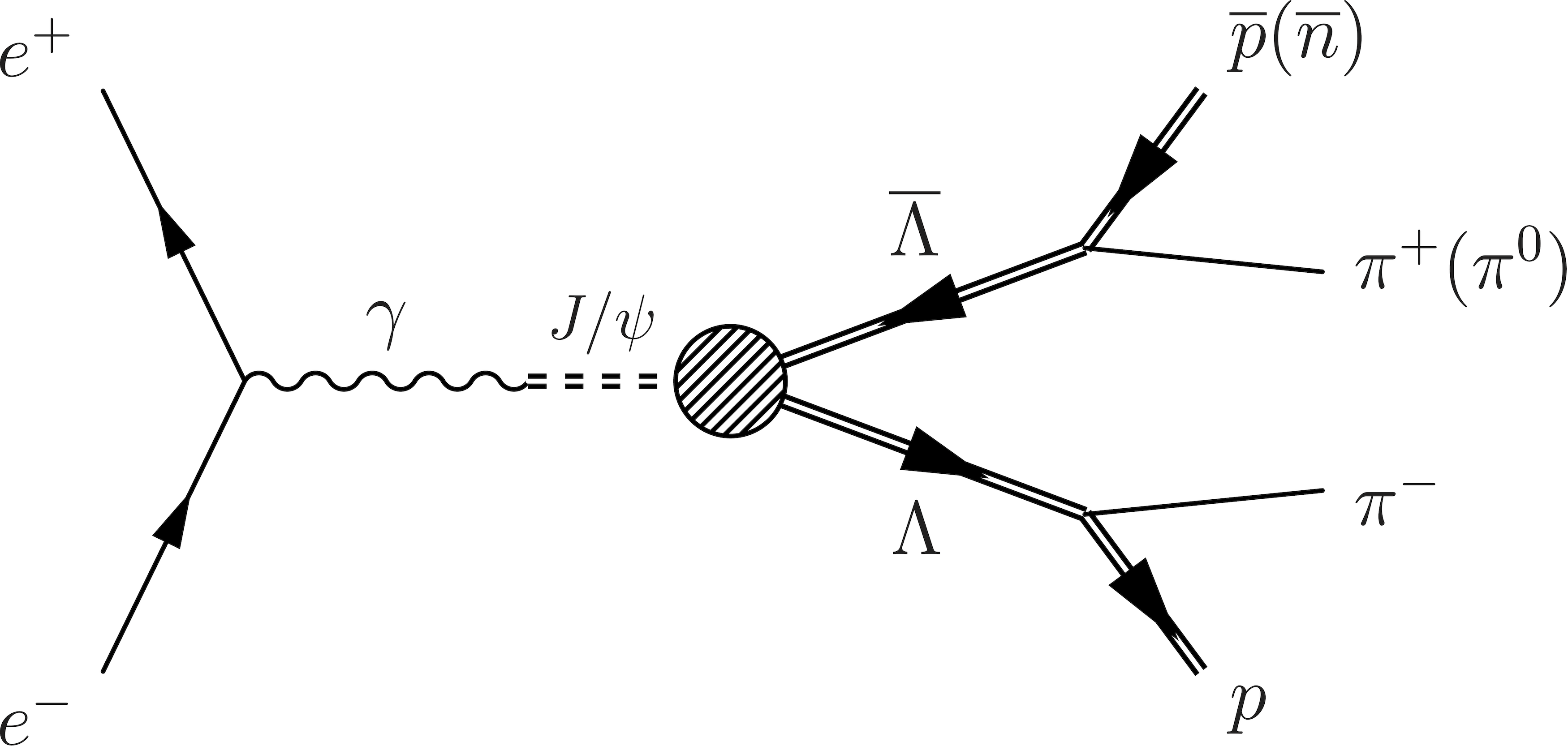}
		\caption{Feynman diagram for a typical process $\ee \to \jp \to \LL$.}\label{fig:feynee}
	\end{center}
\end{figure}\\
}}%
The decays of the $J/\psi$ meson into a $\bb$ pair proceed via strong and EM interactions. The Feynman amplitude, from Eq.~\eqref{eq.3ampl.J}, can be written as a sum of three sub-amplitudes
$$
\mathcal{A}_{\bb} = \mathcal A^{ggg}_{\bb} + \mathcal A^{\gamma}_{\bb} + \mathcal A^{gg\gamma}_{\bb}  \,,
$$
where $\mathcal A^{ggg}_{\bb}$ is the purely strong, $\mathcal A^{\gamma}_{\bb}$ is the purely EM and $\mathcal A^{gg\gamma}_{\bb}$ is the mixed strong-EM sub-amplitude. {\textcolor{\colormod}{The corresponding Feynman diagrams are shown in Fig.~\ref{fig.3.contr.2}}}. %
\begin{figure} [ht!]
	\centering
\subfigure[Purely strong contribution.]{%
  \includegraphics[width=.25\textwidth]{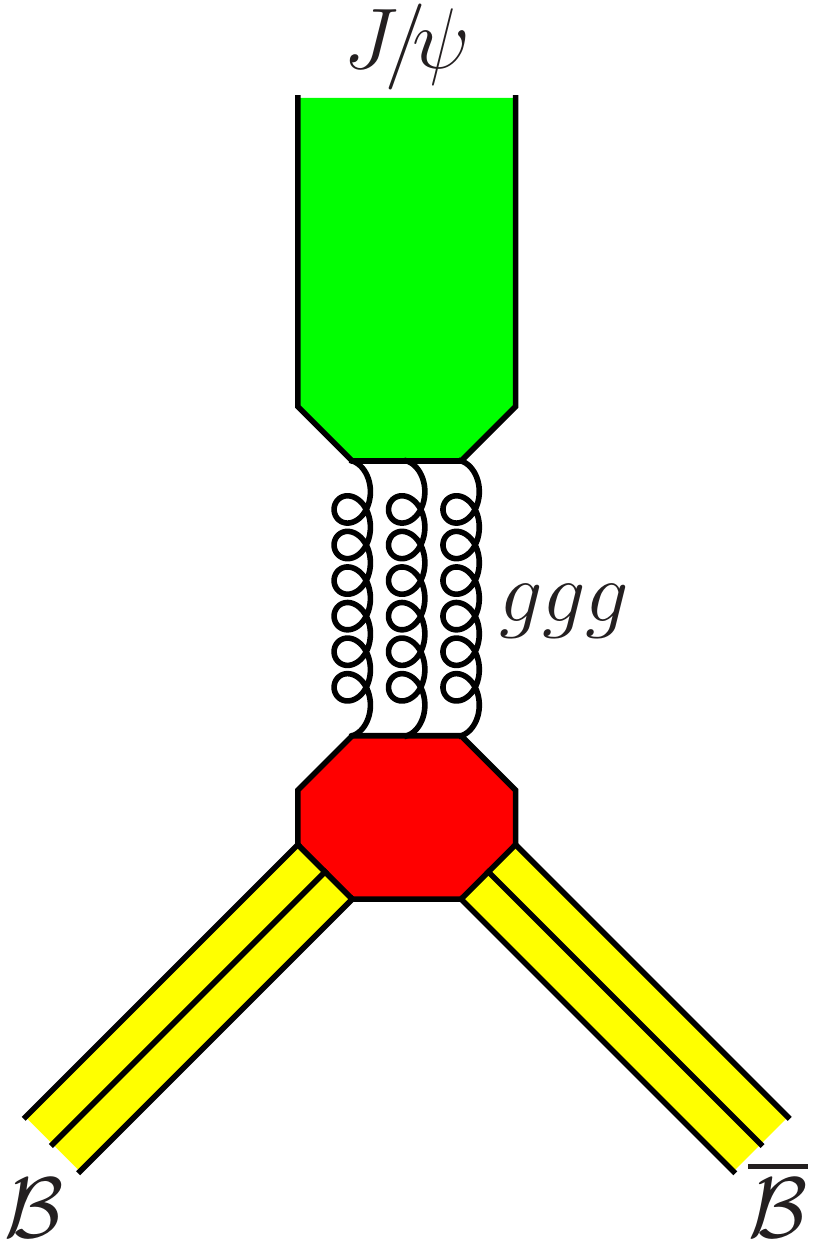}} \quad\quad\quad
\subfigure[Purely EM contribution.]{%
  \includegraphics[width=.25\textwidth]{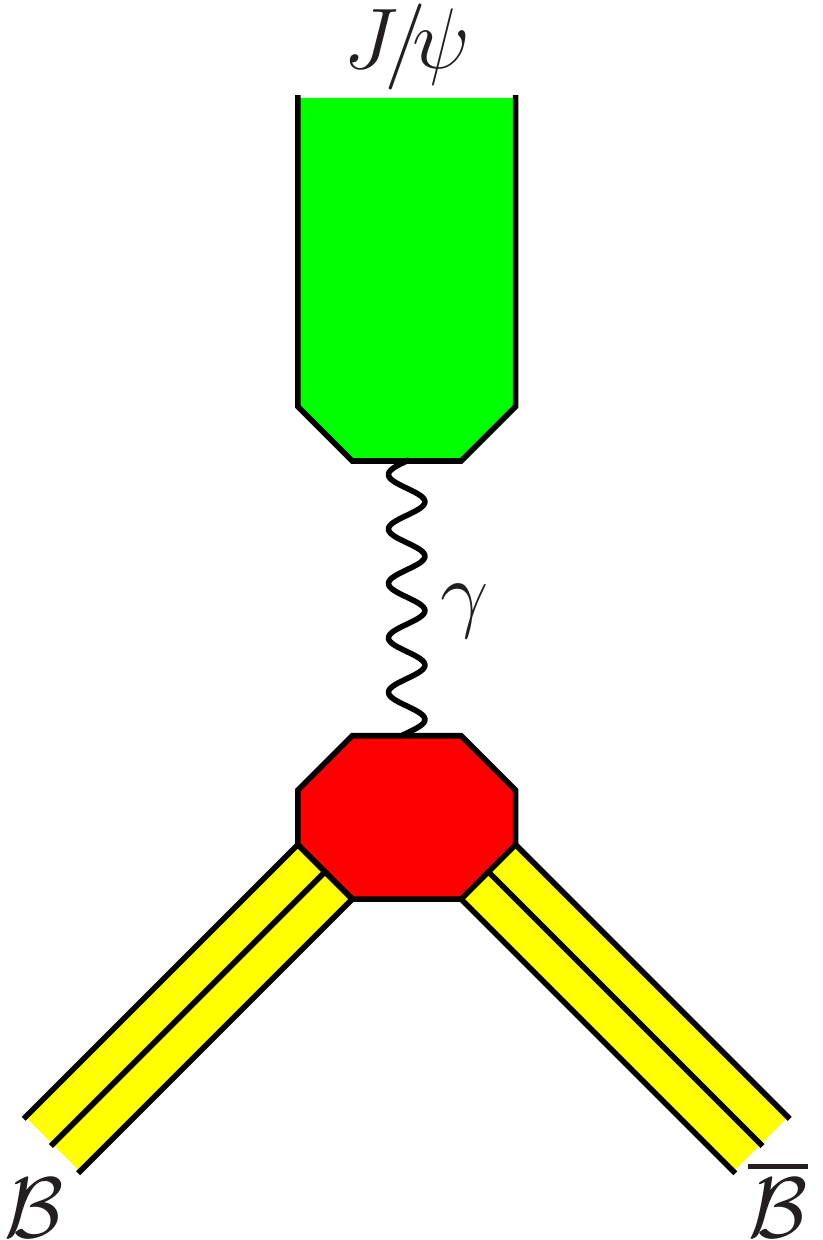}} \quad\quad\quad
\subfigure[Mixed strong-EM contribution.]{%
  \includegraphics[width=.25\textwidth]{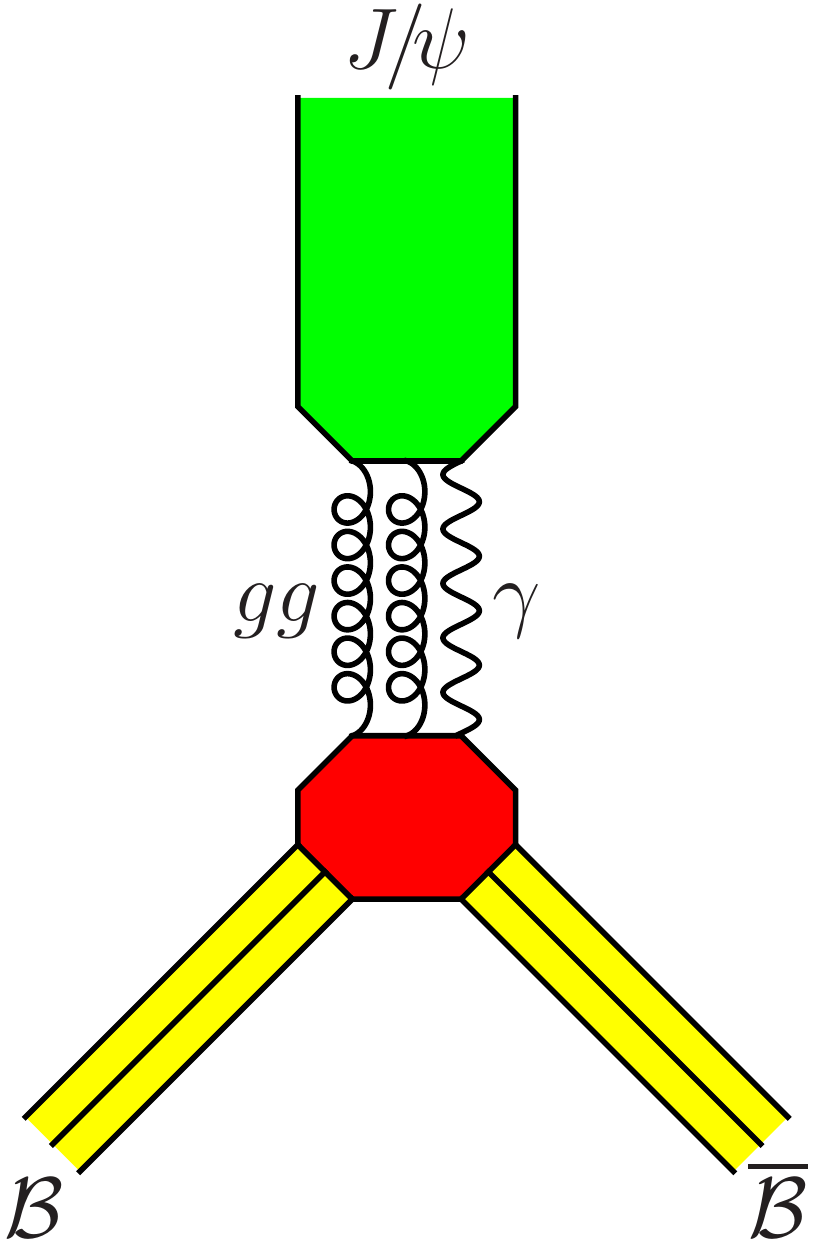}}
	\caption{\label{fig.3.contr.2} Feynman diagrams for the three sub-amplitude of the decay $J/\psi \to \bb$.}
\end{figure}\\
Usually, as in the meson cases, the mixed strong-EM contribution is not considered since it is assumed to be negligible with respect to the strong and EM ones~\cite{Brodsky:1987bb}. The calculation in the framework of QCD of this mixed contribution is a hard task, as stated in sub-section~\ref{thbkg}, also because the hadronization process of the $gg\gamma$ into the final baryon-antibaryon pair does occur at the non-perturbative regime of QCD. \\
Furthermore, as we have shown in section~\ref{secpipi}, in the cases where the purely strong contribution is suppressed, for example in $G$-parity violating decays, the contribution of the mixed strong-EM term (related to $|\mathcal A^{gg \gamma}|$) cannot be neglected~\cite{Ferroli:2016jri}.\\
The BR of the $J/\psi\to\bb$ decays contains the interference term between the sum of the purely strong and mixed sub-amplitudes, $\big(\aggg_{\bb}+\agg_{\bb}\big)$, for which we assume the same phase~\cite{Korner:1986vi}, and the purely EM sub-amplitude, $\ag_{\bb}$. In principle, we can obtain the relative phase between these two contributions $\big(\aggg_{\bb}+\agg_{\bb}\big)$ and $\ag_{\bb}$, called $\varphi$, by studying the $J/\psi$-resonance line-shape in processes as $e^+e^-\to J/\psi\to \bb$.\\
From a theoretical point of view, all sub-amplitudes should become real at a sufficiently high energy~\cite{Brodsky:1981kj,Chernyak:1984bm,Baldini:1996hc}, i.e., at a fully perturbative QCD regime. On the other hand, it is quite difficult to establish whether such a regime, which entails maximum (positive or negative) interference, is attained already at the $J/\psi$ mass. \\
A recent measurement~\cite{Ablikim:2012eu} of the relative phase $\varphi$, performed by exploiting the decay of the $J/\psi$ meson into nucleon-antinucleon, gave
$$
\varphi=(88.7\pm 8.1)^\circ\,,
$$
which is in agreement with the no-interference case.\\
In order to describe the decay amplitude of the $\jp$ meson into baryon-antibaryon pairs of the spin-1/2 flavour SU(3) octet we define a model~\cite{Ferroli:2019nex} based on the effective Lagrangian density discussed in sub-section~\ref{efflagr}. These baryons ($p$, $n$, $\Sigma^\pm,\Sigma^0$, $\Lambda$, $\Xi^-, \Xi^0$) can be organized in the baryon matrix
$$
\b=\begin{pmatrix}
\Lambda/\sqrt{6}+\Sigma^0/\sqrt{2} & \Sigma^+ & p\\
\Sigma^- & \Lambda/\sqrt{6}-\Sigma^0/\sqrt{2} & n\\
\Xi^- & \Xi^0 & -2 \Lambda/\sqrt{6}\end{pmatrix} \,.
$$
\subsection[Theoretical background]{Theoretical background}
The BR for the decay of the $J/\psi$ meson into a baryon-antibaryon pair $\bb$ can be written as
$$
\br_{\bb} = \frac{\lmo\vec{p}\,\rmo}{8\pi  M_{J/\psi}^2 \Gamma_{J/\psi}}\left|\mathcal A^{ggg}_{\bb} + \mathcal A_{\bb}^{gg \gamma} + \mathcal A_{\bb}^\gamma\right|^2 \,,
$$
see Eq.~\eqref{eq.BRBBbar11}, where $\vec{p}$ is the three-momentum of the baryon (antibaryon) in the $\bb$ center of mass frame ($J/\psi$ rest frame). 
\\
As a consequence of the amplitude decomposition, the BR can be written as the sum of four contributions, i.e., 
\begin{equation}
\br_{\bb} \equiv \br^{ggg}_{\bb} + \br^{gg\gamma}_{\bb} + \br^{\gamma}_{\bb} + \br^{\rm int} \,,
\label{eq.strutturaBR}
\end{equation}
where the symbol $\br_{\bb}^{\lambda}$ stands for the BR due to the $\lambda$ intermediate state, with $\lambda=ggg,\,gg\gamma,\,\gamma$, while the last term accounts for the interference among the sub-amplitudes.\\
This term, as already stated, depends only on the relative phase $\varphi$ between the sub-amplitudes $\big(\aggg_{\bb}+\agg_{\bb}\big)$ and $\ag_{\bb}$, because we assume that purely strong, $\aggg_{\bb}$, and the mixed strong-EM sub-amplitude, $\agg_{\bb}$, have the same phase, i.e., they are relatively real and positive~\cite{Korner:1986vi}. Using the effective Lagrangian density of Eq.~\eqref{eq.dens.lagr}, the total decay amplitude can be parametrized in terms of the phase $\varphi$ and the coupling constants belonging to the set $\mathcal{C}=\{G_0,D_e,D_m,F_e,F_m\}$. Such five coupling constants are linear combinations of those appearing explicitly as coefficients of the traces in Eq.~\eqref{eq.dens.lagr}, i.e., $g,\,d,\,f,\,d',\,f'$.\\
In particular, $G_0$ is related to the coupling constant $g$; $D_e,\,F_e$ to $d,\,g$, which refer to the EM breaking effects, and $D_m,\,F_m$ to $d',\,g'$, describing the mass difference breaking effects.\\
We define two more subsets of coupling constants
$$
\mathcal{C}_{\rm strong}=\{G_0,\,D_m,\,F_m\} \,, \ \ \ \mathcal{C}_{\rm EM}=\{D_e,\,F_e\} \,, \ \ \ \mathcal C = \mathcal{C}_{\rm strong} \cup \mathcal{C}_{\rm EM} \,,
$$
with, as said before, the strong and EM coupling constants of these two subsets, respectively relatively real and positive~\cite{Zhu:2015bha}. %
{\textcolor{\colormod}{%
This assumption does not affect the results because the EM and mass difference effects, parametrized by $D_m$, $F_m$, $D_e$ and $F_e$, represent sub-leading contributions with respect to $G_0$, and hence the eventual presence of non-vanishing imaginary parts would leave almost unchanged the dominant strong coupling. A similar hypothesis is made also in other works, see for example Refs.~\cite{Jousset:1988ni,Haber:1985cv}. %
}%
}%
It follows that the only non-zero relative phase is $\varphi$, the one between strong and EM interactions. Moreover, since such a phase is mainly due to the dominant coupling constant $G_0$, we assume that $\varphi$ does not depend on the $\bb$ final state.\\
To account for the presence of the mixed strong-EM sub-amplitude $\agg_{\bb}$, we introduce a new parameter $R$, that represents the ratio between the mixed strong-EM and the purely strong sub-amplitude
\begin{equation}
R = {\mathcal A^{gg \gamma}_{\bb} \over \mathcal A^{ggg}_{\bb}}\,.\label{eq:ratio}
\end{equation}
We assume that the dependence on the hadronization processes of the $gg \gamma$ and the $ggg$ intermediate states cancels out in the ratio of the two sub-amplitudes, so that the new parameter $R$ has the same value for each $\bb$ final state, like the relative phase $\varphi$ and the five coupling constants of the set $\mathcal C$.
Being proportional to the baryon charge~\cite{Claudson:1981fj,Chernyak:1984bm}, the sub-amplitude $\agg_{\bb}$, and hence the parameter $R$ are non-vanishing only for charged baryons.
\\
Asymptotically, when QCD is in perturbative regime, for $q^2\gg\Lambda^2_{\rm QCD}$, from Eq.~\eqref{eq.ratio1} the ratio $R$ becomes real and it can be written as 
\begin{equation}
R_{\rm pQCD} (q^2)
\equiv\left.{\mathcal A^{gg \gamma}_{\bb} \over \mathcal A^{ggg}_{\bb}}\right|_{\rm pQCD} 
= \lim_{q^2 \to +\infty} R(q^2)
= -{4 \over 5} {\alpha \over \alpha_S(q^2)} \,.
\label{eq:RpQCD}	
\end{equation}
The realization of the perturbative regime at the $J/\psi$ mass is not well established, as already discussed, so the value $R_{\rm pQCD}$ is not a good approximation for our parameter $R$. %
{\textcolor{\colormod}{%
We do retain the reality of $R$, following Refs.~\cite{Kopke:1988cs,Chernyak:1984bm,Claudson:1981fj}, as the main working hypothesis. %
}%
}\\
In light of that, we write the amplitude $\mathcal{A}_{\bb}$ for a given decay $J/\psi\to\bb$ as a combination of the coupling constants of the set $\mathcal{C}=\{G_0,D_e,D_m,F_e,F_m\}$, of the parameter $R$ and using the relative phase $\varphi$.
The amplitudes for the $J/\psi\to\bb$ decays for the nine baryon-antibaryon pair of the spin-1/2 SU(3) octet are reported in Table~\ref{tab:A.par}.
\begin{table}
\vspace{-2mm}
\centering
\caption{Amplitudes parametrization.}
\smallskip
\label{tab:A.par} 
\begin{tabular}{r|r}
\hline\hline\noalign{\smallskip}
\bb & $\mathcal A_{\bb}=\mathcal A^{ggg}_{\bb} + \mathcal A^{gg\gamma}_{\bb} + \mathcal A^\gamma_{\bb}$ \\
\noalign{\smallskip}\hline\hline\noalign{\smallskip}%
$\Sigma^0 \overline \Sigma{}^0$ & $(G_0 + 2 D_m) e^{i \varphi} + D_e$ \\
\hline
$\Lambda \overline \Lambda$ & $(G_0 - 2 D_m) e^{i \varphi} - D_e$ \\
\hline
$\Lambda \overline \Sigma{}^0 +$ c.c. & $\sqrt{3}\,D_e$ \\
\hline
$p \overline p$ & $(G_0 - D_m + F_m)(1+R) e^{i \varphi} + D_e + F_e$ \\
\hline
$n \overline n$ & $(G_0 - D_m + F_m) e^{i \varphi} - 2\,D_e$ \\
\hline
$\Sigma^+ \overline \Sigma{}^-$ & $(G_0 + 2 D_m)(1+R) e^{i \varphi} + D_e + F_e$ \\
\hline
$\Sigma^- \overline \Sigma{}^+$ & $(G_0 + 2 D_m)(1+R) e^{i \varphi} + D_e - F_e$ \\
\hline
$\Xi^0 \overline \Xi{}^0$ & $(G_0 - D_m - F_m) e^{i \varphi} - 2\,D_e$ \\
\hline
$\Xi^- \overline \Xi{}^+$ & $(G_0 - D_m - F_m)(1+R) e^{i \varphi} + D_e - F_e$ \\
\noalign{\smallskip}\hline\hline
\end{tabular}
\end{table}\\
The purely EM $\ag_{\bb}$ is the combination of the coupling constants of the subset $\mathcal{C}_{\rm EM}$, the purely strong $\aggg_{\bb}$ is the combination of the coupling constants of the subset $\mathcal{C}_{\rm strong}$ multiplied by the phase $e^{i\varphi}$ and the mixed strong-EM sub-amplitude $\agg_{\bb}$, which is present only for charged baryons, is given by $\agg_{\bb}=R\,\aggg_{\bb}$.\\
Considering the decay of the $\jp$ meson into a pair of nucleons, i.e., $\bb=p\overline{p},n\overline{n}$, we have the following sub-amplitudes
\begin{gather}
\label{eqs.allamp.1}
\ag_{p\overline{p}} = D_e+F_e\,, \ \ \ \ \ \ \ \aggg_{p\overline{p}} = \left( G_0-D_m+F_m\right) e^{i\varphi}\,, \nonumber	\\
\agg_{p\overline{p}} = R\left( G_0-D_m+F_m\right) e^{i\varphi}\,, \ \ \ \ \ \ \ \ag_{n\overline{n}} = -2D_e\,, \\
\aggg_{n\overline{n}} = \left( G_0-D_m+F_m\right) e^{i\varphi} \,, \ \ \ \ \ \ \ \agg_{n\overline{n}} = 0\,, \nonumber
\end{gather}
where it can be seen that the neutron-antineutron final state has the same purely strong sub-amplitude of the proton-antiproton one, different purely EM and, being neutral, vanishing mixed sub-amplitude.\\
For the sigma baryons, i.e., $\bb=\Sigma^+ \overline \Sigma{}^-,\Sigma^- \overline \Sigma{}^+,\Sigma^0 \overline \Sigma{}^0$, we have the following sub-amplitudes
\begin{gather}
\label{eqs.allamp.2}
\ag_{\Sigma^+ \overline \Sigma{}^-} = D_e+F_e\,, \ \ \ \aggg_{\Sigma^+ \overline \Sigma{}^-} = \left( G_0+2D_m\right) e^{i\varphi}\,, \ \ \ \agg_{\Sigma^+ \overline \Sigma{}^-} = R\left( G_0+2D_m\right) e^{i\varphi}\,, \nonumber\\
\ag_{\Sigma^- \overline \Sigma{}^+} = D_e-F_e\,, \ \ \ \aggg_{\Sigma^- \overline \Sigma{}^+} = \left( G_0+2D_m\right) e^{i\varphi}\,, \ \ \ \agg_{\Sigma^- \overline \Sigma{}^+} = R\left( G_0+2D_m\right) e^{i\varphi}\,, \nonumber\\
\ag_{\Sigma^0 \overline \Sigma{}^0} = D_e\,, \ \ \ \aggg_{\Sigma^0 \overline \Sigma{}^0} = \left( G_0+2D_m\right) e^{i\varphi}\,, \ \ \ \agg_{\Sigma^0 \overline \Sigma{}^0} = 0 \,,
\end{gather}
they have, as expected, the same purely strong sub-amplitude, but different purely EM ones.\\
The sub-amplitudes for the $\Lambda$ particle are
\be
\label{eqs.allamp.3}
\ag_{\Lambda \overline \Lambda} = -D_e\,, \ \ \ \ \ \aggg_{\Lambda \overline \Lambda} = \left( G_0-2D_m\right) e^{i\varphi}\,, \ \ \ \ \ \agg_{\Lambda \overline \Lambda} = 0\,,
\en
while those for the $\Xi$ baryons, i.e., $\bb=\Xi^0 \overline \Xi{}^0,\Xi^- \overline \Xi{}^+$ are
\begin{gather}
\label{eqs.allamp.4}
\ag_{\Xi^0 \overline \Xi{}^0} = -2D_e\,, \ \ \ \ \ \aggg_{\Xi^0 \overline \Xi{}^0} = \left( G_0-D_m-F_m\right) e^{i\varphi}\,, \ \ \ \ \ \agg_{\Xi^0 \overline \Xi{}^0} = 0\,, \nonumber \\
\ag_{\Xi^- \overline \Xi{}^+} = D_e-F_e\,, \ \ \ \ \ \aggg_{\Xi^- \overline \Xi{}^+} = \left( G_0-D_m-F_m\right) e^{i\varphi}\,, \\
\agg_{\Xi^- \overline \Xi{}^+} = R\left( G_0-D_m-F_m\right) e^{i\varphi} \,. \nonumber
\end{gather}
The generic total amplitude for the $\jp\to\bb$ decay is parametrized as
\begin{eqnarray}
\mathcal{A}_{\bb}
=\aggg_{\bb}(1+R)e^{i\varphi}+\ag_{\bb}
\equiv
\mathcal{S}_{\bb}\,e^{i\varphi}+\ag_{\bb}
\,,\label{eq:simple-form0}
\end{eqnarray}
where $\mathcal{S}_{\bb}=\aggg_{\bb}(1+R)$ and $\ag_{\bb}$ are real quantities to be determined by a minimization procedure, fitting the model predictions of the BRs to the corresponding experimental values. The total amplitude is defined up to an arbitrary, ineffective overall phase. By setting this phase to have $\mathcal{S}_{\bb}$ always positive, hence $\mathcal{S}_{\bb}=|\mathcal{S}_{\bb}|$, the sub-amplitude $\ag_{\bb}$ could be positive or negative, i.e., $\ag_{\bb}=|\ag_{\bb}|$ or $\ag_{\bb}=|\ag_{\bb}|e^{\pm i\pi}$. So that, the amplitude can be redefined up to an overall sign as
\be
\mathcal{A}_{\bb} \to \mathcal{A}_{\bb}=
|\mathcal{S}_{\bb}|e^{i \varphi_{\bb}}+|\ag_{\bb}|\,,
\label{eq:simple-form}
\en
where $\varphi_{\bb}=\varphi$ if $\ag_{\bb}>0$ and $\varphi_{\bb}=\varphi\pm\pi$ if $\ag_{\bb}<0$.
This form is useful in order to make comparisons with the moduli of sub-amplitudes and the relative phase that the BESIII Collaboration~\cite{Ablikim:2012eu} has obtained by fitting the data with a phenomenological parametrization of the amplitude.
The choice between $\varphi_{\bb}=\varphi+\pi$ and $\varphi_{\bb}=\varphi-\pi$, when $\ag_{\bb}$ is negative, is guided by the request that the total relative phase has to be in a given determination, for instance, $\varphi_{\bb}\in[0,2\pi]$. Actually, since the  experimental observable is the modulus squared of the amplitude $\mathcal{A}_{\bb}$, which depends only on the cosine of the relative phase, being
\be
\left| \mathcal{A}_{\bb}\right|^2=
|\mathcal{S}_{\bb}|^2+|\ag_{\bb}|^2
+2|\mathcal{S}_{\bb}||\ag_{\bb}|\cos\lt
\varphi_{\bb}\rt\,,
\nen
the ambiguity between the two values $\varphi^{\rm exp}_{\bb}$ and $2\pi-\varphi_{\bb}^{\rm exp}$, with $\varphi_{\bb}^{\rm exp}\in[0,\pi]$ and hence $2\pi-\varphi_{\bb}^{\rm exp}\in[\pi,2\pi]$, cannot be resolved. In other words, both values $\pm \varphi_{\bb}$, $|\mathcal{S}_{\bb}|$ and $|\ag_{\bb}|$ being equal, give the same modulus squared, because $|\mathcal{A}_{\bb}|^2=|\mathcal{A}^*_{\bb}|^2$,
where $\mathcal{A}^*_{\bb}$ is the complex conjugate of $\mathcal{A}_{\bb}$.
\subsection[Experimental data]{Experimental data}
At present, data are available for eight out of the nine decays, their values are reported in Table~\ref{tab:BRJpsi.pdg}. The decay $J/\psi\to\Sigma^-\overline{\Sigma}{}^+$ is the only one that has not yet been observed.%
\begin{table} [ht!]
\vspace{-2mm}
\centering
\caption{Branching ratios data from PDG~\cite{Tanabashi:2018oca} and BESIII experiment~\cite{Ablikim:2017tys}.}
\smallskip
\label{tab:BRJpsi.pdg} 
\begin{tabular}{r|r|r} 
\hline\hline\noalign{\smallskip}
Decay process & Branching ratio & Error \\
\noalign{\smallskip}\hline\hline\noalign{\smallskip}%
$J/\psi \to \Sigma^0 \overline \Sigma{}^0$ & $(1.164 \pm 0.004) \times 10^{-3}$ & $0.34 \%$ \\
\hline
$J/\psi \to \Lambda \overline \Lambda$ & $(1.943 \pm 0.003) \times 10^{-3}$ & $0.15 \%$ \\
\hline
$J/\psi \to \Lambda \overline \Sigma{}^0 + {\rm c.c.} $ & $(2.83 \pm 0.23) \times 10^{-5}$ & $8.13 \%$ \\
\hline
$J/\psi \to p \overline p$ & $(2.121 \pm 0.029) \times 10^{-3}$ & $1.37 \%$ \\
\hline
$J/\psi \to n \overline n$ & $(2.09 \pm 0.16) \times 10^{-3}$ & $7.66 \%$ \\
\hline
$J/\psi \to \Sigma^+ \overline \Sigma{}^-$ & $(1.50 \pm 0.24) \times 10^{-3}$ & $16.00 \%$ \\
\hline
$J/\psi \to \Xi^0 \overline \Xi{}^0$ & $(1.17 \pm 0.04) \times 10^{-3}$ & $3.42 \%$ \\
\hline
$J/\psi \to \Xi^- \overline \Xi{}^+$ & $(9.7 \pm 0.8) \times 10^{-4}$ & $8.25 \%$ \\
\noalign{\smallskip}\hline\hline
\end{tabular}
\end{table}\\%
From the BR of the decay $J/\psi \to (\Lambda \overline \Sigma{}^0 + {\rm c.c.})$, which is purely EM, see the third row of Table~\ref{tab:A.par}, we can extract the modulus of $D_e$ as
$$
|D_e|=\sqrt{16 \pi M_{J/\psi} \Gamma_{J/\psi}\br(J/\psi \to \big(\Lambda \overline \Sigma{}^0 + {\rm c.c.}\big) \over 3 \beta_{\Lambda \overline \Sigma{}^0}} \,,
$$
where the velocity of the outgoing baryon in the $J/\psi$-CM system is defined as
$$
\beta_{\Lambda \overline \Sigma{}^0} \equiv \sqrt{1-{2(M_{\Sigma^0}^2+M_{\Lambda}^2) \over M_{J/\psi}^2} + {(M_{\Sigma^0}^2-M_{\Lambda}^2)^2 \over M_{J/\psi}^4}} \,.
$$
Using the experimental value of BR$_{\Lambda \overline \Sigma{}^0}$, given in the third row of Table~\ref{tab:BRJpsi.pdg}, we obtain
$$
|D_e|=(4.52 \pm 0.20) \times 10^{-4} \ \rm GeV \,,
$$
where the data on the masses are taken from PDG~\cite{\pdg}, see Table~\ref{tab:spin1.2.mass.pdg}, the same data are used also in the following. Analogously, the EM BR of the decay of the $J/\psi$ meson into proton-antiproton is given by
$$
\br^\gamma_{p \overline p} = {\beta_{p \overline p} \over 16 \pi M_{J/\psi}\Gamma_{J/\psi}} |D_e+F_e|^2\,,
$$
whit
$$
\beta_{p \overline p}=\sqrt{1-\frac{4M_p^2}{M_{J/\psi}^2}}\,.
$$
The EM BR is related to the $e^+e^-\to p\overline{p}$ non-resonant cross section at the $J/\psi$ mass by the formula, see Eq.\eqref{eq:b-gamma},
\begin{equation}
\label{eq.BRgamma}
\br^\gamma_{\bb} = \br_{\mu\mu} \, {\sigma_{\ee \to \bb}(M_{\jp}^2) \over \sigma^0_{\ee \to \mu^+ \mu^-}(M_{\jp}^2)} \,,
\end{equation}
where $\br_{\mu\mu}$ is the BR of the decay $J/\psi \to \mu^+\mu^-$, and $\sigma^0_{\ee \to \mu^+ \mu^-}(q^2)$ represents the bare $\ee \to \mu^+ \mu^-$ cross section, i.e., the cross section corrected for the vacuum-polarization 
$$
\sigma^0_{\ee \to \mu^+ \mu^-} (q^2)= {4 \pi \alpha^2 \over 3 q^2} \,.
$$
The modulus of the sum of the two parameters $D_e$ and $F_e$ has, therefore, the expression 
$$
|D_e + F_e| = \sqrt{12 \, \br_{\mu\mu} M_{J/\psi}^3 \Gamma_{J/\psi} \,\sigma_{\ee \to p \overline p}(M_{\jp}^2) \over \alpha^2 \beta_{p \overline p}} \,.
$$

The most recent data on the $p \overline p$ cross section obtained by the BESIII Collaboration~\cite{Ablikim:2019njl} give at the $J/\psi$ mass the cross section
\begin{eqnarray}
\sigma_{\ee\to p \overline p} (M_{\jp}^2)&=& {6912 \,\pi \alpha^2 (M_{J/\psi}^2\!\!+\!2M_p^2) \over M_{J/\psi}^{12}\,{\rm GeV^{-8}}} \times\!\left[ \ln^2\left(\!{M_{J/\psi}^2 \over 0.52^2\,\rm GeV^2}\!\right)\!+\!\pi^2 \right]^{\!-2} \,. \no \\
\end{eqnarray}
Using this result together with~\cite{Tanabashi:2018oca}
$$
\br_{\mu\mu} = (5.961 \pm 0.033) \times 10^{-2} \,,
$$
for the $J/\psi \to \mu^+\mu^-$ BR, we obtain
\begin{equation}
\label{eq.brEMpp}
\br^{\gamma,\rm exp}_{p\overline{p}} = (8.46 \pm 0.79) \times 10^{-5} \,,
\end{equation}
from which
$$
|D_e + F_e| = (1.240 \pm 0.061) \times 10^{-3} \ \rm GeV \,.
$$
The errors include both statistical and systematic contributions due to the cross section fit to the BESIII data.
\subsection[Results]{Results}
We need to perform a minimization, fitting the model predictions to the experimental data, in order to obtain the seven free parameters, five of them being part of the $\mathcal C$ set, previously defined, plus the parameter $R$ and the relative phase $\varphi$. We define the $\chi^2$ function
\be
\label{eq.chi.2}
\chi^2\left(\mathcal{C};R,\varphi\right) = \sum_{\bb} \left({\br_{\bb}^{\rm th}-\br_{\bb}^{\rm exp} \over \delta \br_{\bb}^{\rm exp}}\right)^2 + \left({\br_{p\overline{p}}^{\gamma,\rm th}-\br_{p\overline{p}}^{\gamma,\rm exp} \over \delta \br_{p\overline{p}}^{\gamma,\rm exp}}\right)^2 \,,
\en
where the sum runs over the eight baryon-antibaryon pairs, \bb, for which experimental data are available, reported in Table~\ref{tab:BRJpsi.pdg} and the last term imposes the constraint of the EM BR reported in Eq.~\eqref{eq.brEMpp}. 
\\
The numerical minimization is performed with respect to the five coupling constants of the set  $\mathcal C=\{G_0,D_e,D_m,F_e,F_m\}$, the ratio $R$ defined in Eq.~\eqref{eq:ratio}, and the relative phase $\varphi$.\\
The best values of the parameters resulting from the numerical minimization are shown in Table~\ref{tab:results.parQ}. The errors have been obtained by means of a Monte Carlo Gaussian simulation.
\begin{table} [ht!]
\vspace{-2mm}
\centering
\caption{Values of the parameters from the $\chi^2$ minimization.}
\smallskip
\label{tab:results.parQ} 
\begin{tabular}{r|r} 
\hline\hline\noalign{\smallskip}
$G_0$ & $(5.73511 \pm 0.0059) \times 10^{-3} \ \rm GeV$ \\
\hline
$D_e$ & $(4.52 \pm 0.19) \times 10^{-4} \ \rm GeV$ \\
\hline
$D_m$ & $(-3.74 \pm 0.34) \times 10^{-4} \ \rm GeV$ \\
\hline
$F_e$ & $(7.91 \pm 0.62) \times 10^{-4} \ \rm GeV$  \\
\hline
$F_m$ & $(2.42 \pm 0.12) \times 10^{-4} \ \rm GeV$ \\
\hline
$\varphi$ & $1.27 \pm 0.14 = (73 \pm 8)^\circ $  \\
\hline
$R$ & $(-9.7 \pm 2.1) \times 10^{-2}$ \\
\noalign{\smallskip}\hline\hline
\end{tabular}
\end{table}\\
Using the obtained parameters we can calculate the value of the BRs for each baryon-antibaryon final state. The BRs are reported in Table~\ref{tab:BRcalc.J.psi}, where they are compared with the corresponding experimental values (see Table~\ref{tab:BRJpsi.pdg}) used to perform the $\chi^2$ minimization.
\begin{table} [ht!]
\vspace{-2mm}
\centering
\caption{Branching ratios from PDG~\cite{Tanabashi:2018oca} (second column), from parameters of Table~\ref{tab:results.parQ} (third column) end their difference in units of the total error (fourth column).}
\smallskip
\label{tab:BRcalc.J.psi} 
\begin{tabular}{l|l|l|l} 
\hline\hline\noalign{\smallskip}
\bb & BR$^{\rm PDG}_{\bb}\times 10^3$ & BR$_{\bb}\times 10^3$ & ${\Delta {\br} \over \sum \sigma_\br}$ \\
\noalign{\smallskip}\hline\hline\noalign{\smallskip}%
$\Sigma^0 \overline \Sigma{}^0$ & $1.164 \pm 0.004 $ & $1.160 \pm 0.041 $ & $\sim 0.09$ \\
\hline
$\Lambda \overline \Lambda$ & $1.943 \pm 0.003 $ & $1.940 \pm 0.055 $ & $\sim 0.05$ \\
\hline
$\Lambda \overline \Sigma{}^0 +$ c.c. & $0.0283 \pm 0.0023$ & $0.0280 \pm 0.0024$ & $\sim 0.06$ \\
\hline
$p \overline p$ & $2.121 \pm 0.029 $ & $2.10 \pm 0.16 $ & $\sim 0.1$ \\
\hline
$n \overline n$ & $2.09 \pm 0.16 $ & $2.10 \pm 0.12 $ & $\sim 0.04$ \\
\hline
$\Sigma^+ \overline \Sigma{}^-$ & $1.50 \pm 0.24$ & $1.110 \pm 0.086$ & $\sim 1$ \\
\hline
$\Sigma^- \overline \Sigma{}^+$ & $/$ & $0.857 \pm 0.051 $ & $/$ \\
\hline
$\Xi^0 \overline \Xi{}^0$ & $1.17 \pm 0.04 $ & $1.180 \pm 0.072 $ & $\sim 0.09$ \\
\hline
$\Xi^- \overline \Xi{}^+$ & $0.97 \pm 0.08 $ & $0.979 \pm 0.065 $ & $\sim 0.06$ \\
\noalign{\smallskip}\hline\hline
\end{tabular}
\end{table}\\
It is interesting to notice that the obtained value for the BR of the unobserved $J/\psi \to \Sigma^- \overline \Sigma{}^+$ decay, i.e,
\be
\label{eq.pred.BR.sigm-}
{\rm BR}_{\Sigma^- \overline \Sigma{}^+} = (0.857 \pm 0.051) \times 10^{-3} \,,
\en
represents a prediction of our model. Moreover, it is important to notice that only one of the obtained BRs is quite different from its corresponding PDG value, the BR of the $J/\psi \to \Sigma^+ \overline \Sigma{}^-$ decay, in fact we found the value
\be
\label{eq.pred.BR.sigm+}
{\rm BR}_{\Sigma^+ \overline \Sigma{}^-} = (1.110 \pm 0.086) \times 10^{-3} \,,
\en
to be compared with the PDG value
$$
{\rm BR}^{\rm PDG}_{\Sigma^- \overline \Sigma{}^+} = (1.50 \pm 0.24) \times 10^{-3}
$$
that has also a large relative error of about the 16\%. {\textcolor{\colormodtwo}{A recent independent preliminary analysis~\cite{Liangsigma} confirms our prediction for the BR of the $J/\psi \to \Sigma^+ \overline \Sigma{}^-$ decay. The analysis is actually only preliminary and the found value is ${\rm BR}_{\Sigma^+ \overline \Sigma{}^-}=(1.115 \pm 0.005^{\rm stat}) \times 10^{-3}$.}}\\
\begin{figure} [ht!]
	\begin{center}
		\includegraphics[width=.7\columnwidth]{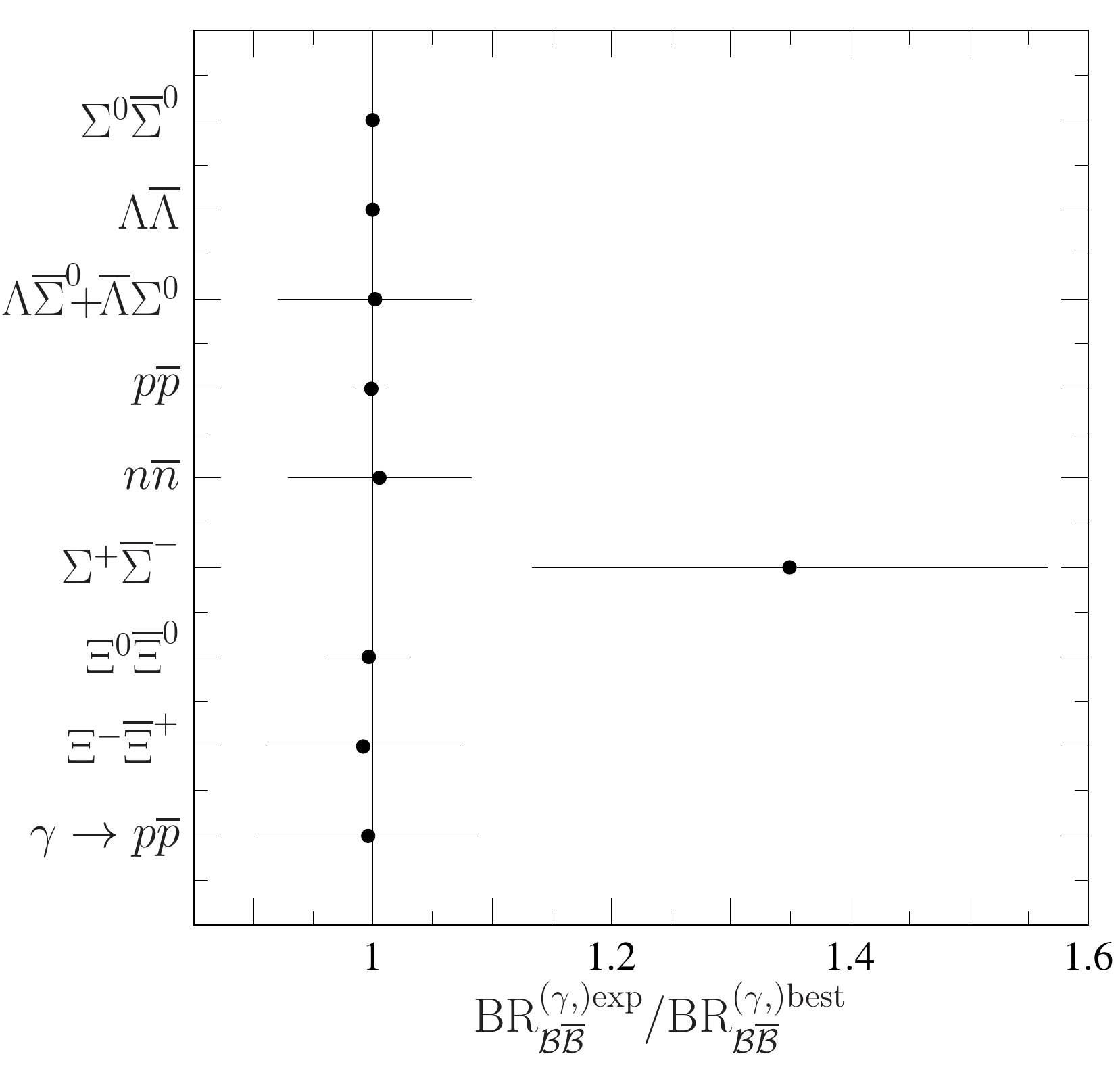}
		\caption{Ratios between the experimental input values of BRs and their best values obtained by minimizing the $\chi^2$ of Eq.~\eqref{eq.chi.2}. The lower point, at the ordinate labelled width $\gamma\to p\overline{p}$, is the contribution due to EM BR of the proton, see Eq.~\eqref{eq.brEMpp}.}\label{fig:rchi2}
	\end{center}
\end{figure}%
The ratios between the input and best values of the BRs, that represent the nine free parameters of the $\chi^2$ given in Eq.~\eqref{eq.chi.2}, are shown in Fig.~\ref{fig:rchi2}. The minimum normalized $\chi^2$ is
\be
\frac{\chi^2\left(\mathcal{C}^{\rm best};R^{\rm best},\varphi^{\rm best}\right)}{N_{
\rm dof}} &=& 1.33\,,
\label{eq:chi2}\en
where the number of degrees of freedom is
$$
N_{\rm dof}=N_{\rm const}-N_{\rm param}=2 \,,
$$
in fact we have nine constraints, $N_{\rm const}=9$, and seven free parameters, $N_{\rm param}=7$. It is possible to verify the significance of the mixed strong-EM contribution in the description of the \jp\ decay mechanism by comparing the normalized $\chi^2$ of Eq.~\eqref{eq:chi2}, obtained in the case where $R$ is considered as a free parameter, to that in which it is fixed at $R=0$, i.e.,
\be
\frac{\chi^2\left(\mathcal{C'}^{\rm best};R=0,\varphi'^{\rm best}\right)}{N_{
\rm dof}}=\frac{16.44}{3} &=& 5.48\,,
\label{eq:chi2-2}
\en
where $\mathcal{C'}^{\rm best}$ and $\varphi'^{\rm best}$ are the set of best values of the coupling constants and the best relative phase obtained in this case. Despite the quite low number of degrees of freedom and also the smallness of the best value obtained for $R$, see Table~\ref{tab:results.parQ}, this large $\chi^2$, see Eq.~\eqref{eq:chi2-2}, represents a clear indication in favor of the necessity of the mixed strong-EM contribution. %
{\textcolor{\colormod}{
In fact the most suitable criterion to compare these two hypotheses, namely: free $R$ and $R=0$, is the one provided by the $p$-value, $p(\chi^2;N_{\rm dof})$. The two p-values are
\be
\label{eq.pvalue1}
p(2.65;2)=0.266\,, \ \ \ \ \ \ \ p(16.44;3)=9.21\times 10^{-4}\,,
\en
and represent the probabilities to obtain by chance $\chi^2=2.65$ and $\chi^2=16.44$, with two and three degrees of freedom respectively, if the model is correct. %
}%
}\\
As mentioned before, the knowledge of the seven coupling constants of Table~\ref{tab:results.parQ} brings important information on the structure of the amplitudes for the considered $J/\psi \to \bb$ decays. In fact, using these values, from Eq.~\eqref{eqs.allamp.1}, Eq.~\eqref{eqs.allamp.2}, Eq.~\eqref{eqs.allamp.3} and Eq.~\eqref{eqs.allamp.4}, we can calculate each individual sub-amplitude and, hence, the corresponding contribution to the total BR, under the assumptions concerning their relative phases. The resulting purely strong, purely EM and mixed strong-EM contributions to the total BR for the nine final states are reported in Table~\ref{tab:BRsepcalc.J.psi}.\\%
\begin{table} [ht!]
\vspace{-2mm}
\centering
\caption{Purely strong (second column), purely EM (third column) and mixed (fourth column) BRs.}
\smallskip
\label{tab:BRsepcalc.J.psi} 
\begin{tabular}{l|l|l|l} 
\hline\hline\noalign{\smallskip}
\bb & $\br^{ggg}_{\bb} \times 10^{3}$ & $\br^{\gamma}_{\bb} \times 10^{5}$ & $\br^{gg\gamma}_{\bb} \times 10^{5}$ \\
\noalign{\smallskip}\hline\hline\noalign{\smallskip}%
$\Sigma^0 \overline \Sigma{}^0$            & $1.100 \pm 0.030$ & $0.902 \pm 0.076$       & $0$             \\
\hline
$\Lambda \overline \Lambda$              & $2.020 \pm 0.042$ & $0.981 \pm 0.083$       & $0$             \\
\hline
$\Lambda \overline \Sigma{}^0+$ c.c.  & $0$               & $2.83 \pm 0.24$         & $0$             \\
\hline
$p \overline p$                          & $2.220 \pm 0.085$ & $8.52 \pm 0.89$         & $2.19 \pm 0.93$ \\
\hline
$n \overline n$                          & $2.220 \pm 0.085$ & $4.50 \pm 0.38$         & $0$             \\
\hline
$\Sigma^+ \overline \Sigma{}^-$            & $1.100 \pm 0.030$ & $6.86 \pm 0.72$         & $1.08 \pm 0.46$ \\
\hline
$\Sigma^- \overline \Sigma{}^+$            & $1.090 \pm 0.030$ & $0.52 \pm 0.20$         & $1.07 \pm 0.46$ \\
\hline
$\Xi^0 \overline \Xi{}^0$                  & $1.260 \pm 0.053$ & $2.99 \pm 0.25$         & $0$             \\
\hline
$\Xi^- \overline \Xi{}^+$                  & $1.240 \pm 0.052$ & $0.43 \pm 0.16$         & $1.22 \pm 0.52$ \\
\noalign{\smallskip}\hline\hline
\end{tabular}
\end{table}%
We calculate also the ratios of the moduli of the sub-amplitudes $|\mathcal A^\gamma_{\bb}/ \mathcal A^{ggg}_{\bb}|$ and $|\mathcal A^{gg\gamma}_{\bb}/ \mathcal A^{ggg}_{\bb}|$, i.e., the moduli of the purely EM and mixed strong-EM sub-amplitudes normalized to the modulus of the purely strong sub-amplitude, the approximate results are shown in Table~\ref{tab:BRsepcalc.J.psi.ratio}.\\%
\begin{table} [ht!]
\vspace{-2mm}
\centering
\caption{Approximate values of moduli of the ratios between sub-amplitudes $\mathcal A^\gamma_{\bb}$ and $\mathcal A^{ggg}_{\bb}$ (second column), and between $\mathcal A^{gg\gamma}_{\bb}$ and $\mathcal A^{ggg}_{\bb}$ (third column).}
\smallskip
\label{tab:BRsepcalc.J.psi.ratio} 
\begin{tabular}{l|l|l} 
\hline\hline\noalign{\smallskip}
\bb $ \ $ & $|\mathcal A^\gamma_{\bb}/ \mathcal A^{ggg}_{\bb}| $ & $|\mathcal A^{gg\gamma}_{\bb}/ \mathcal A^{ggg}_{\bb}| $ \\
\noalign{\smallskip}\hline\hline\noalign{\smallskip}%
$\Sigma^0 \overline \Sigma{}^0$ & $\sim 0.09$ & $0$ \\
\hline
$\Lambda \overline \Lambda$ & $\sim 0.07$& $0$ \\
\hline
$p \overline p$ & $\sim 0.20$ & $\sim 0.1$ \\
\hline
$n \overline n$ & $\sim 0.14$ & $0$ \\
\hline
$\Sigma^+ \overline \Sigma{}^-$ & $\sim 0.25$ & $\sim 0.1$ \\
\hline
$\Sigma^- \overline \Sigma{}^+$ & $\sim 0.07$ & $\sim 0.1$ \\
\hline
$\Xi^0 \overline \Xi{}^0$ & $\sim 0.15$ & $0$ \\
\hline
$\Xi^- \overline \Xi{}^+$ & $\sim 0.06$ & $\sim 0.1$ \\
\noalign{\smallskip}\hline\hline
\end{tabular}
\end{table}%
First of all we observe that, in all cases, since by assumption it does not depend on the $\bb$ final state, the strength of the mixed strong-EM sub-amplitude relative to the dominant three-gluon one represents about the 10\% of the dominant contribution and becomes $\sim 1\%$ for the BR, see the fourth column of Table~\ref{tab:BRsepcalc.J.psi}. More intriguing is the comparison between the mixed and the purely EM contributions, third and fourth columns of Table~\ref{tab:BRsepcalc.J.psi} for the BRs and, second and third columns of Table~\ref{tab:BRsepcalc.J.psi.ratio} for the sub-amplitudes. They are always of the same order, but while for the proton and $\Sigma^+$ the modulus of the purely EM sub-amplitude is about twice the modulus of the mixed sub-amplitude, in the cases of $\Sigma^-$ and $\Xi^-$ the hierarchy is inverted. Such different behavior could be due to the different quark structure of the two pairs of baryons.\\%
\begin{table} [ht!]
\vspace{-2mm}
\centering
\caption{Moduli of sub-amplitudes $\mathcal{S}_{\bb}$, $\ag_{\bb}$ and phase $\varphi_{\bb}$, defined in Eq.~\eqref{eq:simple-form}.\label{tab:simple-form}}
\smallskip
\begin{tabular}{lc|lc|lc|r} 
\hline\hline\noalign{\smallskip}
\bb $ \ $ && $|\mathcal{S}_{\bb}|\times 10^3 $ && $|\ag_{\bb}|\times 10^4 $ && $\varphi_{\bb} $\\
\noalign{\smallskip}\hline\hline\noalign{\smallskip}%
$\Sigma^0 \overline \Sigma{}^0$ && $4.987\pm 0.065$ && $4.52\pm0.19$ && $\varphi$\\
\hline
$\Lambda \overline \Lambda$ && $6.483\pm 0.065$ && $4.52\pm0.19$ && $\pi-\varphi$\\
\hline
$\Lambda\overline{\Sigma}{}^0+$ c.c. && 0 && $7.83\pm 0.33$ && $\varphi$\\
\hline
$p \overline p$ && $5.74\pm 0.14$ && $12.43\pm 0.65$ && $\varphi$\\
\hline
$n \overline n$ && $6.351\pm 0.037$ && $9.04\pm 0.38$ && $\pi-\varphi$\\
\hline
$\Sigma^+ \overline \Sigma{}^-$ && $4.50\pm 0.12$ && $12.43\pm 0.65$ && $\varphi$\\
\hline
$\Sigma^- \overline \Sigma{}^+$ && $4.50\pm 0.12$ && $3.39\pm 0.65$ && $\pi-\varphi$\\
\hline
$\Xi^0 \overline \Xi{}^0$ && $5.867\pm 0.037$ && $9.04\pm 0.38$ && $\pi-\varphi$\\
\hline
$\Xi^- \overline \Xi{}^+$ && $5.30\pm 0.13$ && $3.39\pm 0.65$ && $\pi-\varphi$\\
\noalign{\smallskip}\hline\hline
\end{tabular}
\end{table}%
The moduli of the sub-amplitudes $\mathcal{S}_{\bb}$ and $\ag_{\bb}$, together with the phase $\varphi_{\bb}$, defined in Eq.~\eqref{eq:simple-form0} and Eq.~\eqref{eq:simple-form}, are reported in Table~\ref{tab:simple-form}. The five final states: $\Lambda\ov{\Lambda}$, $n\ov{n}$, $\Sigma^-\ov{\Sigma}^+$, $\Xi^0\ov{\Xi}^0$, $\Xi^-\ov{\Xi}^+$, have negative $\ag_{\bb}$ sub-amplitudes and then $\varphi_{\bb}=\pi-\varphi$, this is a phenomenological finding due to the values that have been obtained for the coupling constants $D_e$ and $F_e$ with the fitting procedure, see Table~\ref{tab:results.parQ}, and to the SU(3) symmetry of the model, that determines the signs of the coupling constants in the definition of the sub-amplitudes, see Table~\ref{tab:A.par}. As an example, let us consider the $p\ov{p}$ and $n\ov{n}$ final states. Using the standard parametrization of Eq.~\eqref{eq:simple-form}, the total relative phase between the two sub-amplitudes $\mathcal{S}_{\bb}e^{i\varphi}$ and $\ag_{\bb}$ differ by $180^\circ$, i.e.,
\be
\arg\lt\frac{\mathcal{S}_{p\ov{p}}e^{i\varphi}}{\ag_{p\ov{p}}}\rt=\varphi\,,\hh \arg\lt\frac{\mathcal{S}_{n\ov{n}}e^{i\varphi}}{\ag_{n\ov{n}}}\rt=\varphi\pm\pi\,.
\nen
The last result is a consequence of the negative value of $\ag_{n\ov{n}}=-2D_e$, with $D_e>0$, see Table~\ref{tab:results.parQ}.%
As shown in the sixth row of Table~\ref{tab:results.parQ} the best value of the relative phase between strong and EM sub-amplitudes is
$$
\varphi = (73\pm 8)^\circ
$$
and it agrees with the result given in Refs.~\cite{Zhu:2015bha,Baldini:1998en} and with the value given in Ref.~\cite{Ablikim:2012eu}, i.e., $(88.7 \pm 8.1)^\circ$, obtained by studying the decays of the $J/\psi$ meson into nucleon-antinucleon. There is also a very good agreement between our result and the value found in Ref.~\cite{Ablikim:2012bw}, i.e., $(76 \pm 11)^\circ$.
More in detail, by considering the relative sign between the sub-amplitudes $|\mathcal{S}_{\bb}|$ and $|\ag_{\bb}|$, defined in Eq.~\eqref{eq:simple-form}, we can distinguish between two values of the relative phase $\varphi_{\bb}$, see Table~\ref{tab:simple-form},
\be
\varphi_{\Sigma^0\overline{\Sigma}{}^0,\Lambda\overline{\Sigma}{}^0,p\overline{p},\Sigma^+\overline{\Sigma}{}^-}&=&(73\pm 8)^\circ\,,
\no\\
\varphi_{\Lambda\overline{\Lambda},n\overline{n},\Sigma^-\overline{\Sigma}{}^+,\Xi^0\overline{\Xi}{}^0,\Xi^-\overline{\Xi}{}^+}&=&(107\pm 8)^\circ\,.
\nen
The fact that the relative phase $\varphi$ is closer to $90^\circ$ rather than to $0^\circ$ or $180^\circ$, as already discussed, disagrees with the pQCD predictions, in fact, in the perturbative regime, all QCD amplitudes should be real. It follows that the obtained relative phase could be interpreted as the indication of a non-complete realization of pQCD at this energy, at least for the examined \jp\ decays. 
\\
Another result that agrees with this conclusion is the value obtained for the ratio $R$, shown in the last row of Table~\ref{tab:results.parQ}, i.e.,
$$
R = -0.097 \pm 0.021 \,.
$$
This value, in addition to confirming that the mixed strong-EM sub-amplitude is negligible with respect to purely strong one, is not compatible with that predicted by pQCD, see Eq.~\eqref{eq:RpQCD}, which is indeed
$$
 R_{\rm pQCD}(M_{J/\psi}^2) = -{4 \over 5} {\alpha \over \alpha_S(M_{J/\psi}^2)} \sim -0.030\,,
$$
where we have used $\alpha_S(M_{J/\psi}^2)\sim 0.2$~\cite{Claudson:1981fj}. Therefore, as anticipated, at the $J/\psi$ mass the perturbative regime of QCD is still not reached. A similar conclusion is also suggested in Ref.~\cite{Chanowitz:1975ee}.\\
Finally, we use the values in the second column of Table~\ref{tab:BRsepcalc.J.psi} to calculate the non-resonant $\ee \to \bb$ Born cross sections at the $\jp$ mass, $q^2=M^2_{J/\psi}$. The results are reported in Table~\ref{tab:sigma.gamma}.%
\begin{table} [ht!]
\vspace{-2mm}
\centering
\caption{Non-resonant $\ee \to \bb$ Born cross sections at $q^2=M_{J/\psi}^2$.}
\smallskip
\label{tab:sigma.gamma} 
\begin{tabular}{l|l} 
\hline\hline\noalign{\smallskip}
$\ee\to\bb$ & Cross section at the $q^2=M_{J/\psi}^2$ \\
\noalign{\smallskip}\hline\hline\noalign{\smallskip}%
$\ee \to \Sigma^0 \overline \Sigma{}^0$ & $(1.37 \pm 0.12) \ \rm pb$ \\
\hline
$\ee \to \Lambda \overline \Lambda$ & $(1.49 \pm 0.13) \ \rm pb$ \\
\hline
$\ee \to (\Lambda \overline \Sigma{}^0 + {\rm c.c.}) \ $ & $(4.30 \pm 0.36) \ \rm pb$ \\
\hline
$\ee \to p \overline p$ & $(12.9 \pm 1.4) \ \rm pb$ \\
\hline
$\ee \to n \overline n$ & $(6.84 \pm 0.58) \ \rm pb$ \\
\hline
$\ee \to \Sigma^+ \overline \Sigma{}^-$ & $(10.4 \pm 1.1) \ \rm pb$ \\
\hline
$\ee \to \Sigma^- \overline \Sigma{}^+$ & $(0.79 \pm 0.30) \ \rm pb$ \\
\hline
$\ee \to \Xi^0 \overline \Xi{}^0$ & $(4.54 \pm 0.38) \ \rm pb$ \\
\hline
$\ee \to \Xi^- \overline \Xi{}^+$ & $(0.65 \pm 0.24) \ \rm pb$ \\
\noalign{\smallskip}\hline\hline
\end{tabular}
\end{table}\\%
%
%
Currently there are no data for the majority of these cross sections, so we cannot make direct comparisons. These results represent a prediction of our model that could be useful for future experiments. We remember that the EM BR for the proton-antiproton final state is the only exception, in fact we used its experimental value, given in Eq.~\eqref{eq.brEMpp} and extracted from the non-resonant $\ee \to p \overline p$ cross section data~\cite{Ablikim:2019njl}, as a constraint in the numerical $\chi^2$ minimization.
\subsection[The case of complex $R$]{The case of complex $R$}
{\textcolor{\colormod}{%
We consider a complex ratio $R$ by introducing a new parameter, the relative phase between the purely strong sub-amplitude and the mixed one, called $\varphi_2$. In this case, from Eq.~\eqref{eq:ratio}, we can write
\begin{equation}
R = {\mathcal A^{gg \gamma}_{\bb} \over \mathcal A^{ggg}_{\bb}} = \left | {\mathcal A^{gg \gamma}_{\bb} \over \mathcal A^{ggg}_{\bb}} \right| e^{i \varphi_2}\,.\label{eq:ratio.2phase}
\end{equation}
\begin{table} [ht!]
\vspace{-2mm}
\centering
\caption{Amplitudes parameterization with a complex ratio $R$.}
\smallskip
\label{tab:A.par.2phase} 
\begin{tabular}{r|r} 
\hline\hline\noalign{\smallskip}
\bb & $\mathcal A_{\bb}=\mathcal A^{ggg}_{\bb} + \mathcal A^{gg\gamma}_{\bb} + \mathcal A^\gamma_{\bb}$ \\
\noalign{\smallskip}\hline\hline\noalign{\smallskip}%
$\Sigma^0 \overline \Sigma{}^0$ & $(G_0 + 2 D_m) e^{i \varphi} + D_e$ \\
\hline
$ \Lambda \overline \Lambda$ & $(G_0 - 2 D_m) e^{i \varphi} - D_e$ \\
\hline
$\Lambda \overline \Sigma{}^0 +$ c.c. & $\sqrt{3}\,D_e$ \\
\hline
$p \overline p$ & $(G_0 - D_m + F_m)(1+|R|e^{i \varphi_2}) e^{i \varphi} + D_e + F_e$ \\
\hline
$n \overline n$ & $(G_0 - D_m + F_m) e^{i \varphi} - 2\,D_e$ \\
\hline
$\Sigma^+ \overline \Sigma^{}-$ & $(G_0 + 2 D_m)(1+|R|e^{i \varphi_2}) e^{i \varphi} + D_e + F_e$ \\
\hline
$\Sigma^- \overline \Sigma^{}+$ & $(G_0 + 2 D_m)(1+|R|e^{i \varphi_2}) e^{i \varphi} + D_e - F_e$ \\
\hline
$\Xi^0 \overline \Xi^0$ & $(G_0 - D_m - F_m) e^{i \varphi} - 2\,D_e$ \\
\hline
$\Xi^- \overline \Xi^+$ & $(G_0 - D_m - F_m)(1+|R|e^{i \varphi_2}) e^{i \varphi} + D_e - F_e$ \\
\noalign{\smallskip}\hline\hline
\end{tabular}
\end{table}%
\begin{table} [ht!]
\vspace{-2mm}
\centering
\caption{Values of the parameters from the $\chi^2$ minimization in the case of a complex ratio $R$.}
\smallskip
\label{tab:results.par.2phase} 
\begin{tabular}{r|r} 
\noalign{\smallskip}\hline\hline\noalign{\smallskip}%
$G_0$ & $(5.73488 \pm 0.0040) \times 10^{-3} \ \rm GeV$ \\
\hline
$D_e$ & $(4.52 \pm 0.15) \times 10^{-4} \ \rm GeV$ \\
\hline
$D_m$ & $(-3.70 \pm 0.19) \times 10^{-4} \ \rm GeV$ \\
\hline
$F_e$ & $(7.88 \pm 0.28) \times 10^{-4} \ \rm GeV$  \\
\hline
$F_m$ & $(2.38 \pm 0.59) \times 10^{-4} \ \rm GeV$ \\
\hline
$\varphi$ & $1.29 \pm 0.11 = (74 \pm 6)^\circ $  \\
\hline
$\varphi_2$ & $3.59 \pm 0.81 = (206 \pm 46)^\circ $  \\
\hline
$|R|$ & $(11.5 \pm 1.7) \times 10^{-2}$ \\
\noalign{\smallskip}\hline\hline
\end{tabular}
\end{table}%
The amplitudes for the decays $J/\psi\to\bb$, under this new hypothesis, are reported in Table~\ref{tab:A.par.2phase}. By performing the same fitting procedure used in the case of real $R$, we obtain the values shown in Table~\ref{tab:results.par.2phase}.%
\begin{table} [ht!]
\vspace{-2mm}
\centering
\caption{Branching ratios from PDG~\cite{Tanabashi:2018oca} (second column), from parameters of Table~\ref{tab:results.par.2phase} (third column).}
\smallskip
\label{tab:BRcalc.J.psi.2phase} 
\begin{tabular}{l|l|l|l} 
\hline\hline\noalign{\smallskip}
\bb & BR$^{\rm PDG}_{\bb}\times 10^3$ & BR$_{\bb}\times 10^3$ \\
\noalign{\smallskip}\hline\hline\noalign{\smallskip}%
$\Sigma^0 \overline \Sigma{}^0$ & $1.160 \pm 0.041 $ & $1.160 \pm 0.028 $ \\
\hline
$\Lambda \overline \Lambda$ & $1.940 \pm 0.055 $ & $1.940 \pm 0.039 $ \\
\hline
$\Lambda \overline \Sigma{}^0 +$ c.c. & $0.0283 \pm 0.0023$ & $0.0280 \pm 0.0019$ \\
\hline
$p \overline p$ & $2.121 \pm 0.029 $ & $2.20 \pm 0.27 $ \\
\hline
$n \overline n$ & $2.09 \pm 0.16 $ & $2.08 \pm 0.08 $ \\
\hline
$\Sigma^+ \overline \Sigma^{}-$ & $1.50 \pm 0.24 $ & $1.20 \pm 0.14 $ \\
\hline
$\Sigma^- \overline \Sigma^{}+$ & $/$ & $0.91 \pm 0.10 $ \\
\hline
$\Xi^0 \overline \Xi^0$ & $1.17 \pm 0.04 $ & $1.180 \pm 0.049 $ \\
\hline
$\Xi^- \overline \Xi^+$ & $0.97 \pm 0.08 $ & $1.00 \pm 0.12 $ \\
\noalign{\smallskip}\hline\hline
\end{tabular}
\end{table}%
\begin{table} [ht!]
\vspace{-2mm}
\centering
\caption{Purely strong (second column), purely EM (third column) and mixed (fourth column) BRs in the case of a complex ratio $R$.}
\smallskip
\label{tab:BRsepcalc.J.psi.2phase} 
\begin{tabular}{l|l|l|l} 
\hline\hline\noalign{\smallskip}
\bb & $\br^{ggg}_{\bb} \times 10^{3}$ & $\br^{\gamma}_{\bb} \times 10^{5}$ & $\br^{gg\gamma}_{\bb} \times 10^{5}$ \\
\noalign{\smallskip}\hline\hline\noalign{\smallskip}%
$\Sigma^0 \overline \Sigma{}^0$            & $1.100 \pm 0.017$ & $0.903 \pm 0.061$       & $0$             \\
\hline
$\Lambda \overline \Lambda$              & $2.010 \pm 0.024$ & $0.982 \pm 0.066$       & $0$             \\
\hline
$\Lambda \overline \Sigma{}^0+$ c.c.  & $0$               & $2.83 \pm 0.19$         & $0$             \\
\hline
$p \overline p$                          & $2.210 \pm 0.043$ & $8.47 \pm 0.43$         & $2.97 \pm 0.87$ \\
\hline
$n \overline n$                          & $2.210 \pm 0.043$ & $4.50 \pm 0.30$         & $0$             \\
\hline
$\Sigma^+ \overline \Sigma^{}-$            & $1.100 \pm 0.017$ & $6.82 \pm 0.35$         & $1.49 \pm 0.44$ \\
\hline
$\Sigma^- \overline \Sigma^{}+$            & $1.090 \pm 0.017$ & $0.500 \pm 0.094$         & $1.47 \pm 0.43$ \\
\hline
$\Xi^0 \overline \Xi^0$                  & $1.260 \pm 0.027$ & $2.99 \pm 0.20$         & $0$             \\
\hline
$\Xi^- \overline \Xi^+$                  & $1.240 \pm 0.026$ & $0.410 \pm 0.077$         & $1.67 \pm 0.49$ \\
\noalign{\smallskip}\hline\hline
\end{tabular}
\end{table}%
The values of the seven parameters $G_0$, $D_e$, $D_m$, $F_e$, $F_m$, $\varphi$ and $|R|$ are very close to those obtained in the case of real $R$, shown in Table~\ref{tab:results.par}. The total BRs calculated using these parameters are shown in Table~\ref{tab:BRcalc.J.psi.2phase}, together with the corresponding PDG values. The purely strong, purely EM and mixed strong-EM BRs are reported in Table~\ref{tab:BRsepcalc.J.psi.2phase}.\\
The numerical minimization gives the normalized $\chi^2$
\be
\frac{\chi^2\left(\mathcal{C}^{\rm best};|R|^{\rm best},\varphi^{\rm best},\varphi_2^{\rm best}\right)}{N_{
\rm dof}} &=& 2.59\,,
\nen
where, in this case,
$$
N_{\rm dof}=N_{\rm const}-N_{\rm param}=1 \,,
$$
having, $N_{\rm const}=9$ and $N_{\rm param}=8$.\\
The obtained value for the relative phase between the purely strong sub-amplitude and the mixed strong-EM one, $\varphi_2=(206 \pm 46)^\circ$ is compatible with our first hypothesis about the reality and negativity of $R$, corresponding to a relative phase $\varphi_2$ of $180^\circ$. %
{\textcolor{\colormod}{
In this case the $p$-value, see Eq.~\eqref{eq.pvalue1}, is
$$
p(2.59;1)=0.108\,.
$$
}}%
In light of such full agreement and of the lower statistical significance we consider as our main results those obtained under the hypothesis of real $R$.
}%
}%
\subsection[Discussion]{Discussion}
We have calculated the purely strong, purely EM and mixed strong-EM contributions to the total BR, see Table~\ref{tab:BRsepcalc.J.psi}, and hence the moduli of the corresponding sub-amplitudes, for each pair of baryons, see Table~\ref{tab:BRsepcalc.J.psi.ratio} and Table~\ref{tab:simple-form}. The mixed strong-EM contribution is determined for the first time and it is proven to be crucial, in the framework of our model, for the correct description of the decay mechanism. We have obtained the relative phase between strong and EM sub-amplitudes, assuming that the strong and mixed strong-EM ones have the same phase and, finally, we have used the purely EM BRs to calculate the Born non-resonant cross sections of the annihilation processes $e^+ e^- \to B \overline B$ at the $J/\psi$ mass. The possibility of disentangling single contributions allows, for the first time, to determine the mixed strong-EM sub-amplitude for each charged $\bb$ final state. In particular, the mixed strong-EM sub-amplitude is about 10\% of the corresponding purely strong sub-amplitude, for the charged final states, while it is, as supposed, zero for the neutral ones, see third column of Table~\ref{tab:BRsepcalc.J.psi.ratio}. On the other hand the purely EM sub-amplitude is between 6\% and 25\% of the corresponding purely strong sub-amplitude, see second column of Table~\ref{tab:BRsepcalc.J.psi.ratio}. Furthermore considering the four charged final states: $p \overline p$, $\Sigma^+ \overline \Sigma{}^-$, $\Sigma^- \overline \Sigma{}^+$, $\Xi^- \overline \Xi{}^+$, for two of them, $\Sigma^- \overline \Sigma{}^+$ and $\Xi^- \overline \Xi{}^+$, the mixed strong-EM sub-amplitudes are larger than the corresponding purely EM ones, while the remaining two show an opposite trend.\\
{\textcolor{\colormod}{The hypothesis of a complex ratio $R$ has been considered and it has been shown that the resulting relative phase, $\varphi_2$, between the purely strong and the mixed strong-EM sub-amplitudes is compatible with $180^\circ$, i.e., with a real and negative value of $R$.}}\\
Finally, our prediction for the neutron cross section, see fifth row of Table~\ref{tab:sigma.gamma}, i.e.,
$$
\sigma_{\ee\to n\overline{n}} \,(M_{\jp}^2)=(6.84\pm 0.58)\,{\rm pb} \,,
$$
is in agreement with the ``natural'' expectation
\be
\sigma^{\rm expected}_{\ee\to n\overline{n}}\,(M_{\jp}^2) &=&\left(\frac{\mu_n}{\mu_p}\right)^2\!\!\! \sigma_{\ee\to p\overline{p}}\,(M_{\jp}^2) = \left(\frac{-1.913}{2.793}\right)^{\!2}(12.9\pm 1.4)\,{\rm pb} \no \\
&=& (6.1\pm 0.7)\,{\rm pb}\,,
\nen
which is obtained by scaling the proton cross section, reported in the fourth row of Table~\ref{tab:sigma.gamma}, by the square value of the ratio between neutron, $\mu_n$, and proton, $\mu_p$, magnetic moment.
%
%
\section[The decays of the $J/\psi$ and $\psi(2S)$ into $\Lambda \overline{\Lambda}$ and $\Sigma^0 \overline {\Sigma}{}^0$]{The decays of the $J/\psi$ and $\psi(2S)$ into $\Lambda \overline{\Lambda}$ and $\Sigma^0 \overline {\Sigma}{}^0$}
The decay mechanisms of the lightest charmonia can be studied almost only by means of effective models, since these decays happen at energy regimes that do not allow the use of pQCD. %
{\textcolor{\colormodfour}{
From Eq.~\eqref{eq.Lagr.dens.sig.lam} it can be seen that the Lagrangian for the particular decay to $\Lambda \overline \Lambda$ and $\Sigma^0 \overline \Sigma{}^0$ can be described in terms of only two parameters. This fact can be seen also by looking at the first two rows of Table~\ref{tab:A.par}, where the number of parameters can be reduced by replacing them with some appropriate linear combination. For this reason in this chapter we pay attention to the $\jp$ and $\psii$ charmonia, and study their decays into baryon-antibaryon pairs with $\BB = \LL$, $\Ss$.
}}
\subsection[Theoretical background]{Theoretical background}
The differential cross section of the process $\ee \to \psi \to \BB$ has the well known $\cos \theta $ dependence~\cite{Brodsky:1981kj}
\be
\frac{dN} {d \cos \theta} \propto 1 + \alpha_B \cos^2 \theta\,,
\nen
where $\alpha_B$ is the so-called polarization parameter and $\theta$ is the baryon scattering angle, i.e., the angle between the outgoing baryon and the beam direction in the $\ee$ center of mass frame.\\
In Fig.~\ref{fig.ang.distr.j} and Fig.~\ref{fig.ang.distr.2s} are shown BESIII data~\cite{Ablikim:2017tys} on the angular distributions of the four decays: $\jp \to \LL$, $\jp \to \Ss$, and $\psii \to \LL$, $\psii \to \Ss$. Only the decay $\jp \to \Ss$ has a negative polarization parameter $\alpha_B$, as was already pointed out in Ref.~\cite{Ablikim:2005cda}. {\textcolor{\colormod}{These results represent a real finding compared with the first experiments where the opposite trend on the angular distribution of the $J/\psi$ into $\LL$ and $\Ss$ was not found, see for example Ref.~\cite{Pallin:1987py}.}} %
\begin{figure}[ht!]
\begin{center}
	\includegraphics[width=0.7\columnwidth]{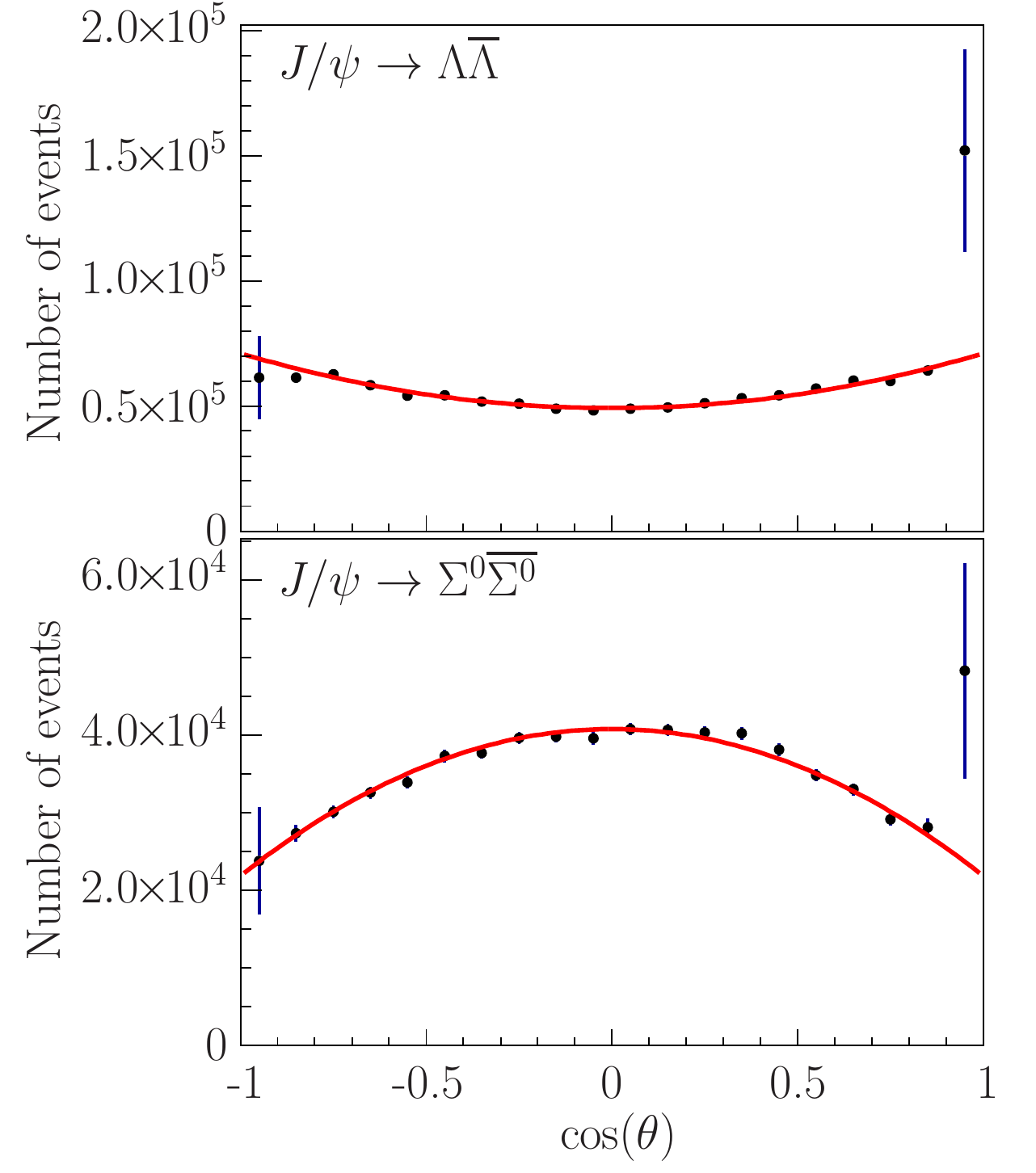}
	\caption{Angular distribution of the baryon for the $\jp$ decays into $\LL$ (upper panel) and $\Ss$ (lower panel). \label{fig.ang.distr.j}}
\end{center}
\end{figure}
\begin{figure}
\begin{center}
	\includegraphics[width=0.7\columnwidth]{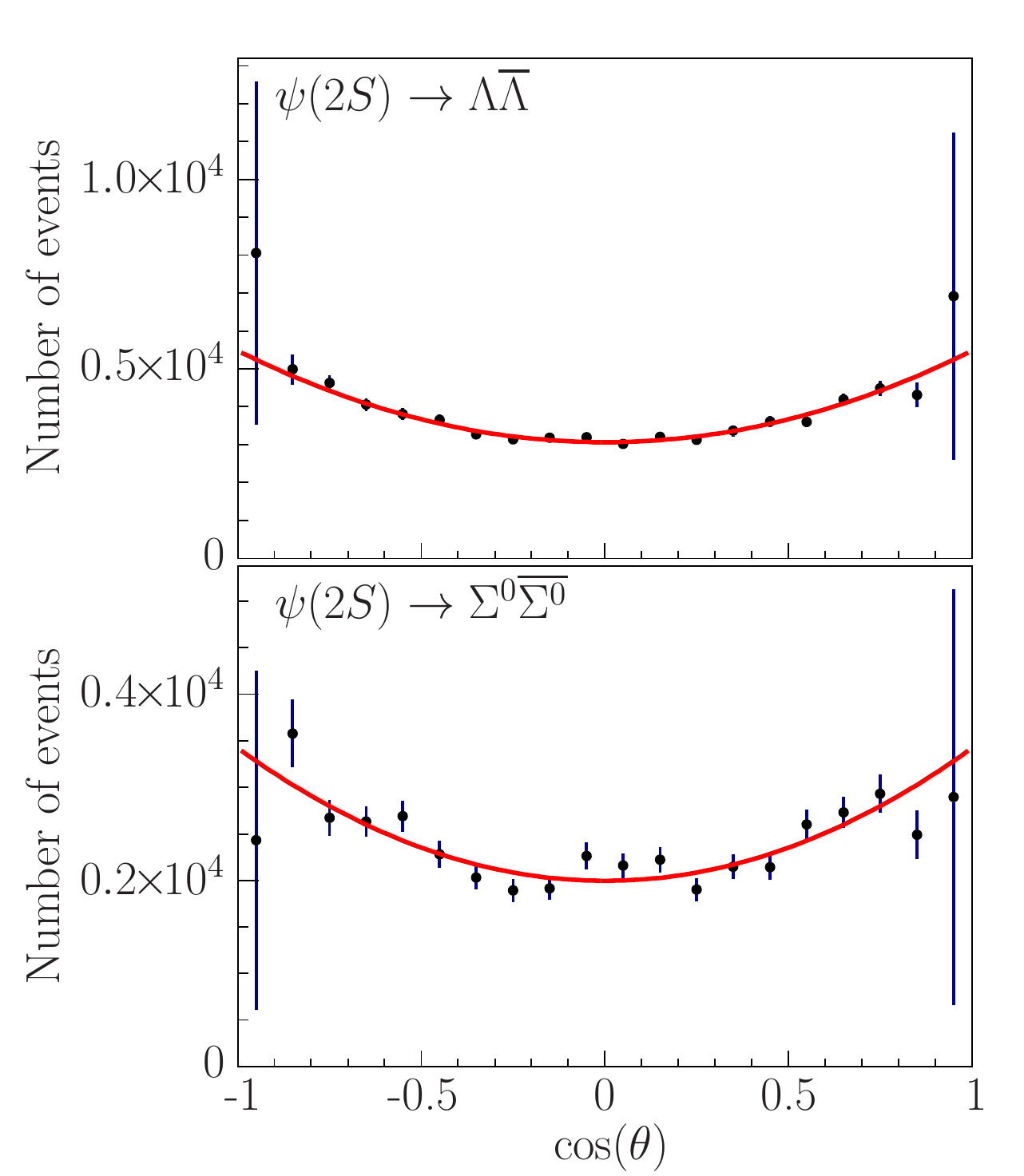}
	\caption{Angular distribution of the baryon for the $\psii$ decays into $\LL$ (upper panel) and $\Ss$ (lower panel).\label{fig.ang.distr.2s}}
\end{center}
\end{figure}\\
Starting from Eq.~\eqref{eq.dens.lagr} and, in particular, considering Eq.~\eqref{eq.Lagr.dens.sig.lam}, it is possible to extract the Lagrangians describing only the $J/\psi$ and $\psi(2S)$ decays into $\Lambda \overline \Lambda$ and $\Sigma^0 \overline \Sigma{}^0$. We can write
\be
\label{Eq.Lagr.S.L}
{\mathcal L}_{\Ss}=(G_0+G_1) \, \Ss \,,   \ \ \ \ \ \
{\mathcal L}_{\Lambda \overline \Lambda} = (G_0-G_1) \, \Lambda \overline \Lambda \,,
\en
where $G_0$ and $G_1$ are the following combinations of coupling constants
\be
G_0 = g \, , \hspace{10mm} G_1 = {d \over 3} \, (2 g_m + g_e) \,.
\nen
By using the same structure of Eq.~\eqref{eq:rate.A}, the BRs 
can be expressed in terms of electric and magnetic amplitudes as
\be
{\rm BR}_{\psi \to \Sigma^0 \overline \Sigma{}^0} = |A_E^\Sigma|^2 + |A_M^\Sigma|^2\,, \ \ \ \ \ \ \ {\rm BR}_{\psi \to \Lambda \overline \Lambda} = |A_E^\Lambda|^2 + |A_M^\Lambda|^2
\,.
\nen
We can also decompose such amplitudes as combinations of leading, $E_0$ and $M_0$, and sub-leading terms, $E_1$ and $M_1$, see Eq.~\eqref{Eq.Lagr.S.L}, with opposite relative signs, i.e.,
\be
\br_{\psi \to\Ss}&=&|E_0+E_1|^2 + |M_0+M_1|^2 = |E_0|^2+|E_1|^2+2 |E_0||E_1|\cos(\rho_E) \no \\
&+& |M_0|^2+|M_1|^2+2 |M_0||M_1|\cos(\rho_M)
\no
\,,\\
\br_{\psi \to\Lambda \overline \Lambda} &=& |E_0-E_1|^2 + |M_0-M_1|^2 = |E_0|^2+|E_1|^2-2 |E_0||E_1|\cos(\rho_E) \no \\
&+& |M_0|^2+|M_1|^2-2 |M_0||M_1|\cos(\rho_M) \,,
\nen
where $\rho_E$ and $\rho_M$ are the phases of the ratios $E_0/E_1$ and $M_0/M_1$.
\subsection[Results]{Results}
We use data from precise measurements~\cite{Ablikim:2017tys,Ablikim:2018zay} of the BRs and polarization parameters, reported in Table~\ref{tab:BRdata}.%
\begin{table}[ht!]
\centering
\caption{Branching ratios and polarization parameters from Ref.~\cite{Ablikim:2017tys}. In particular the value of $\alpha_B$ for the decay $\jp\to\LL$ is from Ref.~\cite{Ablikim:2018zay}.}
\smallskip
\label{tab:BRdata} 
\begin{tabular}{l|l|l}
\hline\hline\noalign{\smallskip}
Decay & BR & Pol. par. $\alpha_B$ \\
\noalign{\smallskip}\hline\hline\noalign{\smallskip}%
$J/\psi \to \Sigma^0 \overline \Sigma{}^0$ &\hfill$(11.64 \pm 0.04) \times 10^{-4}$ &\hfill$-0.449 \pm 0.020$ \\
\hline
$J/\psi \to \Lambda \overline \Lambda$ &\hfill$(19.43 \pm 0.03) \times 10^{-4}$ &\hfill$0.461 \pm 0.009$ \\
\hline
$\psi(2S) \to \Sigma^0 \overline \Sigma{}^0$ &\hfill$(2.44 \pm 0.03) \times 10^{-4}$ &\hfill$0.71 \pm 0.11$ \\
\hline
$\psi(2S) \to \Lambda \overline \Lambda$ &\hfill$(3.97 \pm 0.03) \times 10^{-4}$ &\hfill$0.824 \pm 0.074$ \\
\noalign{\smallskip}\hline\hline
\end{tabular}
\end{table}%
These BESIII data are in agreement with the results of other experiments~\cite{Ablikim:2012pj,Aubert:2007uf,Pedlar:2005px,Ablikim:2006aw,Dobbs:2014ifa}. We have to fix the relative phases $\rho_E$ and $\rho_M$, since we have six free parameters (four moduli and two relative phases) and only four constrains (two BRs and two polarization parameters) for each charmonium state. We find that the values $\rho_E=0$ and $\rho_M=\pi$ are phenomenologically favored by the data themselves. In fact, largely different choices would give negative, and hence unphysical, values for the moduli $|E_0|$, $|E_1|$, $|M_0|$ and $|M_1|$.\\%
\begin{table}[ht!]
\centering
\caption{Moduli of the leading and sub-leading amplitudes.}
\smallskip
\label{tab:results.par} 
\begin{tabular}{l|l|l} 
\hline\hline\noalign{\smallskip}
Ampl. & $J/\psi$  & $\psi(2S)$  \\
\noalign{\smallskip}\hline\hline\noalign{\smallskip}%
$|E_0|$ & $(2.16 \pm 0.02) \times 10^{-2}$ & $(0.42 \pm 0.07) \times 10^{-2}$ \\
\hline
$|E_1|$ & $(0.42 \pm 0.02) \times 10^{-2}$ & $(0.03 \pm 0.05) \times 10^{-2}$ \\
\hline
$|M_0|$ & $(3.15 \pm 0.02) \times 10^{-2}$ & $(1.72 \pm 0.02) \times 10^{-2}$ \\
\hline
$|M_1|$ & $(0.90 \pm 0.02) \times 10^{-2}$ & $(0.23 \pm 0.02) \times 10^{-2}$ \\
\noalign{\smallskip}\hline\hline
\end{tabular}
\end{table}%
\begin{figure}[ht!]
\begin{center}
\includegraphics[width=.7\columnwidth]{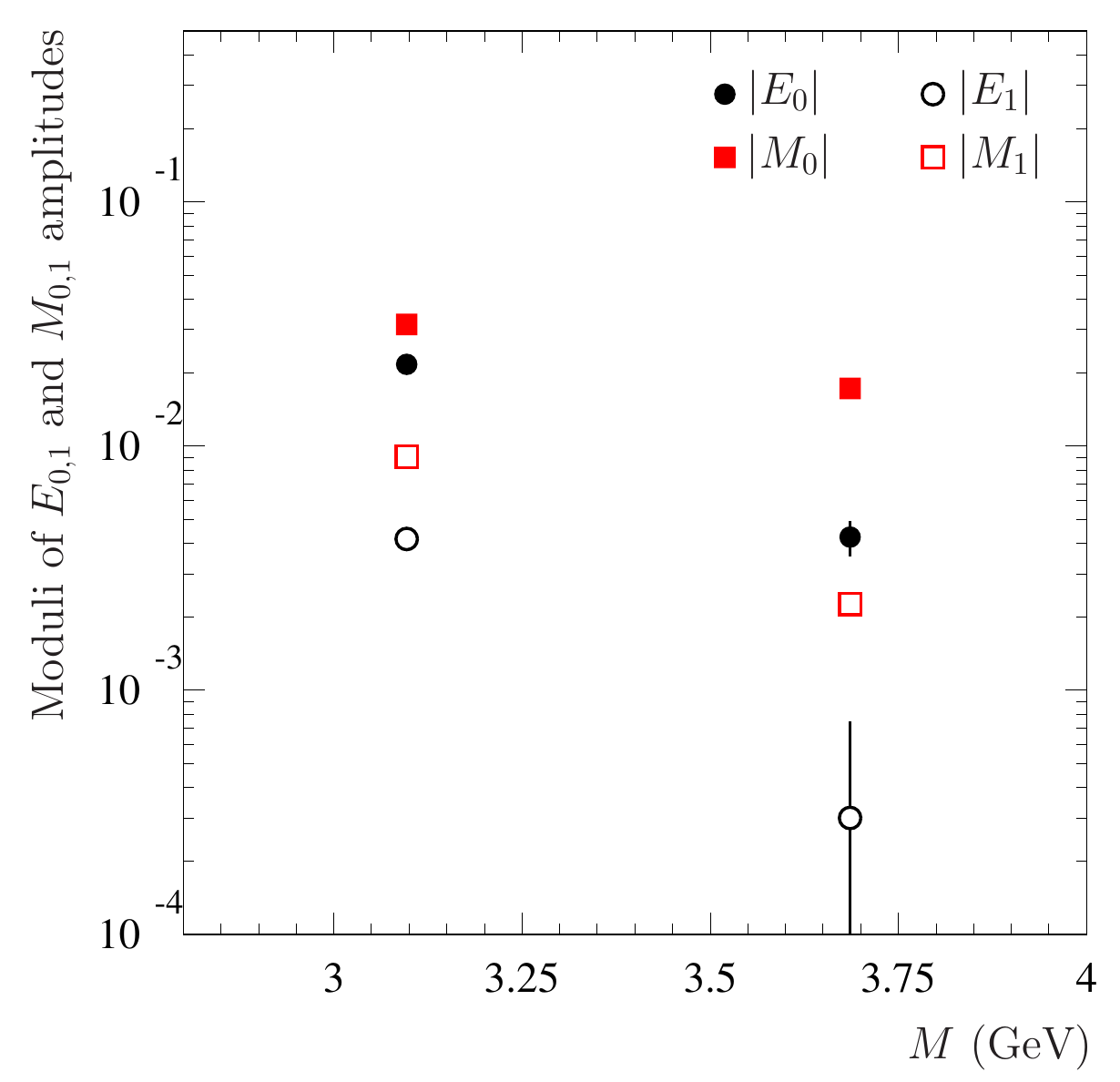}
\caption{\label{fig:EM01} Moduli of the parameters from Table~\ref{tab:results.par} as a function of the charmonium state mass $M$.}
\end{center}
\end{figure}%
\begin{table}[ht!]
\centering
\caption{Moduli of the strong Sachs FFs.}
\smallskip
\label{tab:results.parG} 
\begin{tabular}{l|l|l} 
\hline\hline\noalign{\smallskip}
FFs & $J/\psi$  & $\psi(2S)$  \\
\noalign{\smallskip}\hline\hline\noalign{\smallskip}%
$|g_E^\Sigma|$ & $(1.99 \pm 0.04) \times 10^{-3}$ &
\hfill$(0.6 \pm 0.1) \times 10^{-3}$ \\
\hline
$|g_M^\Sigma|$ &\hfill$(0.94 \pm 0.02) \times 10^{-3}$ & $(0.94 \pm 0.02) \times 10^{-3}$ \\
\hline\hline
$|g_E^\Lambda|$ & $(1.37 \pm 0.04) \times 10^{-3}$ &\hfill$(0.6 \pm 0.1) \times 10^{-3}$ \\
\hline
$|g_M^\Lambda|$ & $(1.64 \pm 0.03) \times 10^{-3}$ &\hfill$(1.20 \pm 0.02) \times 10^{-3}$ \\
\noalign{\smallskip}\hline\hline
\end{tabular}
\end{table}%
\begin{figure}[ht!]
\begin{center}
\includegraphics[width=.7\columnwidth]{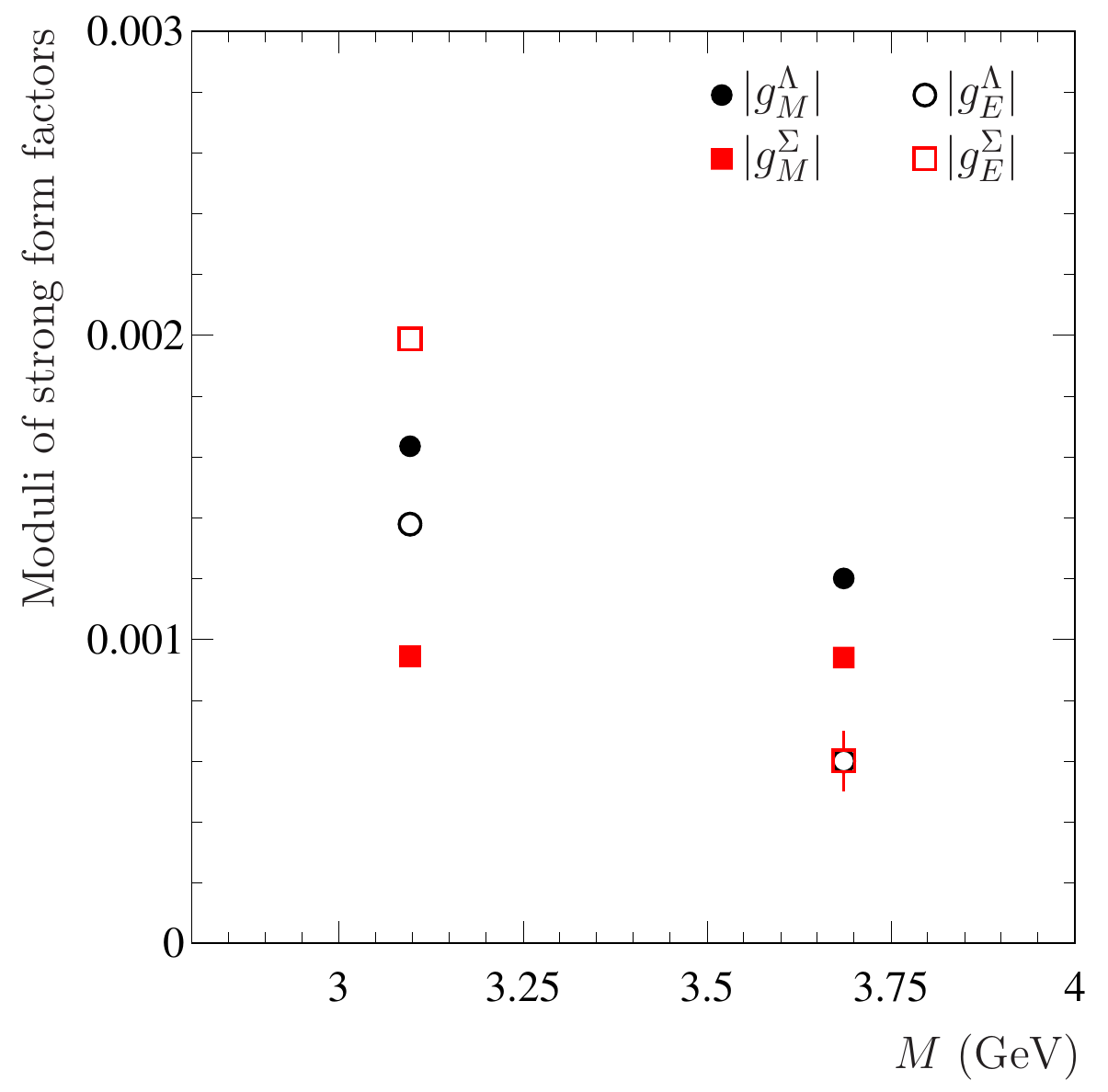}
\caption{\label{fig:gegm} Moduli of the parameters from Table~\ref{tab:results.parG} as a function of the charmonium state mass $M$.}
\end{center}
\end{figure}%
The results of the fitting procedure for $|E_0|$, $|E_1|$, $|M_0|$ and $|M_1|$ are reported in Table~\ref{tab:results.par} and shown in Fig.~\ref{fig:EM01}. We calculate the corresponding values of $|g_{E}|$, $|g_{M}|$, reported in Table~\ref{tab:results.parG} and shown in Fig.~\ref{fig:gegm}. The large sub-leading $\jp$ amplitudes $|E_1|$, $|M_1|$ (see Table~\ref{tab:results.par} and Fig.~\ref{fig:EM01}) are responsible for the inversion of the $|g_E^B|$, $|g_M^B|$ hierarchy (see the upper panel of Fig.~\ref{fig:gegm}).
Different $\Lambda$ and $\Sigma^0$ angular distributions can be explained using an effective model with the SU(3)-driven Lagrangian
\be
\mathcal{L}_{\Ss+\LL} = (G_0+G_1) \Sigma^0 \overline \Sigma{}^0 + (G_0-G_1) \Lambda \overline \Lambda \,.
\nen
The interplay between leading $G_0$ and sub-leading $G_1$ contributions to the decay amplitudes determines signs and values of polarization parameters $\alpha_B$.\\
In particular, the different behavior of the $\jp \to \Ss$ angular distribution is due to the large values of the sub-leading amplitudes $|E_1|$ and $|M_1|$. It implies that the SU(3) mass breaking and the EM effects, which are responsible for these amplitudes, play a different role in the dynamics of the $\jp$ and $\psii$ decays.\\
It is interesting to notice that a similar behavior has been observed also in the angular distributions of $\Sigma^0(1385)$ and $\Sigma^\pm(1385)$ measured by BESIII~\cite{Ablikim:2016sjb,Ablikim:2016iym}.
%
%

%



\chapter*{Conclusions} 
\addcontentsline{toc}{chapter}{Conclusions}
\chaptermark{Conclusions}
{\textcolor{\colormodtwo}{
Concerning the $\jp \to \pipi$ decay BaBar data suggests that it does not proceed only electromagnetically, i.e., $\br^{\rm PDG}\neq \br^{\gamma}$, in particular $\br^{\rm PDG}$ and $\br^{\gamma}$ differ for more than 4 standard deviations. Since the purely strong contribution is suppressed, being a $G$-parity violating decay, we explore the possibility of an unexpected large mixed strong-EM contribution, $gg \gamma$. Using a phenomenological model based on the Cutkosky rule, we calculate the imaginary part of the mixed strong-EM amplitude and hence a lower limit for the contribution to the total BR due to the $gg \gamma$ amplitude, $\bgg$. The result agrees with the hypothesis that, at least for this particular decay with the minimum pion multiplicity in the final state, the mixed strong-EM contribution to the total BR is of the same order of the purely EM one, i.e., $13\% \, \bg \leq \bgg \leq 37\% \, \bg$.\\
We study the theoretical modeling of the baryonic decays of the $J/\psi$ meson, a subject that was modestly investigated about forty years ago and that is being revisited in the light of new data, especially from BESIII. In particular we consider the $\jp \to B \overline B$ decays, being $B$ a spin-1/2 baryon of the SU(3) octet. The model is based on a hadronic Lagrangian with a SU(3)-flavor symmetry that is broken by specific sub-amplitudes of the process. These sub-amplitudes are identified with purely strong and purely EM Feynman diagrams and parameterized by free parameters. The model is completed with a mixed strong-EM sub-amplitude which is parameterized by the ratio $R$ between the mixed and the purely strong sub-amplitudes, only for charged baryons, and with a relative phase between the EM and the strong sub-amplitudes. The obtained relative phase, $\varphi=(73 \pm 8)^\circ$, is compatible with other results~\cite{Ablikim:2012bw,Wei:2009zzh}. For these decays we separate, for the first time, the strong, the EM and the mixed strong-EM contributions to the total BR and we give a prediction for ${\rm BR}_{\Sigma^- \overline \Sigma{}^+}$, see Eq.~\eqref{eq.pred.BR.sigm-}, the only BR not yet measured and some $\ee \to \BB$ cross sections at the $\jp$ mass. In particular the ${\rm BR}_{\Sigma^+ \overline \Sigma{}^-}$, see Eq.~\eqref{eq.pred.BR.sigm+}, is predicted to be smaller than the corresponding PDG value. This fact is confirmed by a new independent preliminary analysis~\cite{Liangsigma}. Finally we find that at the energy $\sqrt{q^2} \sim M_{\jp}$ the regime of QCD is not completely perturbative, in fact we obtain that the ratio $R$ between the mixed strong-EM amplitude and the purely strong one is compatible with the reality hypothesis but is different from its perturbative QCD prevision, $R \sim -0.097 \neq R_{\rm pQCD} \sim -0.030$. %
{\textcolor{\colormodfour}{This fact can be seen also by looking at the behavior of the proton FF at the $J/\psi$ mass, with a trend different from the power-law dependence predicted by perturbative QCD~\cite{TomasiGustafsson:2001za}. Moreover it is possible to explore higher energy values range, with $\sqrt{q^2} > M_{J/\psi}$, by studying particles heavier than the $J/\psi$ meson, such as the $\psi(2S)$ one. In future more data will be available for further investigations, especially from the BESIII experiment that has just reached the World's largest $J/\psi$ data sample.}}\\
Finally, in the case of the $\jp$ and $\psi(2S)$ decays into $\Lambda \overline \Lambda$ and $\Sigma^0 \overline \Sigma{}^0$, the angular distributions, measured by BESIII, show different behaviors. Such a difference can be explained in the framework of a model with an effective Lagrangian $\mathcal{L}_{\Sigma \Lambda} = (G_0+G_1) \Sigma^0 \overline \Sigma{}^0 + (G_0-G_1) \Lambda \overline \Lambda$. We show that the interplay between dominant $G_0$ and sub-dominant $G_1$ contributions to the decay amplitudes is related to signs and values of the different polarization parameters $\alpha$. 
}}


\appendix
\chapter{Notations and experimental data}\label{app:notazioni}
\section{Notations}
In this thesis for the numerical values we consider an error with two significant figures in light of further manipulations.\\
We use also the following notations
\begin{center}
QED $ \ \ \ \ \ \longleftrightarrow \ \ \ \ \ $ {\it Quantum Electrodynamics}\\
QCD $ \ \ \ \ \ \longleftrightarrow \ \ \ \ \ $ {\it Quantum Chromodynamics}\\
pQCD $ \ \ \ \ \ \longleftrightarrow \ \ \ \ \ $ {\it Perturbative QCD}\\
BR $ \ \ \ \ \ \longleftrightarrow \ \ \ \ \ $ {\it Branching Ratio}\\
PS $ \ \ \ \ \ \longleftrightarrow \ \ \ \ \ $ {\it Phase Space}\\
CM $ \ \ \ \ \ \longleftrightarrow \ \ \ \ \ $ {\it Center of Mass}\\
FF $ \ \ \ \ \ \longleftrightarrow \ \ \ \ \ $ {\it Form Factor}\\
\end{center}
We use the so-called natural units
$$
\hbar = c = 1 \,,
$$
so that
$$
[\mbox{length}] = \rm eV^{-1} \, , \ \ \ \ \ [\mbox{time}] = \rm eV^{-1} \, , \ \ \ \ \ [\mbox{speed}] = 1 \, , \ \ \ \ \ [\mbox{energy}] = \rm eV \, ,
$$
$$
[\mbox{mass}] = \rm eV \, , \ \ \ \ \ [\mbox{momentum}] = \rm eV \, , \ \ \ \ \ [\mbox{action}] = 1 \, , \ \ \ \ \ [\mbox{surface}] = \rm eV^{-2}
$$
We use the following Minkowski metric tensor (flat space-time) 
$$
\eta^{\mu \nu} =
\begin{pmatrix} 
1 & 0 & 0 & 0 \\
0 & -1 & 0 & 0 \\
0 & 0 & -1 & 0 \\
0 & 0 & 0 & -1 \\
\end{pmatrix} \,,
$$\\
so that, for example,
$$
x^\mu = (x^0, \vec x) = (x^0,x^1,x^2,x^3) \, ,
$$
$$
x_\mu =\eta_{\mu \nu}x^\nu= (x^0, -\vec x) = (x^0,-x^1,-x^2,-x^3) \, ,
$$
$$
x \cdot y = \eta_{\mu \nu} x^\mu y^\nu = x^\mu y_\mu = x_\mu y^\mu = x^0 y^0 - \vec x \cdot \vec y = x^0 y^0 - x^1 y^1 - x^2 y^2 - x^3 y^3 \,.
$$
We adopt also the following notations
$$
\partial_\mu \equiv {\partial \over \partial x^\mu} \, , \ \ \ \ \ \ \ \ \partial^\mu \equiv {\partial \over \partial x_\mu} \, , \ \ \ \ \ \ \ \ \partial^2 \equiv \square \equiv \partial^\mu \partial_\mu \,,
$$
\section{Experimental data}\label{app:cost.udm.pdg}
In this appendix we report some experimental data, constants and values from PDG~\cite{\pdg}. In particular in Tables~\ref{tab.app.data1} we show masses and quantum numbers for some particles.
\begin{table}[!htb]
    \caption{Data of some particles from PDG~\cite{\pdg}.\label{tab.app.data1}}
    \begin{minipage}{.5\linewidth}
      \centering
\begin{tabular}{c|c|c}
\hline\hline\noalign{\smallskip}
Lepton & Mass (MeV) \\
\noalign{\smallskip}\hline\hline\noalign{\smallskip}%
$e^-$ & $0.510998928 \pm 0.000000011$ \\ \hline
$\mu^-$ & $105.6583715 \pm 0.0000035$ \\ \hline
$\tau^-$ & $1776.86 \pm 0.12$ \\
\noalign{\smallskip}\hline\hline
\end{tabular}
    \end{minipage}%
    \begin{minipage}{.5\linewidth}
      \centering
\begin{tabular}{c|c|c}
\hline\hline\noalign{\smallskip}
Meson & $I^G(J^{PC})$ & Mass (MeV) \\
\noalign{\smallskip}\hline\hline\noalign{\smallskip}%
$\pi^{\pm}$ & $1^-(0^{-})$ & $139.57018 \pm 0.00035$ \\ \hline
$\eta$ & $0^+(0^{-+})$ & $547.862 \pm 0.017$ \\ \hline
$\eta'$ & $0^+(0^{-+})$ & $957.78 \pm 0.006$ \\ \hline
$\rho^0$ & $1^+(1^{--})$ & $775.26 \pm 0.25$ \\ \hline
$f_1$ & $0^+(1^{++})$ & $1281.9 \pm 0.5$ \\
\noalign{\smallskip}\hline\hline
\end{tabular}
    \end{minipage}\\
    \centering
    \smallskip
    \begin{minipage}{.5\linewidth}
      \centering
\begin{tabular}{c|c|c}
\hline\hline\noalign{\smallskip}
Baryon & $I(J^{P})$ & Mass (MeV) \\
\noalign{\smallskip}\hline\hline\noalign{\smallskip}%
$p$ & ${1 \over 2}({1 \over 2}^{+})$ & $938.272081 \pm 0.000006$ \\ \hline
$n$ & ${1 \over 2}({1 \over 2}^{+})$ & $939.565413 \pm 0.000006$ \\ \hline
$\Sigma^+$ & $1({1 \over 2}^{+})$ & $1189.37 \pm 0.07$ \\ \hline
$\Sigma^0$ & $1({1 \over 2}^{+})$ & $1192.642 \pm 0.024$ \\ \hline
$\Sigma^-$ & $1({1 \over 2}^{+})$ & $1197.449 \pm 0.030$ \\ \hline
$\Lambda$ & $0({1 \over 2}^{+})$ & $1115.683 \pm 0.006$ \\ \hline
$\Xi^-$ & ${1 \over 2}({1 \over 2}^{+})$ & $1314.86 \pm 0.20$ \\ \hline
$\Xi^0$ & ${1 \over 2}({1 \over 2}^{+})$ & $1321.71 \pm 0.07$ \\
\noalign{\smallskip}\hline\hline
\end{tabular}   
    \end{minipage}
\end{table}
$$
\alpha = {1 \over 137.035 \, 999 \, 074(44)} = 7.297 \, 352 \, 5698(24) \times 10^{-3} \, .
$$
$$
1 = \hbar c = 197.326 \, 9718(44) \ \rm MeV \, fm  \, ,
$$
$$
1 \ {\rm GeV} = 5.06773094(11) \ \rm fm^{-1} \, ,
$$
$$
1 \ {\rm GeV^2} = 25.6818969(11) \ \rm fm^{-2} \, ,
$$
$$
1 \ {\rm GeV^3} = 130.1489434(85) \ \rm fm^{-3} \, .
$$
In the International System of units we have the following expressions
$$
\mbox{elementary electric charge:} \ \ \ \ \ \ \ \ e = 1.602 \ 176 \ 565(35) \times 10^{-19} \ \rm C \, ,
$$
$$
\mbox{reduced Planck constant:}  \ \ \ \ \ \ \ \ \hbar = 1.054 \ 571 \ 726(47) \times 10^{-34} \ {\rm J \, s} \, ,
$$
$$
\mbox{speed of light in vacuum:}  \ \ \ \ \ \ \ \ c = 2.997 \ 924 \ 58 \times 10^{8} \ \rm m \, s^{-1} \, .
$$
\chapter{Decay width and branching ratio}\label{app:BRDW}
Concerning a particular decay of the $J/\psi$ meson there are two important quantities: its decay width ($\Gamma$) and the corresponding BR. These are related to Feynman total amplitude ($\mathcal A$) and the $n$-body phase space ($d\rho_n$) of the decay.
\section[Phase space]{Phase space}
The phase space for a $n$-body decay is
\be
\label{eq.PSn}
d\rho_n(P;p_1,...,p_n) = (2\pi)^4 \int \delta^4 \left(\sum_{i=1}^n p_i - P \right) \prod_{i=1}^n {d^3 p_i \over (2\pi)^3 2 E_i} \, ,
\en
where $P$ is the momentum of the decaying particle and $p_1,...,p_n$ and $E_1,...,E_n$ are, respectively, the momenta and the energies of the final state particles. Being a Lorentz invariant quantity it is possible to calculate it in the CM system where $P^\mu = (\sqrt s, 0,0,0)$, with $\sqrt s = M$, being $M$ the mass of the decaying particle. Using this formula we have the following particular cases:
\be
d\rho_2(P;p_1,p_2) &=& (2\pi)^4 \int \delta^4 \left(p_1+p_2-P \right) {d^3 p_1 \over (2\pi)^3 2 E_1} {d^3 p_2 \over (2\pi)^3 2 E_2} \notag \\
&=& {1 \over (4\pi)^2} \int \delta(E_1+E_2-M) \, \delta^3 \left(\vec p_1 + \vec p_2 \right) {d^3 p_1 \over E_1} {d^3 p_2 \over E_2} \,, \notag
\en
from which, using the masses of the two final state particles $m_1$ and $m_2$,
\be
d\rho_2(M;m_1,m_2)= {d\Omega \over 32 \pi^2} \sqrt{1 + {\left(m_1^2-m_2^2\right)^2 \over M^4} - {2(m_1^2+m_2^2) \over M^2}} \, ,
\nen
where $p_{1,2}=(E_{1,2}, \vec p_{1,2})$ and $\vec p_1 = -\vec p_2$, with the following expressions for the energies
\be
E_1={M \over 2} \left( 1+{m_1^2 \over M^2}-{m_2^2 \over M^2} \right) \, , \ \ \ \ \ \ E_2={M \over 2} \left( 1+{m_2^2 \over M^2}-{m_1^2 \over M^2} \right) \,.
\nen
Moreover in the case of two particle in the final state with the same mass $m \equiv m_1=m_2$, we put $p \equiv |\vec p_1| = |\vec p_2|$ and the phase space become
\be
d\rho_2(M;m,m)= {|\vec p| \over M}{d\Omega \over 16 \pi^2} \,, \ \ \ \ \ |\vec p| = {M \over 2} \beta\,, \ \ \ \ \ \beta = \sqrt{1-{4m^2 \over M^2}}\,,
\nen
from which
\be
\label{eq.PS2m}
d\rho_2(M;m,m)= {\beta \over 32 \pi^2}\,d\Omega \,.
\en
For a three-body decay
$$
d \rho_3 = \int {d^3 k_1 \over (2 \pi)^3 2E_1} {d^3 k_2 \over (2 \pi)^3 2E_2} {d^3 k_3 \over (2 \pi)^3 2E_3} (2 \pi)^4 \delta^4 \big(P - p_1 - p_2 - p_3 \big) \,,
$$
where $p_1,p_2,p_3$ are the four-momenta and $E_1,E_2,E_3$ the energies of the particles of the final state, in the CM frame, and $P$ is the total four-momenta. \\
For the decay of the $J/\psi$ meson into three particles of masses $m_1,m_2,m_3$ we can use the following expression~\cite{\pdg} (valid in the case of a mean over spin states)
$$
d\rho_3 = {1 \over (2\pi)^3} {1 \over 16 M_{\jp}^2} dp_{12}^2 dp_{23}^2 \,,
$$
where $M_{\jp}$ is the $\jp$ mass, $p_{ij} \equiv p_i+p_j$ and with the following limits on $p_{23}^2$ and $p_{12}^2$
\be
(p_{23}^2)_{{\small \mbox{min}}} &=& {\left(M_{\jp}^2-m_1^2+m_2^2-m_3^2\right)^2 \over 4 p_{12}^2} - \Bigg( \sqrt{{(p_{12}^2-m_1^2+m_2^2)^2 \over 4 p_{12}^2} -m_2^2} \no \\
&+& \sqrt{{(M_{\jp}^2-p_{12}^2-m_3^2)^2 \over 4 p_{12}^2} -m_3^2} \Bigg)^2 \,,
\nen
\be
(p_{23}^2)_{{\small \mbox{max}}} &=& {\left(M_{\jp}^2-m_1^2+m_2^2-m_3^2\right)^2 \over 4 p_{12}^2} - \Bigg( \sqrt{{(p_{12}^2-m_1^2+m_2^2)^2 \over 4 p_{12}^2} -m_2^2} \no \\
&-& \sqrt{{(M_{\jp}^2-p_{12}^2-m_3^2)^2 \over 4 p_{12}^2} -m_3^2} \Bigg)^2 \,,
\nen
$$
(p_{12}^2)_{{\small \mbox{min}}} = (m_1+m_2)^2 \, , \ \ \ \ \ \ \ \ (p_{12}^2)_{{\small \mbox{max}}} = (M_{\jp}-m_3)^2 \,.
$$
In the case of a decay into three massless particles ($m_1=m_2=m_3=0$) the previous results become
\begin{equation}
\label{eq.drho3.sp.fasi.3.inE1E2}
\int d \rho_3 = {1 \over 32 \pi^3} \int_0^{M_{\jp}} dE_1 \int_{M_{\jp}-E_1}^{M_{\jp}} dE_2 \,,
\end{equation}
with $E_1+E_2+E_3 = M_{\jp}$.
%
\section[Decay width]{Decay width}
The decay width for the decay of a particle of mass $M$ into $n$ particles can be calculated in its CM system and has the form
\be
\Gamma(M \to n) = {1 \over 2M} \int d\rho_n \, \overline{|\mathcal A (M \to n)|^2} \,,
\nen
where $\mathcal A (M \to n)$ is the Feynman amplitude of the decay. In particular for the decay of a bound state, as the $J/\psi$ meson, we can calculate the decay width using two equivalent approach. Consider the case of the $J/\psi$ meson ($c \overline c$ bound state) into a generic final state $|f\rangle$ of $n$ particles. We can write
\be
\label{eq.DW.cmfin}
\Gamma\big(\jp \to |f\rangle\big) = {1 \over 2 M_{\jp}} \int d \rho_n \, \overline{\big|\mathcal A\big(\jp \to |f\rangle\big)\big|^2} \,,
\en
by using directly the decay amplitude of $J/\psi \to |f\rangle$ or
\be
\label{eq.DW.J.ccfin}
\Gamma\big(\jp \to |f\rangle\big) = {4|\psi_{J/\psi}(0)|^2 \over M_{\jp}^2} \int d \rho_n \, \overline{\big| \mathcal A_{\sqrt{s} = M_{\jp}}\big(c \overline c \to |f\rangle\big) \big|^2} \, .
\en
by using the amplitude of the scattering process $c \overline c \to |f\rangle$ and the absolute value of the radial wave function of the $\jp$ at the origin $|\psi_{J/\psi}(0)|$ (related to probability that $c$ and $\overline c$ are at the origin). For a two-body decay into particles of the same mass $m$ we have
\be
\label{eq.DW.cm.2mfin}
\Gamma(\jp \to m,m) = {\beta \over 64 \pi^2 M_{\jp}} \int d\Omega \, \overline{\big|\mathcal A(\jp \to m,m)\big|^2} \,,
\en
\be
\label{eq.DW.J.cc2mfin}
\Gamma(\jp \to m,m) = {|\psi_{J/\psi}(0)|^2 \beta \over 8 \pi^2 M_{\jp}^2} \int d\Omega \, \overline{\big| \mathcal A_{\sqrt{s} = M_{\jp}}(c \overline c \to m,m) \big|^2}
\en
and
\be
\label{eq.Ampl.cc.J.conf}
\big|\mathcal A(\jp \to m,m)\big|^2 = {8|\psi_{J/\psi}(0)|^2 \over M_{\jp}} \big| \mathcal A_{\sqrt{s} = M_{\jp}}(c \overline c \to m,m) \big|^2 \,.
\en
%
\section[Branching ratio]{Branching ratio}
The BR for a general decay of a particle of mass $M$ into a generic final state $|f\rangle$ is defined as
$$
\br\big(M \to |f\rangle\big) = {\Gamma\big(M \to |f\rangle\big) \over \Gamma_M}
$$
where $\Gamma_M$ is the total decay width of the decaying particle, i.e., the sum of all its decay widths. In the case of the $\jp$ meson, using Eq.~\eqref{eq.DW.cm.2mfin}, we have the following useful result for the decay into two particle with the same mass $m$
\be
\label{eq.BR.cm.2mfin}
\br(\jp \to m,m) = {\beta \over 64 \pi^2 M_{\jp} \Gamma_{\jp}} \int d\Omega \, \overline{\big|\mathcal A(\jp \to m,m)\big|^2} \,.
\en

\listoffigures

\listoftables

\chapter*{Acknowledgements} 
\addcontentsline{toc}{chapter}{Acknowledgements}
\chaptermark{Acknowledgements}
I would like to warmly thank my mentor and advisor, Prof.~Simone Pacetti for guiding and supporting me over many years.\\
I would like to thank my advisor, Prof.~Livio Fanò for all the support in these years.\\
I would like to thank my girlfriend Dr.~Claudia Meazzini, for all her love and support.\\
I would like to thank my family for the support I have gotten over the years, especially my parents Alba and Giovanni.\\
I would like to thank all my colleagues and friends, the professors and researchers of the Physics and Geology Department and INFN section of Perugia, the BESIII collaboration members, especially the Italian group, for the time devoted to discussing physics and not.\\
I would also like to thank: Prof.~Rinaldo Baldini Ferroli, for his precious support, Dr.~Giulio Mezzadri for all useful discussions, and Dr.~Ilaria Balossino, Prof.~Monica Bertani, Dr.~Diego Bettoni, Dr.~Gianluigi Cibinetto, Dr.~Francesca De~Mori, Dr.~Marco Destefanis, Dr.~Riccardo Farinelli, Dr.~Isabella Garzia, Dr.~Michela Greco, Dr.~Lia Lavezzi, Prof.~Marco Maggiora, Prof.~Simonetta Marcello, Dr.~Stefano Spataro, Prof.~Egle Tomasi-Gustafsson, Dr.~Liang Yan, Dr.~Kai Zhu.

\end{document}